\documentclass[11pt,a4paper]{article} % JHEP
\pdfoutput=1

\usepackage{jheppub}                  % JHEP

\usepackage{graphicx}
\usepackage[T1]{fontenc}
\usepackage[normalem]{ulem}	% Part of the standard distribution

\newcommand{\blue}[1]{\color{blue} #1 \color{black}}

\def\lsim{\raise0.3ex\hbox{$\;<$\kern-0.75em\raise-1.1ex
\hbox{$\sim\;$}}}
\def\gsim{\raise0.3ex\hbox{$\;>$\kern-0.75em\raise-1.1ex
\hbox{$\sim\;$}}}

\title{Non-unitary evolution of neutrinos in matter and the leptonic unitarity test \\  
}
\author{Chee Sheng Fong$^{1}$}
\author{Hisakazu Minakata$^{2,3}$}
\author{Hiroshi Nunokawa$^{4}$}
\affiliation{
$^1$Instituto de F\'{\i}sica, Universidade de S\~ao Paulo, C.\ P.\
66.318, 05315-970 S\~ao Paulo, Brazil \\
$^2$Instituto F\'{\i}sica Te\'{o}rica, UAM/CSIC, Calle Nicol\'as Cabrera 13-15, Cantoblanco E-28049 Madrid, Spain \\
$^3$Research Center for Cosmic Neutrinos, Institute for Cosmic Ray Research, University of Tokyo, Kashiwa, Chiba 277-8582, Japan \\ 
$^4$Departamento de F\'{\i}sica, Pontif{\'\i}cia Universidade Cat{\'o}lica 
do Rio de Janeiro, C. P. 38097, 22451-900, Rio de Janeiro, Brazil  \\
}
\emailAdd{fong@if.usp.br}
\emailAdd{hisakazu.minakata@gmail.com}
\emailAdd{nunokawa@puc-rio.br}
\date{\today}

\abstract{ 
We present a comprehensive study of the three-active plus $N$ sterile neutrino model as a framework for constraining leptonic unitarity violation induced at energy scales much lower than the electroweak scale. We formulate a perturbation theory with expansion in small unitarity violating matrix element $W$ while keeping (non-$W$ suppressed) matter effect to all orders. 
We show that under the same condition of sterile state masses $0.1\, \text{eV}^2 \lsim m^2_{J} \lsim (1-10)\, \text{GeV}^2$ as in vacuum, assuming typical accelerator based long-baseline neutrino oscillation experiment, one can derive a very simple form of the oscillation probability which consists only of zeroth-order terms with the unique exception of probability leaking term $\mathcal{C}_{\alpha \beta}$ of $\mathcal{O} (W^4)$. We argue, based on our explicit computation to fourth-order in $W$, that all the other terms are negligibly small after taking into account the suppression due to the mass condition for sterile states, rendering the oscillation probability {\em sterile-sector model independent}.
Then, we identify a limited energy region in which this suppression is evaded and the effects of order $W^2$ corrections may be observable. Its detection would provide another way, in addition to detecting $\mathcal{C}_{\alpha \beta}$, to distinguish between low-scale and high-scale unitarity violation. We also solve analytically the zeroth-order system in matter with uniform density to provide a basis for numerical evaluation of non-unitary neutrino evolution. 

}

\begin{document} % JHEP 

\begin{flushright}
IFT-UAM/CSIC-17-117
\end{flushright}

\maketitle

%\newpage

\section{Introduction}
\label{sec:introduction}

Studies of neutrino oscillation entered into a ``matured phase'' after
the structure of the three-flavour lepton mixing \cite{Maki:1962mu} is
elucidated. The long-lasted discovery phase of neutrino oscillation has
been unambiguously concluded by the Super-Kamiokande (Super-K)
atmospheric neutrino observation which discovered neutrino oscillation
and hence neutrino mass \cite{Fukuda:1998mi}. It was followed by the
KamLAND reactor and the solar neutrino experiments which uncovered the
three-flavour nature of the mixing by observing oscillation 
and/or adiabatic flavour conversion of neutrinos in matter \cite{Mikheev:1986gs,Wolfenstein:1977ue} in the 1-2 sector \cite{Eguchi:2002dm,Ahmad:2002jz}.\footnote{
%%%%%%%%%%%%%% footnote %%%%%%%%%%%%%%
The unique citation of the solar neutrino measurement here must be understood as the representative of all the foregoing solar neutrino experiments \cite{Cleveland:1998nv,Hirata:1991ub,Hampel:1998xg,Abdurashitov:2002nt,Fukuda:2001nj,Aharmim:2011vm}. 
}
The last step of understanding the three-flavour structure of neutrino oscillation was carried out by the reactor \cite{An:2016ses,RENO:2015ksa,Schoppmann:2016iww} and the accelerator \cite{Abe:2017vif,Adamson:2016tbq} measurement of $\theta_{13}$. It lefts only the two remaining unknowns in the standard three-flavour mixing paradigm, that is, measurement of CP violating phase and determination of neutrino mass ordering.\footnote{
%%%%%%%%%%%%%% footnote %%%%%%%%%%%%%%
Recently, however, there exists accumulating indication that CP phase $\delta$ takes value around $\sim\frac{3\pi}{2}$ \cite{Hartz-KEK-colloquium}.
}

The completion of the theory of the three-flavour neutrino mixing,
however, necessitates the paradigm test. A well-known example of such
efforts is to verify unitarity of the quark CKM matrix~\cite{Olive:2016xmw}. We have argued in ref.~\cite{Fong:2016yyh} that we
may need a different strategy to test leptonic unitarity. That is, first
prepare a generic framework which describes unitarity violation at
certain energy scale, and then confront it to experimental data. We
contrasted the two typical alternatives, unitarity violation by new
physics at high ($E \gg m_{W}$) and low ($E \ll m_{W}$) energy scales,
which are dubbed as high-scale and low-scale unitarity violation, respectively.
They differ in certain characteristic features, such as absence (low-scale) or presence (high-scale) of violation of flavour universality and zero-distance flavour transition.

The scenario of high-scale unitarity violation, based on the orthodox view of new physics at high-energy scales, has been studied extensively in the literature \cite{Antusch:2006vwa,FernandezMartinez:2007ms,Antusch:2009pm,Antusch:2009gn,Antusch:2014woa,Escrihuela:2015wra,Fernandez-Martinez:2016lgt,Blennow:2016jkn,Escrihuela:2016ube}.\footnote{
	%%%%%%%%%%%%%% footnote %%%%%%%%%%%%%%
	Works have also been done on unitarity violation by sterile sector from somewhat different point of view, e.g.,  if it exists, how it could disturb measurement of lepton Kobayashi-Maskawa phase $\delta_{\rm CP}$, or mass ordering. See for example, \cite{Klop:2014ima,Gandhi:2015xza,Agarwalla:2016mrc,Miranda:2016wdr,Ge:2016xya,Abe:2017jit,Dutta:2016vcc,Dutta:2016czj,Pas:2016qbg,Rout:2017udo}. 
}
On the other hand, there exist a good amount of activities in recent years hinting the possibilities of new physics at low energies. For scenarios which involve light sterile neutrino(s), see, e.g., \cite{Nelson:2007yq,Pospelov:2012gm,Harnik:2012ni}, and the references therein. 
In the former case, due to the preserved $SU(2) \times U(1)$ symmetry at high scales, it is conceivable that the constraints from measurements using probes in the charged lepton sector play a dominant role. On the other hand, in the case of low-scale unitarity violation, neutrino oscillation experiments will play key role in constraining unitarity violation. 

In a previous paper \cite{Fong:2016yyh}, we have proposed a model-independent framework for testing low-scale unitarity violation. It is based on the three active plus $N$ sterile lepton (called neutrino) system, which is unitary in the whole $(3+N)$ dimensional state space but restriction to observables in the active neutrino subspace renders the theory non-unitary in that subspace. It is referred to as the ``$(3+N)$ space unitary model''. 
We have shown in the context of accelerator and reactor neutrino measurement that the restriction on the masses of sterile states\footnote{
%%%%%%%%%%%%% footnote %%%%%%%%%%%%%% 
To be more precise, ``masses of sterile states'' implies hereafter masses of neutrinos which are mostly sterile. In this paper, for brevity, we use this simplified terminology in most of the places.} 
to $0.1 \,\text{eV}^2 \lsim m^2_{J} \lsim 1 \,\text{MeV}^2$ (with $J$ being sterile state index) is sufficient to make the observables \emph{sterile-sector model independent}. That is,
the neutrino oscillation probability can be written
in such a way that it is independent of details of 
the sterile neutrino mass spectrum and mixing with active neutrinos. The model-independent nature of the framework will be translated into that of the constraints obtained, thereby making leptonic unitarity test more powerful.

As an outcome of our formulation we have pointed out a new way of distinguishing low-scale unitarity violation from high-scale one by observing the probability leaking term in the oscillation probability. The term signals existence of energetically accessible sterile states, which is characteristic to low-scale unitarity violation, and it has been included for the first time in the analysis of unitarity violation in \cite{Fong:2016yyh} which uses a JUNO~\cite{JUNO}-like setting.
See refs.~\cite{Parke:2015goa,Blennow:2016jkn} for a comprehensive analysis of the currently available neutrino data with the active plus sterile framework, and \cite{Blennow:2016jkn,Tang:2017khg} for analyses of the future experiments. 

In this paper, we give a comprehensive treatment of the $(3+N)$ space unitary model. We formulate a \emph{novel} perturbative framework with small unitarity violating matrix element $W$ as the unique perturbing parameter, which we call ``{\em small unitarity-violation perturbation theory}''. It allows us to calculate the oscillation probability in the
presence of matter effect comparable in size to 
the vacuum mixing effect.
It must be remarked that the sterile sector model-independent nature of the $(3+N)$ space unitary model is demonstrated in 
ref.~\cite{Fong:2016yyh} only in vacuum and in matter to first order in matter perturbation theory. 
Hence, the first goal of this paper is to show that the model-independence holds after inclusion of sizeable matter effect. In fact, we observe that the same condition on the sterile neutrino masses guarantees this property, and the extremely simple expressions of the oscillation probabilities result even with our computation to fourth order. 

The second goal of this paper is to utilize the oscillation probability formulas to uncover in which region of energies and baselines unitarity violating effect is large, and to examine the possibility of sizeable $W^2$ corrections which distinguishes between high- and low-scale unitarity violation. 
These exercises may be useful in the application of our framework to some of the ongoing and next generation neutrino oscillation experiments \cite{Abe:2017vif,Adamson:2016tbq,Abe:2017aap,Collaboration:2011ym,Abe:2015zbg,Abe:2016ero,Acciarri:2015uup,TheIceCube-Gen2:2016cap,Adrian-Martinez:2016zzs}. To carry it out, we derive an exact expression of the oscillation probability in leading order in perturbation theory for uniform matter density. In summary, the framework can be used in dual modes: It serves (1) as a suitable framework for leptonic unitarity test in neutrino oscillation experiments, and (2) as a hunting tool for unitarity violation effects, which could serve for another way of distinguishing low-scale unitarity violation from high-scale one. 

\section{Essence of the present and the previous papers}
\label{sec:essence}

In this section, we present essence of the present and the previous~\cite{Fong:2016yyh} papers, in which an adequate formulation is given to describe neutrino oscillations with unitarity violation caused by new physics at energies much lower than $m_{W}$. 
In section~\ref{sec:3+N-system} we define the system, section~\ref{sec:nonunitarity-vacuum-small-matter} serves for reviewing the content of ref.~\cite{Fong:2016yyh}, and section \ref{sec:nonunitarity-matter} is to summarize the key points of this paper. 

\subsection{Unitary 3 active + $N$ sterile neutrino system with partial decoherence}
\label{sec:3+N-system}

The system we are considering consists of 3 active + $N$ sterile neutrinos which is unitary in the whole state space, but serves for a model of non-unitarity when restricted to observables in the active neutrino subspace. 
The sterile-sector model independence is realized due to decoherence between active-sterile and sterile-sterile states, which essentially wipes out detailed informations of sterile sector such as mass spectrum and mixing structure with active neutrinos. 
Generically, the decoherence condition associated with energy resolution reads (see \cite{Fong:2016yyh})
\begin{eqnarray}
|\Delta m_{Ja}^2| \gtrsim \frac{4 \pi E}{L} \left( \frac{\delta E}{E} \right)^{-1}
\approx 
2.5 \times 10^{-2} \text{eV}^2 
\left( \frac{ E }{ 1 \,\mbox{GeV} } \right) 
\left( \frac{ L }{ 1000 \,\mbox{km} } \right)^{-1}
\left( \frac{ \delta E / E }{ 0.1 } \right)^{-1}
\label{decoherence-mass-condition}
\end{eqnarray}
where $\Delta m_{Ja}^2$ denote either active-sterile or sterile-sterile mass squared difference and $L$ is a baseline. It simplifies to 
$|\Delta m_{Ja}^2| \gsim 2 |\Delta m^2_{31}| \left( \delta E / E \right)^{-1}$ assuming the conventional setting of accelerator long-baseline (LBL) experiments, i.e., a detector at around the oscillation maximum.\footnote{
%%%%%%%%%%%% footnote %%%%%%%%%%%%%%
In medium-baseline reactor neutrino experiments which utilize the solar oscillation maximum, such as JUNO~\cite{JUNO}, the condition becomes 
$\Delta m_{Ja}^2 \gsim 2 \Delta m^2_{21} \left( \delta E / E \right)^{-1} \approx 5 \times 10^{-3} \, \text{eV}^2$, assuming 3\% energy resolution.
}
It leads to $|\Delta m_{Ja}^2| \gsim 5 \times 10^{-2} \,\text{eV}^2$ assuming 10\% energy resolution, which implies the lower limit of (mostly) sterile neutrino mass, $m^2_{J} \gsim 0.1 \,\text{eV}^2$, to ensure partial decoherence \cite{Fong:2016yyh}. 
Though we sometimes quote the lower limit as the reference value in this paper, we have to rely on the formula (\ref{decoherence-mass-condition}) for the condition of partial decoherence in more generic setting off the oscillation maximum.

We also restrict the sterile neutrino mass range from above such that they can be produced energetically from a given source and participate to the neutrino oscillation together with active neutrinos.
It yields the upper bound, typically, $m^2_{J} \lsim 1\, \text{MeV}^2$ for reactor neutrinos and $m^2_{J} \lsim (1 - 10) \,\text{GeV}^2$ for accelerator neutrinos. Thus, a minimal range $0.1\, \text{eV}^2 \lsim m^2_{J} \lsim 1\, \text{MeV}^2$ results as quoted in ref.~\cite{Fong:2016yyh}. 
For more energetic neutrino sources one can take the upper limit of $m_{J}$ as the kinematical limit of production. 

Throughout this paper, we assume for validity of our discussion, the sterile neutrino mass condition (\ref{decoherence-mass-condition}) and that it is below production threshold. When appropriate we may quote the reference $m^2_{J}$ range, $0.1\, \text{eV}^2 \lsim m^2_{J} \lsim 1\, \text{MeV}^2$, or $\lsim (1 - 10) \,\text{GeV}^2$, but otherwise the readers must assume that $m_J$ obeys the general conditions above.

\subsection{Non-unitary evolution of neutrinos in vacuum or with small matter effect}
\label{sec:nonunitarity-vacuum-small-matter}

Here we summarize the main findings of ref.~\cite{Fong:2016yyh}. 
Thanks to partial decoherence, fast oscillations in active-sterile and sterile-sterile channels are averaged out, which leads to a very simple form of the active neutrino oscillation probability in vacuum
\begin{eqnarray}
P(\nu_\beta \rightarrow \nu_\alpha) &=& 
\mathcal{C}_{\alpha \beta} + 
\left| \sum_{j=1}^{3} U_{\alpha j} U^{*}_{\beta j} \right|^2 - 
2 \sum_{j \neq k} 
\mbox{Re} 
\left( U_{\alpha j} U_{\beta j}^* U_{\alpha k}^* U_{\beta k} \right) 
\sin^2 \frac{ ( \Delta_{k} - \Delta_{j} ) x  }{ 2 }
\nonumber\\
&-&
\sum_{j \neq k} \mbox{Im} 
\left( U_{\alpha j} U_{\beta j}^* U_{\alpha k}^* U_{\beta k} \right) 
\sin ( \Delta_{k} - \Delta_{j} ) x, 
\label{P-beta-alpha-ave-vac}
\end{eqnarray}
where $x$ denotes baseline and 
\begin{eqnarray} 
\mathcal{C}_{\alpha \beta} \equiv 
\sum_{J=4}^{3+N}
\vert W_{\alpha J} \vert^2 \vert W_{\beta J} \vert^2, 
\label{Cab} 
\end{eqnarray}
in the appearance ($\alpha \neq \beta$) as well as in the disappearance ($\alpha = \beta$) channels with $\alpha,\beta = e,\mu,\tau$.
In eq.~\eqref{P-beta-alpha-ave-vac}, the indices $i,j,k=1,2,3$ and
$J=4,5,\cdot \cdot \cdot, N+3$ are, respectively, for (mostly) active
and (mostly) sterile neutrino mass eigenstates. 
The active neutrino flavour states $\nu_\alpha$ are connected to mass eigenstates ($\nu_i,\nu_J$) through
\begin{eqnarray} 
	 \nu_{\alpha} &=& \sum_{i=1}^{3} (U)_{\alpha i} \nu_{i} + \sum_{J=4}^{3+N} (W)_{\alpha J} \nu_{J}, 
\end{eqnarray} 
that is, the $(3 \times 3)$ non-unitary $U$ matrix describes mixing in the active neutrino space, whereas the $(3 \times N)$ $W$ matrix elements bridge between active and sterile state spaces.
We have defined the kinematical phase factors $\Delta_{j} \equiv \frac{m^2_{j}}{2E}$ and $\Delta_{J} \equiv \frac{m^2_{J}}{2E}$ where $m_{j}$ and $m_{J}$ denote the active and sterile neutrino masses, respectively, and $E$ denotes neutrino energy. 

The characteristic features of the oscillation probability in (\ref{P-beta-alpha-ave-vac}) are:

\begin{enumerate}

\item
The non-unitary matrix $U$ replaces the standard unitary three-flavour mixing matrix often parametrized with Particle Data Group convention $U_{\text{\tiny PDG}}$ \cite{Olive:2016xmw}. 

\item
Probability leakage term $\mathcal{C}_{\alpha \beta} > 0$ appears reflecting the nature of low-energy unitarity violation in which the probability can flow out from active neutrino space to the sterile state space, and vice versa.

\item
Due to non-unitarity of the $U$ matrix, $\delta_{\alpha \beta}$ term in the unitary case is modified to $\left| \sum_{j=1}^{3} U_{\alpha j} U^{*}_{\beta j} \right|^2$.

\end{enumerate}
\noindent
Notice that each term of $\mathcal{C}_{\alpha \beta}$ in (\ref{Cab}) allows interpretation that ``probability leaking from active to sterile state spaces'' and coming back. The simple terminology of probability leaking assumes that the latter process must also exist which is ensured by generalized T invariance. 
Another aspect of the probability leaking term, which has a form of incoherent sum of the products of probabilities of active-to-sterile and sterile-to-active transitions clearly illustrates the decoherence associated to the sterile states. For instance, near the upper end of the sterile state mass region 
quoted in section~\ref{sec:3+N-system}, it describes effect of decoherence caused by separation of wave packets between active and sterile neutrinos.

The points 2 and 3 above are important ones and the clarifying remarks about them are in order: 

\begin{itemize}

\item
Presence or absence of the probability leakage term $\mathcal{C}_{\alpha \beta}$ distinguishes between low-energy and high-energy unitarity violation~\cite{Fong:2016yyh}. Nevertheless, $\mathcal{C}_{\alpha \beta}$ may be small because it is of fourth order in $W$. 

\item
Difference in normalization factor, the second term in (\ref{P-beta-alpha-ave-vac}), between unitary and non-unitary cases is of order $\sim W^4$ ($\sim W^2$) in the appearance (disappearance) channels. 

\end{itemize}
\noindent
To understand the latter point, we notice that unitarity in the $(3+N)$ space unitary model can be written as 
\begin{eqnarray} 
\delta_{\alpha \beta} = 
\sum_{j=1}^{3} U_{\alpha j} U^{*}_{\beta j} + \sum_{J=4}^{N+3} W_{\alpha J} W^{*}_{\beta J}. 
\label{unitarity0}
\end{eqnarray}
Then, 
$\left| \sum_{j=1}^{3} U_{\alpha j} U^{*}_{\beta j} \right|^2 = \left| \sum_{J=4}^{N+3} W_{\alpha J} W^{*}_{\beta J} \right|^2$ in the appearance channel ($\alpha \neq \beta$), and $\left( \sum_{j=1}^{3} \vert U_{\alpha j} \vert^2 \right)^2 = \left( 1 -  \sum_{J=4}^{N+3} \vert W_{\alpha J} \vert^2 \right)^2 = 1 - \mathcal{O} (W^2)$ in the disappearance channel ($\alpha = \beta$), which justifies the above statement.

We emphasize, therefore, that the probability leaking term $\mathcal{C}_{\alpha \beta}$ and the another constant term $\left| \sum_{j=1}^{3} U_{\alpha j} U^{*}_{\beta j} \right|^2$ in the oscillation probabilities are the same order, $\mathcal{O} (W^4)$, in the appearance channels. Hence, we do not see any good reasons why the former can be ignored, as was done in the existing literatures. It is also worth to note that $\mathcal{O} (W^2)$ difference in normalization in the disappearance channel would make detection of unitarity violation more feasible. 
It is one of the reasons for high sensitivity to unitarity violation that could be reached in disappearance measurement in the JUNO-like setting \cite{Fong:2016yyh}. 

In the same work, by including small matter effect up to first order, we have found that as far as we remain in the region of unitarity violating element $\vert W \vert \simeq 0.1$,\footnote{
	%%%%%%%%%%%%% footnote %%%%%%%%%%%%
	Speaking more precisely, we mean that all the $W$ matrix elements are assumed to be small, of the order of $\simeq 0.1$. 
} 
or somewhat larger, the matter effect does not alter the above features
of the oscillation probability in (\ref{P-beta-alpha-ave-vac}) under the
same restriction on sterile neutrino masses. 
Notice that $\vert W \vert \simeq 0.1$ implies that the unitarity violating effect in the probability is of the order of $\vert W \vert^4 \sim 10^{-4}$, 
except for the $\mathcal{O} (W^2)$ difference in normalization constant in the disappearance probability. 
It is practically the limit of order of magnitude that can be explored by the next generation neutrino oscillation experiments.

\subsection{Non-unitary evolution of neutrinos in matter to all orders}
\label{sec:nonunitarity-matter}

Given the fact that setup of some of the next generation accelerator LBL experiments require consideration of the matter effect comparable with the vacuum mixing one, it is clear that a better treatment is necessary to understand the influence of the matter effect in the $(3+N)$ model. 
Then, we formulate in this paper the small unitarity-violation
perturbation theory, a systematic and controlled way of treating small unitarity violation effect while including all order matter effect. 
We derive a simple expression of the oscillation probability in matter which retains the favourable feature of the vacuum formula
(\ref{P-beta-alpha-ave-vac}), the sterile sector model independence
under the same sterile neutrino mass condition as in vacuum. That is, the model-dependent terms are either averaged out, or made small due to large sterile state mass denominator suppression. 
We must note here that our treatment of the matter effect in this and the previous papers is restricted to the case of uniform matter density. 

The resulting oscillation probability in matter between active flavour neutrinos in the $(3+N)$ space unitary model to fourth order in $W$ can be written as 
\begin{eqnarray}
P(\nu_\beta \rightarrow \nu_\alpha) &=& 
\mathcal{C}_{\alpha \beta} + 
\left| \sum_{j=1}^{3} U_{\alpha j} U^{*}_{\beta j} \right|^2 
\nonumber\\
&-&
2 \sum_{j \neq k} 
\mbox{Re} 
\left[ (UX)_{\alpha j} (UX)_{\beta j}^* (UX)_{\alpha k}^* (UX)_{\beta k} \right] 
\sin^2 \frac{ ( h_{k} - h_{j} ) x  }{ 2 }
\nonumber\\
&-&
\sum_{j \neq k} \mbox{Im} 
\left[ (UX)_{\alpha j} (UX)_{\beta j}^* (UX)_{\alpha k}^* (UX)_{\beta k} \right] 
\sin ( h_{k} - h_{j} ) x, 
\label{P-beta-alpha-final}
\end{eqnarray}
where $h_{i}$ $(i=1,2,3)$ denote the energy eigenvalues of zeroth-order states of active neutrinos in matter, and $X$ is the unitary matrix which diagonalizes the zeroth-order Hamiltonian used to formulate our perturbation theory. $\mathcal{C}_{\alpha \beta}$ is the same as we have in the vacuum case 
%\sout{\red{as given}} 
in (\ref{Cab}). 
The expression is valid under the same restriction on sterile neutrino masses we have in vacuum, $0.1\, \text{eV}^2 \lsim m^2_{J} \lsim (1 - 10) \,\text{GeV}^2$ for $\vert W \vert^4 \gsim 10^{-4}$ assuming neutrino energy and baseline (and the associated matter density) which correspond to accelerator LBL experiments. For more precise conditions we require and for the restriction needed on the sterile state masses for smaller $W$, see section~\ref{sec:energy-denominator}.

The expression (\ref{P-beta-alpha-final}) is a very transparent result in the sense that 
(1) the vacuum non-unitary mixing matrix $U$ is ``dressed'' in a simple way by the matter effect represented by $X$, and 
(2) the probability leaking term $\mathcal{C}_{\alpha \beta}$ and the normalization term stay as they are in vacuum. The latter feature is perfectly natural, given the nature of these terms as probability leaking and (mis-) normalization at zero distance.\footnote{
%%%%%%%%%%%%% footnote %%%%%%%%%%%%%%%
A comment is ready for the normalization term, the second term in (\ref{P-beta-alpha-final}). Its original form is $\sum_{j=1}^{3} (UX)_{\alpha j} (UX)^{*}_{\beta j}$, which is natural because it comes from the contribution of zeroth-order Hamiltonian with all orders effect of the matter potential. It is easily reduced to the vacuum form in (\ref{P-beta-alpha-final}) (or in (\ref{P-beta-alpha-ave-vac})) 
by using unitarity relation $\sum_{j=1}^3 X_{kj} X^*_{lj} = \delta_{kl}$. 
}
The detailed derivation of eq.~\eqref{P-beta-alpha-final} is carried out in section \ref{sec:formulation}. While in section \ref{sec:analytical-numerical}, we derive an exact analytic expression for the matter dependent part of the oscillation probability \eqref{P-beta-alpha-final}. The combinations of $X$ matrix elements that used in the derivation can also be utilized to calculate higher order corrections in $W$. In section~\ref{sec:with-without-UV}, the regions of visible effect of unitarity violation is illuminated by plotting the probabilities with/without unitarity violation in wide ranges of $E$ and baseline $L$. 

After understanding the general feature of perturbative series based on
explicit calculation to order $W^4$, we postulate the ``Uniqueness
theorem'' which states that the oscillation probability formula
eq.~\eqref{P-beta-alpha-final} is valid to all orders in $W$ expansion under the same conditions on the sterile state mass and the kinematical region as used in the discussion of fourth-order formulas. See section~\ref{sec:U-theorem}.
The reasons for this interesting feature, the same mass conditions as in vacuum to guarantee the sterile sector model independence prevail in matter, will be partially explained at the end of section~\ref{sec:energy-denominator}.

Finally, but probably most importantly, we point out that outside
the region of validity of our above theorem, there are regions of
neutrino energy and baseline that condition for suppression due to the large sterile state mass denominators is not fully effective. 
We show that in such region, second order correction terms in $W$, together with the leaking term $\mathcal{C}_{\alpha \beta}$, may not be totally negligible, and it could be detectable. It would offer yet another way of distinguishing low-scale unitarity violation from high-scale one. These {\em new} terms are derived in section \ref{sec:probability-2nd} and their effects are quantified in section \ref{sec:correction-terms}.

\section{Small unitarity-violation perturbation theory of neutrino oscillation in matter }
\label{sec:formulation}

We formulate a perturbation theory of the $(3+N)$ state unitary model using an expansion parameter of matrix elements of $W$ signifying unitarity violation effect, assuming it small. 
It will be done aiming at constructing a model-independent framework for leptonic unitarity test. It necessitates the conditions on the sterile neutrino mass as discussed in section~\ref{sec:3+N-system}. 
In most of the discussions in this section we presume, as an appropriate setting for unitarity test, terrestrial neutrino experiments, i.e., accelerator LBL experiments, and/or atmospheric neutrino measurement. Use of reactor and accelerator neutrinos at short baselines offers an alternative way for testing leptonic unitarity but with only minor matter effect.

In the main text we mostly confine ourselves to the formulas to second order in $W$, but include fourth order terms whenever it is necessary. We take for simplicity the uniform number density approximation for electrons and neutrons in matter. However, extension to the varying density case is, in principle, straightforward as far as adiabaticity holds. 
Usage of the same probability formula as a hunting tool of unitarity violation and discriminator between low-scale and high-scale unitarity violation will be discussed in section~\ref{sec:where-UV}.

\subsection{3 active plus $N$ sterile neutrino system in the flavour basis}
\label{sec:flavor-basis}

The $S$ matrix describes possible flavour changes after traversing 
a distance $x$ 
\begin{eqnarray} 
\nu_{\alpha} (x) = S_{\alpha \beta} \nu_{\beta} (0), 
\label{def-S}
\end{eqnarray}
and the oscillation probability is given by 
\begin{eqnarray} 
P(\nu_{\beta} \rightarrow \nu_{\alpha}; x)= 
\vert S_{\alpha \beta} \vert^2.  
\label{def-P}
\end{eqnarray}
The neutrino evolution in flavour basis in the $(3+N)$ space unitary model is governed by the Schr\"odinger equation 
\begin{eqnarray}
i \frac{d}{dx} \nu = H \nu. 
\label{evolution}
\end{eqnarray}
Given the flavour basis Hamiltonian $H$, the $S$ matrix is given by  
\begin{eqnarray} 
S = T \text{exp} \left[ -i \int^{x}_{0} dx^{\prime} H(x^{\prime}) \right], 
\label{S-matrix-def}
\end{eqnarray}
where $T$ symbol indicates the ``time ordering''  
(in fact ``space ordering'' here). 
The right-hand side of (\ref{S-matrix-def}) may be written as 
$e^{-i H x}$ for the case of constant matter density. 

The flavour basis Hamiltonian $H$ is $(3+N) \times (3+N)$ matrix: 
\begin{eqnarray} 
H = {\bf U}
\left[
\begin{array}{cccccc}
\Delta_{1} & 0 & 0 & 0 & 0 & 0 \\
0 & \Delta_{2} & 0 & 0 & 0 & 0 \\
0 & 0 & \Delta_{3} & 0 & 0 & 0 \\
0 & 0 & 0 & \Delta_{4} & 0 & 0 \\
0 & 0 & 0 & 0 & \cdot \cdot \cdot & 0 \\
0 & 0 & 0 & 0 & 0 & \Delta_{3+N} \\
\end{array}
\right] 
{\bf U}^{\dagger} 
+
\left[
\begin{array}{cccccc}
\Delta_{A} - \Delta_{B} & 0 & 0 & 0 & 0 & 0 \\
0 & - \Delta_{B} & 0 & 0 & 0 & 0 \\
0 & 0 & - \Delta_{B} & 0 & 0 & 0 \\
0 & 0 & 0 & 0 & 0 & 0 \\
0 & 0 & 0 & 0 & \cdot \cdot \cdot & 0 \\
0 & 0 & 0 & 0 & 0 & 0 \\
\end{array}
\right],
\label{hamiltonian}
\end{eqnarray}
where 
\begin{eqnarray}
\Delta_{i}  \equiv \frac{ m^2_{i} }{2E} 
\hspace{4mm}
(i = 1,2,3),
\hspace{8mm}
\Delta_{J}  \equiv \frac{ m^2_{J} }{2E}
\hspace{4mm}
(J = 4, \cdot \cdot \cdot, 3+N).
\label{Delta-def}
\end{eqnarray}
Here, $m_{i}$ ($m_{J}$) denote the mass of mostly active (sterile) neutrinos and $E$ is the neutrino energy. 
$\Delta_{A}$ and $\Delta_{B}$ are related to Wolfenstein's matter
potential \cite{Wolfenstein:1977ue} due to charged current (CC) and neutral current (NC) reactions, $a$ and $b$, as
\begin{eqnarray} 
\Delta_{A}  \equiv \frac{ a }{2E}, 
\hspace{10mm}
\Delta_{B}  \equiv \frac{ b }{2E}, 
\label{Delta-ab-def}
\end{eqnarray}
where  
\begin{eqnarray} 
a &=&  
2 \sqrt{2} G_F N_e E \approx 1.52 \times 10^{-4} \left( \frac{Y_e \rho}{\rm g\,cm^{-3}} \right) \left( \frac{E}{\rm GeV} \right) {\rm eV}^2, 
\nonumber \\
b &=& \sqrt{2} G_F N_n E = \frac{1}{2} \left( \frac{N_n}{N_e} \right) a.
\label{matt-potential}
\end{eqnarray}
In the above, both $a$ and $b$ are positive. For antineutrinos, we take $\Delta_A \to -\Delta_A$ and $\Delta_B \to -\Delta_B$.
Here, $G_F$ is the Fermi constant, $N_e$ and $N_n$ are, respectively, the electron and neutron number densities in matter. $\rho$ and $Y_e$ denote, respectively, the matter density and number of electron per nucleon in matter. 
In (\ref{hamiltonian}), ${\bf U}$ denotes the flavour mixing matrix which relates $(3+N)$ dimensional flavour neutrino states to the vacuum mass eigenstate basis as $\nu_{\zeta} = {\bf U}_{\zeta z} \tilde{\nu}_{z}$, where $\zeta$ runs over active flavour $\alpha = e,\mu,\tau$ and sterile flavour $s = s_1,\cdot \cdot \cdot,s_N$ indices, $z$ runs over mostly active $i=1,2,3$ and mostly sterile mass eigenstate $J=4,5,\cdot \cdot \cdot, N+3$ indices. 

For simplicity, we introduce a compact notation which writes $(3+N) \times (3+N)$ matrix in a form of $2 \times 2$ matrix. By defining the active $3 \times 3$ matter potential matrix 
\begin{eqnarray} 
A =
\left[
\begin{array}{ccc}
\Delta_{A} - \Delta_{B} & 0 & 0 \\
0 & - \Delta_{B} & 0 \\
0 & 0 & - \Delta_{B} \\
\end{array}
\right] 
\label{matter-pot2}
\end{eqnarray} 
the flavour basis Hamiltonian is written as 
\begin{eqnarray}
H = 
{\bf U} 
\left[
\begin{array}{cc}
{\bf \Delta_{a} } & 0 \\
0 & {\bf \Delta_{s} } \\
\end{array}
\right] 
{\bf U}^{\dagger} 
+
\left[
\begin{array}{cc}
A & 0 \\
0 & 0 \\
\end{array}
\right] 
\equiv H_{ \text{vac} } + H_{ \text{matt} } 
\label{flavor-hamiltonian}
\end{eqnarray}
where ${\bf \Delta_{a} } = \text{diag} ( \Delta_{1}, \Delta_{2}, \Delta_{3})$ and 
${\bf \Delta_{s} } = \text{diag} (\Delta_{4}, \Delta_{5}, \cdot \cdot \cdot, \Delta_{N+3})$. 

As an application of our framework, we anticipate leptonic unitarity test in the LBL accelerator neutrino experiments which utilize atmospheric-scale neutrino oscillations. Therefore, we assume that the system satisfies the following conditions in formulating our perturbation theory
\begin{eqnarray}
\frac{ \Delta m^2_{31} L }{2E} \sim \frac{ \Delta m^2_{32} L }{2E} \sim \mathcal{O} (1), 
\hspace{10mm}
\frac{ a L }{2E} \sim \frac{ b L }{2E} \sim \mathcal{O} (1), 
\label{assumption}
\end{eqnarray}
where $L$ denotes baseline, $\Delta m^2_{ji} \equiv m_j^2 - m_i^2$, and 
\begin{eqnarray}
\frac{ a L }{2E} 
&=& \sqrt{2} G_F N_e L
= 0.58  
\left(\frac{\rho}{3\ \text{g/cm}^3}\right)
\left(\frac{L}{1000\ \mbox{km}}\right). 
\label{aL-2E}
\end{eqnarray}
They probably ensure that our oscillation probability formulas have applicability to the terrestrial LBL and atmospheric neutrino experiments with baseline up to $\sim 10^4$ km and energies from low to high, up to $E \sim 100$ GeV. More precise discussions on where our formulas are valid will be given in sections~\ref{sec:energy-denominator} and \ref{sec:higher-order}.

\subsection{Vacuum mass eigenstate basis, or tilde basis}
\label{sec:tilde-basis}

To formulate perturbative treatment it is convenient to consider the vacuum mass eigenstate basis, the tilde basis, introduced in the previous section 
\begin{eqnarray}
\tilde{\nu}_{z} = ({\bf U}^{\dagger})_{z \zeta} \nu_{\zeta}.
\label{tilde-basis}
\end{eqnarray}
The tilde basis Hamiltonian is related to the flavour basis one as 
\begin{eqnarray} 
\tilde{H} = {\bf U}^{\dagger} H {\bf U}. 
\label{tilde-hamiltonian}
\end{eqnarray}
The explicit form of $\tilde{H}$ is given by 
\begin{eqnarray} 
\tilde{H} &=& \tilde{H}_{ \text{vac} } + \tilde{H}_{ \text{matt} } = 
\left[
\begin{array}{cc}
{\bf \Delta_{a} } & 0 \\
0 & {\bf \Delta_{s} } \\
\end{array}
\right] 
+ 
{\bf U}^{\dagger} 
\left[
\begin{array}{cc}
A & 0 \\
0 & 0 \\
\end{array}
\right] 
{\bf U}. 
\label{H-tilde-3+N}
\end{eqnarray}

We parameterize the $(3+N) \times (3+N)$ dimensional flavour mixing matrix ${\bf U}$ as
\begin{eqnarray} 
{\bf U} = \left[
\begin{array}{cc}
U & W \\
Z & V \\
\end{array}
\right]. 
\label{U-parametrize}
\end{eqnarray}
The matrix $U$ and $V$ are $3 \times 3$ and $N \times N$ matrices, respectively, and $W$ and $Z$ have sizes that just fill in the space. 
In our $(3+N)$ model, unitarity is obeyed in the whole $(3+N)$ state space: 
\begin{eqnarray} 
{\bf U} {\bf U}^{\dagger} &=& 
\left[
\begin{array}{cc}
U U^{\dagger} + W W^{\dagger} & U Z^{\dagger} + W V^{\dagger} \\ 
Z U^{\dagger} + V W^{\dagger}  & Z Z^{\dagger} + V V^{\dagger} \\
\end{array}
\right] = 
\left[
\begin{array}{cc}
{\bf 1}_{3 \times 3} & 0 \\
0 & {\bf 1}_{N \times N} \\
\end{array}
\right], 
\nonumber \\
{\bf U}^{\dagger} {\bf U} &=&  
\left[
\begin{array}{cc}
U^{\dagger} U + Z^{\dagger} Z & U^{\dagger} W + Z^{\dagger} V  \\
W^{\dagger} U + V^{\dagger} Z & W^{\dagger} W + V^{\dagger} V \\
\end{array}
\right] =
\left[
\begin{array}{cc}
{\bf 1}_{3 \times 3} & 0 \\
0 & {\bf 1}_{N \times N} \\
\end{array}
\right].
\label{unitarity}
\end{eqnarray}

Then, the Hamiltonian $\tilde{H}$ in vacuum mass eigenstate basis is given by 
\begin{eqnarray} 
\tilde{H} &=& 
\left[
\begin{array}{cc}
{\bf \Delta_{a} } & 0 \\
0 & {\bf \Delta_{s} } \\
\end{array}
\right] 
+
\left[
\begin{array}{cc}
U^{\dagger} A U & U^{\dagger} A W \\
W^{\dagger} A U & W^{\dagger} A W \\
\end{array}
\right]. 
\label{tilde-H}
\end{eqnarray}
As in vacuum, the neutrino oscillation is governed only by the $U$ and $W$ matrices, and is independent of $Z$ and $V$ matrices. It is natural that $V$ matrix does not show up in physical Hamiltonian matrix because the rotations inside sterile basis does not have any physical meaning, if we observe the system only by the Standard Model interactions. 
However, the flavour basis Hamiltonian $H$ in (\ref{hamiltonian}) obviously depends on $Z$ and $V$. The apparent puzzle will be resolved in appendix~\ref{sec:flavor-basis-evolution}. 

\subsection{Formulating small unitarity-violation perturbation theory}
\label{sec:UV-perturb}

We now construct the small unitarity-violation perturbation theory. It is natural to consider the framework in which the tilde-basis Hamiltonian $\tilde{H}$ is decomposed into the un-perturbed and perturbed parts, $\tilde{H}_{0} + \tilde{H}_{1}$, as follows: 
\begin{eqnarray} 
\tilde{H}_{0} = 
\left[
\begin{array}{cc}
{\bf \Delta_{a} } + U^{\dagger} A U & 0 \\
0 & {\bf \Delta_{s} } \\
\end{array}
\right], 
\hspace{10mm}
%\nonumber \\
\tilde{H}_{1} = 
\left[
\begin{array}{cc}
0 & U^{\dagger} A W \\
W^{\dagger} A U & W^{\dagger} A W \\
\end{array}
\right]. 
\label{tilde-H0+H1} 
\end{eqnarray}
Therefore, what we mean by ``expansion by unitarity violation effect'' is an expansion by the $W$ matrix elements.\footnote{
%%%%%%%%%%%%%% footnote %%%%%%%%%%%%%%%
Through unitarity (\ref{unitarity}), $U$ matrix elements have some dependence on $W$ matrix elements. We choose not to expand $U$ matrix elements by this $W$ dependence. In this sense, we use a ``renormalized basis'' (in the same sense as in ref.~\cite{Minakata:2015gra}) in which some higher order effects are absorbed into the zeroth-order state. 
}
We assume, for simplicity, that all the $W$ matrix elements are small and have the same order $\epsilon_{s}$. Then, $3 \times N$ ($N \times 3$) sub-matrix elements in $\tilde{H}_{ \text{matt} }$ are of order $\epsilon_{s}$, while the pure sterile space $N \times N$ sub-matrix elements are of order $\epsilon_{s}^2$. For simplicity, we often use the expression ``expanding to order $W^n$'' which means to order $\epsilon_{s}^n$ in this paper. 

\subsubsection{Hat basis}

To formulate perturbation theory with $\tilde{H}_{0}$ and $\tilde{H}_{1}$ given above we transform to a basis in which the un-perturbed part of the Hamiltonian is diagonal, which we call the ``hat basis''. Since the $3 \times 3$ sub-matrix ${\bf \Delta_{a} } + U^{\dagger} A U$ in $\tilde{H}_{0}$ is Hermitian, it can be diagonalized by the unitary transformation 
\begin{eqnarray} 
X^{\dagger} \left(  {\bf \Delta_{a} } + U^{\dagger} A U \right) X = 
\left[
\begin{array}{ccc}
h_{1} & 0 & 0 \\
0 & h_{2} & 0 \\
0 & 0 & h_{3} \\
\end{array}
\right] \equiv {\bf h} 
\label{H0-diag}
\end{eqnarray}
with $X$ being the $3 \times 3$ unitary matrix. 
Then, $\tilde{H}_{0}$ can be diagonalized by using 
\begin{eqnarray} 
{\bf X} \equiv 
\left[
\begin{array}{cc}
X & 0 \\
0 & 1 \\
\end{array}
\right] 
\label{bfX-def}
\end{eqnarray}
as 
\begin{eqnarray} 
&& 
{\bf X}^{\dagger} \tilde{H}_{0} {\bf X}
= \left[
\begin{array}{cc}
X^{\dagger} \left(  {\bf \Delta_{a} } + U^{\dagger} A U \right) X & 0 \\
0 & {\bf \Delta_{s} } \\
\end{array}
\right] 
%\nonumber \\
%&=&
= \left[
\begin{array}{cc}
{\bf h} & 0 \\
0 & {\bf \Delta_{s} } \\
\end{array}
\right] 
\equiv \hat{H}_{0}, 
\label{hat-H0}
\end{eqnarray}
the zeroth-order Hamiltonian in the hat basis. 
Since $\hat{H}_{0}$ is diagonal it is easy to compute $e^{ \pm i \hat{H}_{0} x}$:
\begin{eqnarray} 
e^{ \pm i \hat{H}_{0} x} = 
\left[
\begin{array}{cc}
e^{ \pm i {\bf h} x } & 0 \\
0 & e^{ \pm i {\bf \Delta_{s} x }   } \\
\end{array}
\right]. 
\label{exp-H0}
\end{eqnarray}
Then, the perturbed Hamiltonian is given by
\begin{eqnarray} 
\hat{H}_{1} &=& {\bf X}^{\dagger} \tilde{H}_{1} {\bf X} 
=
\left[
\begin{array}{cc}
0 & (UX)^{\dagger} A W \\
W^{\dagger} A (UX) & W^{\dagger} A W \\
\end{array}
\right]. 
\label{hat-H1}
\end{eqnarray}
The eigenvalues of $\tilde{H}_{0}$ is therefore $h_{1}$, $h_{2}$,
$h_{3}$, and $\Delta_{J}$ ($J=4, \cdot \cdot \cdot, 3+N$). Therefore,
the sterile neutrino masses are affected neither by the active states
nor the matter potential in our zeroth-order unperturbed basis. It must
be a good approximation because we have assumed that the sterile 
neutrino masses are much heavier than the active ones, and we are
interested in the energy region implied by $a \sim \Delta m^2_{31}$.

To do real calculations of the $S$ matrix elements we must solve the zeroth order Hamiltonian $\tilde{H}_{0}$. This task will be carried out in section~\ref{sec:exact-solution-zeroth}, in which we derive explicit expressions of the eigenvalues $h_{i}$ and the unitary matrix $X$. 

Now, we formulate perturbation theory with the hat basis Hamiltonian, $\hat{H}_{0}$ in (\ref{hat-H0}) and $\hat{H}_{1}$ in (\ref{hat-H1}) after a clarifying note in the next subsection.

\subsubsection{The relationship between quantities in various bases}

So far we have introduced the tilde- and the hat-basis:
\begin{eqnarray} 
\tilde{H} = {\bf U}^{\dagger} H {\bf U}, 
\hspace{10mm}
\hat{H} = {\bf X}^{\dagger} \tilde{H}  {\bf X}, 
\label{tilde-hat-relation}
\end{eqnarray}
where ${\bf X}$ is given by eq.~(\ref{bfX-def}).
Therefore, 
\begin{eqnarray} 
\hat{H} 
= \left( {\bf U}  {\bf X} \right)^{\dagger} H \left( {\bf U}  {\bf X} \right).
\label{flavor-hat-relation}
\end{eqnarray}
Or 
\begin{eqnarray} 
H = \left( {\bf U}  {\bf X} \right) \hat{H} \left( {\bf U}  {\bf X} \right)^{\dagger}, 
\hspace{10mm}
S = \left( {\bf U}  {\bf X} \right) \hat{S} \left( {\bf U}  {\bf X} \right)^{\dagger}.
\label{flavor-hat-relation2}
\end{eqnarray}
Notice that both ${\bf U}$ and ${\bf X}$ are unitary, and hence ${\bf U} {\bf X}$ is unitary too. The relationship between wave functions of various basis are given by
\begin{eqnarray} 
\hat{\nu}_{y} &=& 
{\bf X}^{\dagger}_{y z} \tilde{\nu}_{z} 
= \left( {\bf U} {\bf X} \right)^{\dagger}_{y \zeta} \nu_{\zeta},
\nonumber \\
\nu_{\zeta} &=& 
\left( {\bf U} {\bf X} \right)_{\zeta y} 
\hat{\nu}_{y}. 
\label{hat-basis}
\end{eqnarray}
where $y$ denote the hat-basis indices. 
Using the explicit parametrization of the ${\bf U}$ matrix we have 
\begin{eqnarray} 
{\bf U}  {\bf X} &=& 
\left[
\begin{array}{cc}
U & W \\
Z & V \\
\end{array}
\right] 
\left[
\begin{array}{cc}
X & 0 \\
0 & 1 \\
\end{array}
\right] 
= 
\left[
\begin{array}{cc}
U X & W \\
Z X & V \\
\end{array}
\right], 
%
%\nonumber \\
\hspace{5mm}
\left( {\bf U}  {\bf X} \right)^{\dagger} =
\left[
\begin{array}{cc}
(UX)^{\dagger} & \left( Z X \right)^{\dagger} \\
W^{\dagger} & V^{\dagger} \\
\end{array}
\right].
\label{UX-parametrize}
\end{eqnarray}

It may be helpful for our discussions later to understand the relationship between $S$ and $\hat{S}$ matrix elements. For this purpose, we denote them in the block form 
\begin{eqnarray} 
S = 
\left[
\begin{array}{cc}
S_{aa} & S_{aS}  \\
S_{Sa} & S_{SS} \\
\end{array}
\right], 
\hspace{10mm}
\hat{S} = 
\left[
\begin{array}{cc}
\hat{S}_{aa} & \hat{S}_{aS}  \\
\hat{S}_{Sa} & \hat{S}_{SS} \\
\end{array}
\right],
\label{S-hatS}
\end{eqnarray}
where the subscripts $a$ and $S$ indicate that they act (for the right index) to the active or the sterile subspaces. Notice that $S_{aa}$ and $S_{aS}$, for example, are $3 \times 3$ and $3 \times N$ matrices, respectively. Then, the relationship between $S$ and $\hat{S}$ matrix elements can be written explicitly as
\begin{eqnarray} 
S_{aa} &=& 
(U X) \hat{S}_{aa} (UX)^{\dagger} + (U X) \hat{S}_{aS} W^{\dagger} + 
W \hat{S}_{Sa} (UX)^{\dagger} + W \hat{S}_{SS} W^{\dagger},
\nonumber \\
S_{aS} &=& 
(U X) \hat{S}_{aa} \left( Z X \right)^{\dagger} + (U X) \hat{S}_{aS} V^{\dagger} + 
W \hat{S}_{Sa} \left( Z X \right)^{\dagger} + W \hat{S}_{SS} V^{\dagger},
\nonumber \\
S_{Sa} &=& 
\left( Z X \right) \hat{S}_{aa} (U X)^{\dagger} + \left( Z X \right) \hat{S}_{aS} W^{\dagger}  + 
V \hat{S}_{Sa} (U X)^{\dagger} + V \hat{S}_{SS} W^{\dagger},
\nonumber \\
S_{SS} &=& 
\left( Z X \right) \hat{S}_{aa} \left( Z X \right)^{\dagger} + \left( Z X \right) \hat{S}_{aS} V^{\dagger}  + 
V \hat{S}_{Sa} \left( Z X \right)^{\dagger} + V \hat{S}_{SS} V^{\dagger}. 
\label{S-Shat-relation}
\end{eqnarray}

\subsubsection{Computation of $\hat{S}$ matrix elements}
\label{sec:hatS-matrix}

To calculate $\hat {S} (x) = \exp \left[ -i \int^{x}_{0} dx \hat{H} (x)  \right] $ we define $\Omega(x)$ as 
\begin{eqnarray} 
\Omega(x) = e^{i \hat{H}_{0} x} \hat{S} (x).
\label{def-omega}
\end{eqnarray}
$\Omega(x)$ obeys the evolution equation 
\begin{eqnarray} 
i \frac{d}{dx} \Omega(x) = H_{1} \Omega(x),
\label{omega-evolution}
\end{eqnarray}
where 
\begin{eqnarray} 
H_{1} \equiv e^{i \hat{H}_{0} x} \hat{H}_{1} e^{-i \hat{H}_{0} x}. 
\label{def-H1}
\end{eqnarray}
Then, $\Omega(x)$ can be computed perturbatively as 
\begin{eqnarray} 
\Omega(x) &=& 1 + 
(-i) \int^{x}_{0} dx' H_{1} (x') + 
(-i)^2 \int^{x}_{0} dx' H_{1} (x') \int^{x'}_{0} dx'' H_{1} (x'') 
\nonumber \\
&+& 
(-i)^3 \int^{x}_{0} dx' H_{1} (x') \int^{x'}_{0} dx'' H_{1} (x'') \int^{x''}_{0} dx''' H_{1} (x''') 
\nonumber \\
&+& 
(-i)^4 \int^{x}_{0} dx' H_{1} (x') \int^{x'}_{0} dx'' H_{1} (x'') \int^{x''}_{0} dx''' H_{1} (x''') \int^{x'''}_{0} dx'''' H_{1} (x'''') +  \cdot \cdot \cdot, 
\nonumber \\
\label{Omega-expand}
\end{eqnarray}
where the ``space-ordered'' form in (\ref{Omega-expand}) is essential 
because of the highly nontrivial spatial dependence in $H_{1}$. 
Upon obtaining $\Omega(x)$, $\hat{S}$ matrix can be obtained as 
\begin{eqnarray} 
\hat{S} (x) = e^{- i \hat{H}_{0} x} \Omega(x). 
\label{hatS-Omega}
\end{eqnarray}
By knowing $\hat{S}$ matrix elements, the $S$ matrix is obtained by
using (\ref{flavor-hat-relation2}), or (\ref{S-Shat-relation}). 

The perturbing Hamiltonian $H_{1}$ defined in (\ref{def-H1}) has a structure 
\begin{eqnarray} 
H_{1} = \left[
\begin{array}{cc}
0 & e^{ i {\bf h} x} (UX)^{\dagger} A W e^{ - i {\bf \Delta_{s} } x}  \\
e^{ i {\bf \Delta_{s} } x} W^{\dagger} A (UX) e^{ - i {\bf h} x}  & e^{ i {\bf \Delta_{s} } x} W^{\dagger} A W e^{ - i {\bf \Delta_{s} } x} \\
\end{array}
\right]. 
\label{H1-matrix}
\end{eqnarray}
That is, $(H_{1})_{i j} = 0$ in the whole active neutrino subspace.
The non-vanishing elements of $H_{1}$ are as follows:
\begin{eqnarray} 
(H_{1})_{i J} &=& 
e^{- i ( \Delta_{J} - h_{i} ) x} \left\{ (UX)^{\dagger} A W \right\}_{i J}, 
\nonumber \\
(H_{1})_{J i} &=& 
e^{ - i ( h_{i} - \Delta_{J} ) x} \left\{ W ^{\dagger} A (UX) \right\}_{J i}, 
\nonumber \\
(H_{1})_{J K} &=& 
e^{- i ( \Delta_{K} - \Delta_{J} ) x} \left\{ W ^{\dagger} A W \right\}_{J K}. 
\label{H1-elements}
\end{eqnarray}
Inserting eq.~(\ref{H1-elements}) into (\ref{Omega-expand}), we can compute all the $\Omega$ matrix elements. 
The simplest ones in first order in $H_{1}$, the second term in (\ref{Omega-expand}), are given by 
\begin{eqnarray} 
\Omega_{i j} [1] &=& 0, 
\nonumber \\
\Omega_{i J} [1] &=& 
\frac{e^{- i ( \Delta_{J} - h_{i} ) x} - 1 }{ ( \Delta_{J} - h_{i} )  }
\left\{ (UX)^{\dagger} A W \right\}_{i J}, 
\nonumber \\
\Omega_{J i} [1] &=&
- \frac{e^{ i ( \Delta_{J} - h_{i} ) x} - 1 }{ ( \Delta_{J} - h_{i} ) }
\left\{ W ^{\dagger} A (UX) \right\}_{J i}, \ 
\nonumber \\
\Omega_{J K} \vert_{J \neq K} [1] &=& 
\frac{e^{- i ( \Delta_{K} - \Delta_{J} ) x} - 1 }{ ( \Delta_{K} - \Delta_{J} )  }
\left\{ W ^{\dagger} A W \right\}_{J K}, 
\nonumber \\
\Omega_{J J} [1] &= &
(-i x) \left\{ W ^{\dagger} A W \right\}_{J J}, 
\label{Omega-1st-order}
\end{eqnarray}
which serve as a building block of the perturbation series because of the structure in (\ref{Omega-expand}). The notation ``[1]'' implies that the terms come from first order perturbation with $H_{1}$. For more about notations, see appendix~\ref{sec:hatS-elements}. 

We need to compute up to fourth order in $H_{1}$ because we want to keep all the order $W^4$ terms. The requirement arises because the probability leaking term, whose observation is crucial to distinguish between low-energy and high-energy unitarity violation, is of order $W^4$. The other normalization term, the second term in (\ref{P-beta-alpha-ave-vac}), also deviates from the one in unitary case by a quantity of order $W^4$ in the appearance channels, but in an implicit way. The resulting expressions of $\hat{S}$ matrix elements to order $W^4$ are summarized in appendix~\ref{sec:hatS-elements}. 

There exists important consistency check in the calculation. That is, the identity relation between $\hat{S}$ matrix elements that follows from generalized T invariance:\footnote{
%%%%%%%%%%%%%% footnote %%%%%%%%%%%%%%%
As in the Standard Model in particle physics T invariance is broken in our system only by complex numbers in the mixing matrix. 
}
 \begin{eqnarray} 
% \hat{S}_{i j} (U, W, X, A) &=& \hat{S}_{j i} (U^*, W^*, X^*, A^*),
% \nonumber \\
% \hat{S}_{i J} (U, W, X, A) &=& \hat{S}_{J i} (U^*, W^*, X^*, A^*),
% \nonumber \\\hat{S}_{I J} (U, W, X, A) &=& \hat{S}_{J I} (U^*, W^*, X^*, A^*), 
 \hat{S}_{AB} (U, W, X, A) &=& \hat{S}_{BA} (U^*, W^*, X^*, A^*),
\;\;\; AB = \{ ij, iJ, IJ \} 
\label{T-invariance}
\end{eqnarray}
where $\hat{S}_{J i}$ is obtained by performing the exchange $h_{i}
\leftrightarrow \Delta_{J}$ in $\hat{S}_{i J}$. 
The generalized T invariance relation is explicitly verified by the computed results of $\hat{S}$ matrix elements to fourth order in $W$ given in appendix~\ref{sec:hatS-elements}.\footnote{
%%%%%%%%%%%%%% footnote %%%%%%%%%%%%%%%%
Since $\hat{H}$ system is a consistent dynamical system it is legitimate
and easier to verify generalized T invariance in the $\hat{S}$ level,
though it can be done in the $S$ matrix level as well. A pedagogical
treatment for proving generalized T invariance is given in version 1 
of this work, arXiv ePrint: 1712.02798. 
}

\subsubsection{Computation of $S$ matrix elements}
\label{sec:S-matrix}

Given the results of $\hat{S}$ matrix elements it is straightforward to calculate $S$ matrix elements by using the formulas in eq.~(\ref{S-Shat-relation}). 
The active neutrino space $S$ matrix elements can be written in perturbative forms, $S_{\alpha \beta} = S_{\alpha \beta}^{(0)} + S_{\alpha \beta}^{(2)} + S_{\alpha \beta}^{(4)}$, where 
\begin{eqnarray} 
S_{\alpha \beta}^{(0)} &=& 
\sum_{k l} (UX)_{\alpha k} (UX)^*_{\beta l} 
\hat{S}_{kl}^{(0)}, 
\nonumber \\
S_{\alpha \beta}^{(2)} &=&
\sum_{k l} (UX)_{\alpha k} (UX)^*_{\beta l} 
\hat{S}_{kl}^{(2)} 
+
\sum_{k L} (UX)_{\alpha k} W^*_{\beta L} 
\hat{S}_{kL}^{(1)} 
\nonumber \\
&+&
\sum_{K l} W_{\alpha K} (UX)^*_{\beta l} 
\hat{S}_{K l}^{(1)}
+
\sum_{K L} W_{\alpha K} W^*_{\beta L}
\hat{S}_{KL}^{(0)}, 
\nonumber \\
S_{\alpha \beta}^{(4)} &=&
\sum_{k l} (UX)_{\alpha k} (UX)^*_{\beta l} 
\hat{S}_{kl}^{(4)} 
+
\sum_{k L} (UX)_{\alpha k} W^*_{\beta L} 
\hat{S}_{kL}^{(3)} 
\nonumber \\
&+&
\sum_{K l} W_{\alpha K} (UX)^*_{\beta l} 
\hat{S}_{K l}^{(3)}
+
\sum_{K L} W_{\alpha K} W^*_{\beta L}
\hat{S}_{KL}^{(2)}.
\label{Sab-hatSab}
\end{eqnarray}

Using (\ref{Sab-hatSab}) the explicit expressions of $S$ matrix elements
can be easily obtained with use of $\hat{S}$ matrix elements given in
appendix~\ref{sec:hatS-elements}. 
For example, $S_{\alpha \beta}$ in zeroth and second orders in $W$ are given, respectively, by 
\begin{eqnarray} 
S_{\alpha \beta}^{(0)} &=& 
\sum_{k} (UX)_{\alpha k} (UX)^*_{\beta k} 
e^{-i h_{k} x}, 
\label{S-alpha-beta-0th}
\end{eqnarray}
and
\begin{eqnarray} 
S_{\alpha \beta}^{(2)} &=  &
\sum_{k, K} 
%\hat{S}_{kk}^{(2)} 
\frac{ 1 }{ \Delta_{K} - h_{k} } 
\left[
(ix) e^{- i h_{k} x} + \frac{e^{- i \Delta_{K} x} - e^{- i h_{k} x} }{ ( \Delta_{K} - h_{k} )  } 
\right]
(UX)_{\alpha k} (UX)^*_{\beta k} 
\left\{ (UX)^{\dagger} A W \right\}_{k K} 
\nonumber \\ 
&\times&
\left\{ W ^{\dagger} A (UX) \right\}_{K k} 
%\left\{ (UX)^{\dagger} \right\}_{k \beta} 
%
%\nonumber \\
%&-& 
-\sum_{k \neq l} \sum_{K}  
%\hat{S}_{kl}^{(2)} 
\frac{[
\left( \Delta_{K} - h_{k} \right) e^{- i h_{l} x} 
- \left( \Delta_{K} - h_{l} \right) e^{- i h_{k} x} 
- ( h_{l}  - h_{k} )  e^{- i \Delta_{K} x} 
]
}{ ( h_{l}  - h_{k} ) (\Delta_{K} - h_{k}) (\Delta_{K} - h_{l}) } 
%
% \nonumber \\
% &\times& 
%
\nonumber \\
&\times&
(UX)_{\alpha k} (UX)^*_{\beta l} 
\left\{ (UX)^{\dagger} A W \right\}_{k K} 
\left\{ W ^{\dagger} A (UX) \right\}_{K l} 
%\left\{ (UX)^{\dagger} \right\}_{l \beta} 
%
%\nonumber \\
%&+& 
+
\sum_{k, K} 
%\hat{S}_{kL}^{(1)} 
\frac{e^{- i \Delta_{K} x} - e^{- i h_{k} x} }{ ( \Delta_{K} - h_{k} ) } 
\nonumber \\ 
&\times&
\biggl[ 
(UX)_{\alpha k} W^*_{\beta K} 
\left\{ (UX)^{\dagger} A W \right\}_{k K} 
%(W^{\dagger})_{K \beta} 
+
W_{\alpha K} (UX)^*_{\beta k} 
\left\{ W ^{\dagger} A (UX) \right\}_{K k}
%\left\{ (UX)^{\dagger} \right\}_{k \beta} 
\biggr]
\nonumber \\
&+&
\sum_{K} 
%\hat{S}_{KK}^{(0)} 
e^{- i \Delta_{K} x} 
W_{\alpha K} W^*_{\beta K}. 
%(W^{\dagger})_{K \beta} 
%
\label{S-alpha-beta-2nd}
\end{eqnarray}

\subsection{The oscillation probability to second order in $W$}
\label{sec:probability-2nd}

In this section, we discuss the oscillation probability to second order in $W$. It is to illuminate the principle of calculation, how averaging over the fast oscillation works, and to show which constraints are obtained on the sterile state masses by the requirement of suppression by the large sterile state mass denominators to make these sterile-sector model dependent terms negligible. 

Of course, we will calculate in this paper all the oscillation probabilities $P(\nu_\beta \rightarrow \nu_\alpha)$ in matter to fourth order in $W$ to keep the necessary term, the probability leaking term $\mathcal{C}_{\alpha \beta}$, as mentioned earlier.  The key features of the fourth-order terms will be described in the next section~\ref{sec:probability-4th}. 

The oscillation probability $P(\nu_\beta \rightarrow \nu_\alpha)$ is given to second order in $W$ as 
\begin{eqnarray} 
&&P(\nu_\beta \rightarrow \nu_\alpha)^{(0+2)} = 
\left| S^{(0)}_{\alpha \beta} \right|^2 
+ 2 \mbox{Re} \left[ \left( S^{(0)}_{\alpha \beta} \right)^{*} S^{(2)}_{\alpha \beta} \right] 
\nonumber \\
&=&
\sum_{k} (UX)_{\alpha k} (UX)^*_{\beta k} 
(UX)^*_{\alpha k} (UX)_{\beta k} 
%
%\nonumber \\
%&+&
+ \sum_{k \neq l} (UX)_{\alpha k} (UX)^*_{\beta k} 
(UX)^*_{\alpha l} (UX)_{\beta l} 
e^{-i ( h_{k} - h_{l} ) x}
\nonumber \\ 
&+& 
2 \mbox{Re} 
\biggl\{
\sum_{m}
\sum_{k, K} 
\frac{ 1 }{ \Delta_{K} - h_{k} } 
\left[
(ix) e^{- i ( h_{k} - h_{m} ) x} + \frac{e^{- i ( \Delta_{K} - h_{m} ) x} - e^{- i ( h_{k} - h_{m} ) x} }{ ( \Delta_{K} - h_{k} )  } 
\right]
\nonumber \\ 
&\times&
(UX)_{\alpha k} (UX)^*_{\beta k} 
(UX)^*_{\alpha m} (UX)_{\beta m} 
\left\{ (UX)^{\dagger} A W \right\}_{k K} 
\left\{ W ^{\dagger} A (UX) \right\}_{K k} 
\nonumber \\
&-& 
\sum_{m}
\sum_{k \neq l} \sum_{K} 
\frac{ 
\left( \Delta_{K} - h_{k} \right) e^{- i ( h_{l} - h_{m} ) x} 
- \left( \Delta_{K} - h_{l} \right) e^{- i ( h_{k} - h_{m} ) x} 
- ( h_{l}  - h_{k} )  e^{- i ( \Delta_{K} - h_{m} ) x} 
}
{ ( h_{l}  - h_{k} ) (\Delta_{K} - h_{k}) (\Delta_{K} - h_{l}) } 
\nonumber \\
%&\times& 
%
\nonumber \\
&\times&
(UX)_{\alpha k} (UX)^*_{\beta l} 
(UX)^*_{\alpha m} (UX)_{\beta m} 
\left\{ (UX)^{\dagger} A W \right\}_{k K} 
\left\{ W ^{\dagger} A (UX) \right\}_{K l} 
\nonumber \\
&+& 
\sum_{m}
\sum_{k, K} 
\frac{e^{- i (\Delta_{K} - h_{m} ) x} - e^{- i ( h_{k} - h_{m} ) x} }{ ( \Delta_{K} - h_{k} ) } 
\biggl[ 
(UX)_{\alpha k} W^*_{\beta K} 
(UX)^*_{\alpha m} (UX)_{\beta m} 
\left\{ (UX)^{\dagger} A W \right\}_{k K} 
\nonumber \\
&+& 
W_{\alpha K} (UX)^*_{\beta k} 
(UX)^*_{\alpha m} (UX)_{\beta m} 
\left\{ W ^{\dagger} A (UX) \right\}_{K k}
\biggr]
\nonumber \\ 
&+& 
\sum_{m}
\sum_{K} 
e^{- i (\Delta_{K} - h_{m} ) x} 
W_{\alpha K} W^*_{\beta K} 
(UX)^*_{\alpha m} (UX)_{\beta m} 
\biggr\}. 
\label{P-beta-alpha-0th+2nd}
\end{eqnarray}
The following formulas include the cases of both disappearance ($\alpha=\beta$) and appearance ($\alpha \neq \beta$) channels. 

Notice that there is no matter dependent terms without suppression either by high-frequency oscillations $\propto \cos (\Delta_{K} - h_{m}) x$ (or $\sin$), or by large sterile state mass denominators $\propto \frac{ 1 }{ \Delta_{K} - h_{k} }$. 

We take averaging over fast oscillations due to active-sterile and sterile-sterile mass squared differences which leads to 
\begin{eqnarray}
\left\langle \sin \Delta_{J i} x \right\rangle 
\approx
\left\langle \sin \Delta_{J K} x \right\rangle \approx 0, 
%\nonumber \\ 
\hspace{5mm}
\left\langle \cos \Delta_{J i} x 
\right\rangle \approx
\left\langle \sin \Delta_{J K} x
\right\rangle \approx 0,
\label{average-out}
\end{eqnarray} 
where $\langle ... \rangle$ stands for averaging over neutrino energy 
within the uncertainty of energy resolution, as well as averaging over uncertainty of distance between production and detection points of neutrinos.\footnote{
%%%%%%%%%%%%%% footnote %%%%%%%%%%%%%%
To check the point of how the ``averaging out the fast oscillation'' procedure works, we numerically solved the $3+1$ system explicitly and confirmed that it does, as it should be.
}   
The second approximate equalities in \eqref{average-out} assume that there is no accidental degeneracy among the sterile state masses. That is, we assume that the relation $|\Delta m^2_{JK}| \gg |\Delta m^2_{31}|$ always holds.
%\newpage

After averaging out the fast oscillations, $P(\nu_\beta \rightarrow \nu_\alpha)$ is given to second order in $W$ as 
\begin{eqnarray} 
&& P(\nu_\beta \rightarrow \nu_\alpha)^{(0+2)} 
=  P(\nu_\beta \rightarrow \nu_\alpha)^{(0)} 
+ P(\nu_\beta \rightarrow \nu_\alpha)^{(2)}. 
\end{eqnarray}
The zeroth-order term $P(\nu_\beta \rightarrow \nu_\alpha)^{(0)}$ is nothing but the one in eq.~(\ref{P-beta-alpha-final}) except for dropping the probability leaking term 
\begin{eqnarray} 
P(\nu_\beta \rightarrow \nu_\alpha)^{(0)} &=& 
\left| \sum_{j=1}^{3} U_{\alpha j} U^{*}_{\beta j} \right|^2 
- 2 \sum_{j \neq k} 
\mbox{Re} 
\left[ (UX)_{\alpha j} (UX)_{\beta j}^* (UX)_{\alpha k}^* (UX)_{\beta k} \right] 
\sin^2 \frac{ ( h_{k} - h_{j} ) x  }{ 2 }
\nonumber\\
&-&
\sum_{j \neq k} \mbox{Im} 
\left[ (UX)_{\alpha j} (UX)_{\beta j}^* (UX)_{\alpha k}^* (UX)_{\beta k} \right] 
\sin ( h_{k} - h_{j} ) x, 
\label{P-beta-alpha-2nd-averaged}
\end{eqnarray}
while the $W^2$ correction terms are given by 
%collected in 
%
\begin{eqnarray}
P(\nu_\beta \to \nu_\alpha)^{(2)} & = & 
2 \mbox{Re} 
\biggl\{
\sum_{m}
\sum_{k, K} 
\frac{ 1 }{ \Delta_{K} - h_{k} } 
\left[
(ix) e^{- i ( h_{k} - h_{m} ) x} 
- \frac{ e^{- i ( h_{k} - h_{m} ) x} }{ ( \Delta_{K} - h_{k} )  } 
\right]
\nonumber \\ 
&\times&
(UX)_{\alpha k} (UX)^*_{\beta k} 
(UX)^*_{\alpha m} (UX)_{\beta m} 
\left\{ (UX)^{\dagger} A W \right\}_{k K} 
\left\{ W ^{\dagger} A (UX) \right\}_{K k} 
\nonumber \\
&-& 
\sum_{m}
\sum_{k \neq l} \sum_{K} 
\frac{
\left( \Delta_{K} - h_{k} \right) e^{- i ( h_{l} - h_{m} ) x} 
- \left( \Delta_{K} - h_{l} \right) e^{- i ( h_{k} - h_{m} ) x} 
}{ ( h_{l}  - h_{k} ) (\Delta_{K} - h_{k}) (\Delta_{K} - h_{l}) } 
\nonumber \\
&\times&
(UX)_{\alpha k} (UX)^*_{\beta l} 
(UX)^*_{\alpha m} (UX)_{\beta m} 
\left\{ (UX)^{\dagger} A W \right\}_{k K} 
\left\{ W ^{\dagger} A (UX) \right\}_{K l} 
\nonumber \\
&-& 
\sum_{m}
\sum_{k, K} 
\frac{ e^{- i ( h_{k} - h_{m} ) x} }{ ( \Delta_{K} - h_{k} ) } 
\biggl[ 
(UX)_{\alpha k} W^*_{\beta K} 
(UX)^*_{\alpha m} (UX)_{\beta m} 
\left\{ (UX)^{\dagger} A W \right\}_{k K} 
\nonumber \\
&+& 
W_{\alpha K} (UX)^*_{\beta k} 
(UX)^*_{\alpha m} (UX)_{\beta m} 
\left\{ W ^{\dagger} A (UX) \right\}_{K k}
\biggr]
\biggr\} . 
\label{W2_correction_terms}
\end{eqnarray}
The $W^2$ correction terms in $P(\nu_\beta \to \nu_\alpha)^{(2)}$, together with the probability leaking term $\mathcal{C}_{\alpha \beta}$ in eq.~(\ref{P-beta-alpha-final}), will be utilized in section~\ref{sec:correction-terms} to explore the possibility of distinguishing between low- and high-scale unitarity violation. If such terms are detected, the sterile sector model-dependence in $P(\nu_\beta \to \nu_\alpha)^{(2)}$ would serve for identifying the structure of the sterile sector. 

\subsection{Suppression by the large sterile state mass denominator} 
\label{sec:energy-denominator}

In this section, we study the conditions under which $P(\nu_\beta \to \nu_\alpha)^{(2)}$ in eq.~\eqref{W2_correction_terms} can become negligibly small. 
It would allow us to use $P(\nu_\beta \to \nu_\alpha)^{(0)} + \mathcal{C}_{\alpha \beta}$ (= eq.~(\ref{P-beta-alpha-final})) for leptonic unitarity test in a sterile sector model-independent manner. 

We start by examining the effect of suppression by the large sterile state mass denominator which characterizes transition between active-sterile states, $1/ ( \Delta_{K} - h_{k} )$. 
We demand that the matter dependent terms in (\ref{W2_correction_terms}) be smaller than the probability leaking and the normalization terms of order $\sim W^4$. It leads to 
\begin{eqnarray} 
\biggl | \frac{ AA L }{ ( \Delta_{J} - h_{i} ) } \biggr | 
< |W|^2, 
\hspace{2.5mm}
\biggl | \frac{ AA }{ ( h_{k} - h_{j} ) ( \Delta_{J} - h_{i} ) } \biggr | 
< |W|^2, 
\hspace{2.5mm}
\text{and}
\hspace{2.5mm}
\biggl | \frac{ A }{ ( \Delta_{J} - h_{i} ) } \biggr | 
< |W|^2, 
\label{denominator-size}
\end{eqnarray}
where $L$ is the baseline distance and $i$ and $J$ denote, respectively, generic indices for active and sterile states. 
For notational convenience, we define $\lambda_{i}$ $(i=1,2,3)$ to be the eigenvalues of $3 \times 3$ submatrix $2E \tilde{H}_0$ in (\ref{tilde-H0+H1}) corresponding to the active neutrino mass squared in matter and hence $\lambda_{i} = 2 E h_i$. 

In region $\lambda_{i} \sim |\Delta m^2_{31}|$ and near the atmospheric oscillation maximum, $L \sim \frac{ 2E }{ |\Delta m^2_{31}| } \sim \frac{1}{ | h_{k} - h_{j} | }$ holds. Then, the left-hand side of the first two inequalities in \eqref{denominator-size} receive an extra factor $ |LA|  \sim \left| \frac{A}{ h_{k} - h_{j} } \right| \sim \frac{ a }{ | \Delta m^2_{31} | } \simeq 0.1 \left(\frac{\rho}{2.8 \,\text{g/cm}^3}\right) \left(\frac{E}{1~\mbox{GeV}}\right)$, which further suppresses the first and the second items in (\ref{denominator-size}) 
unless $\rho E \gsim 10\, \text{ (g/cm}^3) \text{GeV}$. 
Therefore, in this region the last one in (\ref{denominator-size}) gives the severest constraint (taking the matter potential due to CC in $A$ and removing the factor $\frac{1}{2E}$)
\begin{eqnarray} 
 \frac{ a }{ | m^2_{J} - \lambda_{i} | }  
\approx 
 \frac{ a }{ \Delta m^2_{J i}  }  
< |W|^2.
\label{suppression-cond}
\end{eqnarray}
Notice that, in order for the first inequality in
(\ref{suppression-cond}) to be valid, we have restricted the energy
region for a given matter density such that $\lambda_{i}$ remain in the
order of active neutrino masses. Roughly speaking, it corresponds to 
$- 50 \,\text{ (g/cm}^3) \text{GeV} \lsim Y_{e} \rho E \lsim 50 \,
\text{ (g/cm}^3) \text{GeV}$ where the negative sign is relevant for antineutrinos. 
See e.g., figure 3 of ref.~\cite{Minakata:2015gra}. Clearly, it excludes the interesting region of ``IceCube resonance'' due to sterile neutrino mass of eV scales \cite{Nunokawa:2003ep}, for which an entirely different theoretical framework would be necessary. 

Then, we notice that in a regime $|W|^2 \sim 10^{-2}$, the condition in (\ref{suppression-cond}) is valid given the estimation (assuming $Y_{e} = 0.5$) 
\begin{eqnarray} 
 \frac{ a }{ \Delta m^2_{J i} } = 2.13 \times 10^{-3} 
\left(\frac{ \Delta m^2_{J i} }{ 0.1~\mbox{eV}^2}\right)^{-1}
\left(\frac{\rho}{2.8 \,\text{g/cm}^3}\right) \left(\frac{E}{1~\mbox{GeV}}\right), 
\label{rA-def-value}
\end{eqnarray}
unless $\rho E \gsim 10 \, \text{ (g/cm}^3) \text{GeV}$. 
That is, the second-order matter dependent correction terms can be ignored in comparison with $\mathcal{O} (W^4)$ terms if $\Delta m^2_{J k} \gsim 0.1$ eV$^2$, which is already required in vacuum. If we want to treat the regime $|W|^2 \gsim 10^{-n}$, we need to limit the sterile masses to $\Delta m^2_{J k} \simeq m^2_{J} \gsim 10^{(n-3)}$ eV$^2$ to keep our $(3+N)$ space unitary model insensitive to details of the sterile sector \cite{Fong:2016yyh}. We note, however, that terms of order $|W|^4 \sim 10^{-4}$ may be the limit of exploration for near future neutrino oscillation experiments. 

The condition (\ref{suppression-cond}) is identical with the one obtained using the first order matter perturbation theory \cite{Fong:2016yyh}, which may look strange to the readers. Let us understand the reason why taking care of all order matter effect does not alter the condition obtained by first-order treatment in matter perturbation theory. The matter-dependent term in the zeroth-order Hamiltonian $\tilde{H}_{0}$ only involves $U$ matrix, but no $W$ matrix. Since we treat $\tilde{H}_{0}$ in an unperturbed fashion it produces all-order effect of the matter potential which is however independent of $W$ matrix elements. 
On the other hand, perturbative effects that come from single or double powers in $W$ in $\hat{H}_{1}$ are always accompanied by the matter potential in the form of $WA$ or $W^{\dagger} A$, as in eq.~(\ref{H1-matrix}). That is, perturbative effect of $W$ is always accompanied by matter potential, and hence can always be dealt with matter perturbation theory.\footnote{
%%%%%%%%%%%%%% footnote %%%%%%%%%%%%%%
An example of this feature can be observed in eq.~(7.13) in ref.~\cite{Fong:2016yyh}. 
We must remark, however, that this reasoning does not prove that the first order in matter perturbation theory is sufficient to obtain all the necessary conditions on the sterile state masses. 
}
It is the reason why the matter perturbation theory is able to yield the same condition on sterile masses as obtained in a fuller treatment of matter effect done in this paper.

\subsection{The oscillation probability in fourth order in $W$}
\label{sec:probability-4th}

The oscillation probability in fourth order in $W$ contains the two terms 
\begin{eqnarray} 
&&P(\nu_\beta \rightarrow \nu_\alpha)^{(4)} = 
\left| S^{(2)}_{\alpha \beta} \right|^2 
+ 2 \mbox{Re} \left[ \left( S^{(0)}_{\alpha \beta} \right)^{*} S^{(4)}_{\alpha \beta} \right]. 
\label{P-beta-alpha-4th-def}
\end{eqnarray}
We will show in appendix~\ref{sec:second-order-square} that the first term 
in (\ref{P-beta-alpha-4th-def}), after averaging over the fast oscillations 
and using the suppression by large sterile state mass denominator as discussed in the previous section, leaves the unique term, the probability leaking term $\mathcal{C}_{\alpha \beta}$ in eq.~\eqref{P-beta-alpha-final}, 
which can be seen in eq.~(\ref{S(2)-squared-1}). 
An interesting feature of $\mathcal{C}_{\alpha \beta}$ in matter is that it is identical to the one in vacuum, eq.~(\ref{Cab}) without any matter effect dressing. In our computation the term comes from the hat basis $S$ matrix in zeroth order, the first term in the last line of eq.~(\ref{hat-S-elements-1st}), and hence it is free from the matter potential.\footnote{
%%%%%%%%%%%%% footnote %%%%%%%%%%%%%%
One may suspect that including higher order corrections could alter the feature of matter potential independence of $\mathcal{C}_{\alpha \beta}$. However, one can show (see section~\ref{sec:U-theorem}) that higher order $W$ corrections to the piece of $S$ matrix elements relevant to $\mathcal{C}_{\alpha \beta}$ organize themselves as a phase factor, so that $\mathcal{C}_{\alpha \beta}$ has no matter effect dressing. The rest of the correction terms are suppressed due to the dimensional reason, an extra matter potential must be accompanied by an energy denominator.
}
We will also show in appendix~\ref{sec:interference} that the second term in (\ref{P-beta-alpha-4th-def}), under the same treatment for the first term, gives vanishing contribution. Therefore, no matter-dependent fourth order term survives after large sterile state mass denominator suppression is used and averaging over the fast oscillations is performed.

In conclusion, the oscillation probability in matter between active flavour neutrinos in the $(3+N)$ space unitary model to fourth order in $W$ in our small unitarity-violation perturbation theory can be written as in
eq.~(\ref{P-beta-alpha-final}) in section~\ref{sec:essence}. We hope that it serves as a useful tool to test leptonic unitarity in various ongoing and future neutrino oscillation experiments. 

\section{Analytical and numerical methods for solving non-unitary evolution in matter }
\label{sec:analytical-numerical}

In this section, we describe the numerical and analytical methods for calculating the neutrino oscillation probability by solving non-unitary evolution in matter. 

\subsection{Numerical method for calculating neutrino oscillation probability }
\label{sec:numerical}

We describe a numerical method for computing the oscillation probability in matter. This method can be used, assuming adiabaticity, in cases with varying matter density. We show that in zeroth order in $W$ the system simplifies to an evolution equation in the $3 \times 3$ active subspace.

We solve the Schr\"odinger equation in the vacuum mass eigenstate basis (``tilde basis''), $\tilde{\nu}_{z} = ({\bf U}^{\dagger})_{z \zeta} \nu_{\zeta}$ with Hamiltonian $\tilde{H}$ in (\ref{tilde-H}): 
\begin{eqnarray}
i \frac{d}{dx} 
\left[
\begin{array}{c}
\tilde{\nu}_{i} \\
\tilde{\nu}_{J} \\
\end{array}
\right] = 
\left[
\begin{array}{cc}
{\bf \Delta_{a} } + U^{\dagger} A U & U^{\dagger} A W \\
W^{\dagger} A U & {\bf \Delta_{s} } + W^{\dagger} A W \\
\end{array}
\right] %
\left[
\begin{array}{c}
\tilde{\nu}_{i} \\
\tilde{\nu}_{J} \\
\end{array}
\right],
\label{Schroedinger-eq}
\end{eqnarray}
where $i = 1,2,3$ and $J= 4,5,\cdot \cdot \cdot,3+N$ denote mostly
active and mostly sterile neutrino mass eigenstate labels,
respectively. 
The initial condition with only active component implies 
\begin{eqnarray}
%\tilde{\nu}_{i} (0) &=& \sum_{\alpha} (U^{\dagger})_{i \alpha} \nu_{\alpha} (0), \nonumber \\
%\tilde{\nu}_{J} (0) &=& \sum_{\alpha} (W^{\dagger})_{J \alpha} \nu_{\alpha} (0). 
\tilde{\nu}_{i} (0) = \sum_{\alpha} (U^{\dagger})_{i \alpha} \nu_{\alpha} (0), \hspace{5mm}
\tilde{\nu}_{J} (0) = \sum_{\alpha} (W^{\dagger})_{J \alpha} \nu_{\alpha} (0). 
\label{initial-condition}
\end{eqnarray}
Using the solution of equation (\ref{Schroedinger-eq}), we need the wave function of active flavour component to calculate the probability at baseline $x=L$. 
\begin{eqnarray}
\nu_{\alpha} (L) &=& 
\sum_{i} U_{\alpha i} \tilde{\nu}_{i} (L) 
+ \sum_{J} W_{\alpha J} \tilde{\nu}_{J} (L).
\label{final-condition}
\end{eqnarray}
Therefore, in the mass-basis formulation only $U$ and $W$ are involved, which is consistent with our experience in $W$ perturbation theory. An apparent contradiction to this property that one faces in the evolution equation in the flavour basis is resolved in appendix~\ref{sec:flavor-basis-evolution}. 

A drawback of this method is that we have to solve explicitly the evolution of the sterile states which are coupled to the active states. Then, we need to specify the sterile sector model, and have to know how to deal with averaging over the fast modes.

We notice, however, that in the zeroth-order in $W$ the system simplifies. Since the Hamiltonian $\tilde{H}$ is block-diagonal it suffices to solve the equation only in the $3 \times 3$ active neutrino subspace: 
\begin{eqnarray} 
i \frac{d}{dx} \nu_{i} = 
\sum_{j}
\left( {\bf \Delta_{a} } + U^{\dagger} A U \right)_{ij} 
\nu_{j}. 
\label{Schroedinger-eq-0th}
\end{eqnarray}
The initial condition (\ref{initial-condition}) and final reverse-back formula (\ref{final-condition}) involve only $U$ matrix elements. Therefore, the oscillation probability in the zeroth-order in $W$ can be calculable in a manner independent of sterile sector models.\footnote{
%%%%%%%%%%%%% footnote %%%%%%%%%%%%%%
As we remarked in footnote~8 the non-unitary mixing matrix $U$ has some $W$ dependence through unitarity of the ${\bf U}$ matrix in the whole $(3+N)$ space. Therefore, the nature of the eq. (\ref{Schroedinger-eq-0th}) as the zeroth-order in $W$ is ambiguous. However, following \cite{Fong:2016yyh}, we remain in the treatment with this ``$W$ effect renormalized basis'' in this paper.
}

\subsection{An exact solution of zeroth-order oscillation probability }
\label{sec:exact-solution-zeroth}

Here, we describe a method for obtaining the analytical solution of the zeroth-order Hamiltonian. The exact solution, as well as the numerical one described in the previous section, provides the basis for computing the higher order corrections in $W$. 

We calculate an exact form of the oscillation probability $P(\nu_\beta \rightarrow \nu_\alpha)$ in leading order in our perturbative framework, the one in (\ref{P-beta-alpha-final}) except for $\mathcal{C}_{\alpha \beta}$, in the case of uniform matter density.  

The zeroth-order $S$ matrix element $S_{\alpha \beta}^{(0)}$ in (\ref{S-alpha-beta-0th}) can be written as 
\begin{eqnarray} 
S_{\alpha \beta}^{(0)} &=& 
\sum_{i, j} U_{\alpha i} U^*_{\beta j} 
\left( 
\sum_{k} X_{i k} X^*_{j k} e^{-i h_{k} x}
\right),
\label{S-alpha-beta-0th-KTY}
\end{eqnarray}
and the factor in parenthesis can be calculated by the KTY technique \cite{Kimura:2002wd}. 
We want to diagonalize the Hamiltonian 
\begin{eqnarray} 
H_{0} \equiv 
\frac{1}{2E} 
\left\{  
\left[
\begin{array}{ccc}
m^2_{1} & 0 & 0 \\
0 & m^2_{2} & 0 \\
0 & 0 & m^2_{3} \\
\end{array}
\right] + 
U^{\dagger} \left[
\begin{array}{ccc}
a - b & 0 & 0 \\
0 & -b & 0 \\
0 & 0 & -b \\
\end{array}
\right] U 
\right\} ,
\label{H0-def}
\end{eqnarray}
the active $3 \times 3$ block of $\tilde{H}_0$ in (\ref{tilde-H0+H1}). We have defined in eq.~(\ref{H0-diag}) the unitary matrix $X$ which diagonalize $H_{0}$ as 
\begin{eqnarray} 
H_{0} = 
\frac{1}{2E} 
X \left[
\begin{array}{ccc}
\lambda_{1} & 0 & 0 \\
0 & \lambda_{2} & 0 \\
0 & 0 & \lambda_{3} \\
\end{array}
\right] X^{\dagger}
\equiv 
H_{d}. 
\label{H0-diagonal}
\end{eqnarray}
For our notational convenience we call this form of $H_{0}$ as $H_{d}$. Note that $h_{i} = \frac{ \lambda_{i} }{2E}$ where
\begin{eqnarray}
\lambda_{1,2} & = & \frac{\cal T}{3}\mp\frac{1}{3}{\cal F}
\cos {\cal G}  -\frac{1}{\sqrt{3}}{\cal F}\sin{\cal G}, \hspace{5mm}
\lambda_{3}  =  \frac{\cal T}{3}+ \frac{2}{3}{\cal F} \cos{\cal G},
\end{eqnarray}
% \begin{eqnarray}
% \lambda_{1} & = & \frac{\cal T}{3}-\frac{1}{3}\sqrt{{\cal T}^{2}-3{\cal
%  A}}\cos {\cal G}  -\frac{1}{\sqrt{3}}\sqrt{{\cal T}^{2}-3{\cal A}}\sin{\cal G},\\
% \lambda_{2} & = & \frac{\cal T}{3}-\frac{1}{3}\sqrt{{\cal T}^{2}-3{\cal
%  A}}\cos{\cal G}
% +\frac{1}{\sqrt{3}}\sqrt{{\cal T}^{2}-3{\cal A}}\sin{\cal G},\\
% \lambda_{3} & = & \frac{\cal T}{3}+\frac{2}{3}\sqrt{{\cal T}^{2}-3{\cal A}}\cos{\cal G},
% \end{eqnarray}
%
where 
\begin{eqnarray}
{\cal F} \equiv \sqrt{{\cal T}^{2}-3{\cal  A}}, \hspace{5mm}
{\cal G} \equiv
\frac{1}{3}\arccos\left\{\frac{2{\cal T}^{3}-9{\cal AT}+27{\cal D}}{2\left({\cal T}^{2}-3{\cal A}\right)^{3/2}}\right\},
\end{eqnarray}
with
\begin{eqnarray}
%\cal{T} & = & (2E)\, {\rm Tr} H_0,\\
%\cal{A} & = & (2E)^2\, {\rm Tr}\left({\rm Adj} H_0 \right),\\
%\cal{D} & = & (2E)^3 \det H_0.
{\cal T} =  (2E)\, {\rm Tr} H_0, \hspace{5mm}
{\cal A} =  (2E)^2\, {\rm Tr}\left({\rm Adj} H_0 \right), \hspace{5mm}
{\cal D}  = (2E)^3 \det H_0.
\end{eqnarray}
The adjugate of $H_{0}$ is defined as $\text{Adj} H_{0} \equiv (H_{0})^{-1} \text{det} H_{0}$.
Notice that $\cal T$, $\cal A$ and $\cal D$ are invariant under unitary transformation
of $ H_0\to K H_0 K^{\dagger}$ with $K$ any unitary matrix and so are $\lambda_i$.

Following the notation in \cite{Kimura:2002wd} we define $p_{ij}$ and $q_{ij}$ as ($i,j=1,2,3$)
\begin{eqnarray} 
\frac{ p_{ij} }{ 2E } \equiv \left( H_{0} \right)_{ij}, \hspace{5mm}
\frac{ q_{ij} }{ (2E)^2 } \equiv \left( \text{Adj} H_{0} \right)_{ij}. 
\label{pq-def}
\end{eqnarray}
Notice that $p_{ij}$ and $q_{ij}$ are written only by the known (or given) quantities. Then, the equations 
\begin{eqnarray} 
\left( H_{d} \right)_{ij} = \frac{ p_{ij} }{ 2E }, \hspace{5mm}
%
%\nonumber \\
\left( \text{Adj} H_{d} \right)_{ij} = \frac{ q_{ij} }{ (2E)^2 }, 
\label{KTY-eq}
\end{eqnarray}
together with unitarity of $X$, become the equations to determine $X X^{\dagger}$: 
\begin{eqnarray} 
X_{i 1} X_{j 1}^* + X_{i 2} X_{j 2}^* + X_{i 3} X_{j 3}^* &=& \delta_{ij}, 
\nonumber \\
\lambda_{1} X_{i 1} X_{j 1}^* + \lambda_{2} X_{i 2} X_{j 2}^* + \lambda_{3}X_{i 3} X_{j 3}^* &=& p_{ij}, 
\nonumber \\
\lambda_{2} \lambda_{3} X_{i 1} X_{j 1}^* + \lambda_{3} \lambda_{1} X_{i 2} X_{j 2}^* + \lambda_{1} \lambda_{2} X_{i 3} X_{j 3}^* &=& q_{ij}. 
\label{KTY-eq-explicit}
\end{eqnarray}
They lead to the solution ($k=1,2,3$)
\begin{eqnarray} 
X_{i k} X_{j k}^* = 
\frac{ q_{ij} + p_{ij} \lambda_{k} - \delta_{ij} \lambda_{k} ( \lambda_{l} + \lambda_{m} ) }{ (\lambda_{l} -\lambda_{k} ) (\lambda_{m} -\lambda_{k} ) },
\label{KTY-eq-solution}
\end{eqnarray}
where $k,l,m$ is cyclic, and sum over $k$ is not implied in (\ref{KTY-eq-solution}). 

Therefore, to zeroth-order in $W$ expansion, the $S$ matrix elements are given by 
\begin{eqnarray} 
S_{\alpha \beta}^{(0)} &=& 
\sum_{k} 
\left(
\sum_{i, j} 
U_{\alpha i} 
\left[ q_{ij} + p_{ij} \lambda_{k} - \delta_{ij} \lambda_{k} ( \lambda_{l} + \lambda_{m} ) \right] 
U^*_{\beta j} 
\right)
\frac{ e^{-i h_{k} x} }{ (\lambda_{l} -\lambda_{k} ) (\lambda_{m} -\lambda_{k} ) }, 
\label{S-alpha-beta-0th-final}
\end{eqnarray}
and the oscillation probability by $P(\nu_\beta \rightarrow \nu_\alpha) = \vert S_{\alpha \beta}^{(0)} \vert^2$.  

Finally, armed with the solution \eqref{KTY-eq-solution}, we can also calculate all higher order terms in oscillation probability for e.g. 
those in eq.~\eqref{W2_correction_terms} since only such combination $X_{ik} X_{jk}^*$ (no sum over $k$ implied) can appear.

\section{ Where are the unitarity violation and  $W^2$ corrections? } 
\label{sec:where-UV}

Having formulated the small unitarity violation perturbation theory, we now utilize it to answer the following questions: (1) Where is the regions of energy $E$ and baseline $L$ in which the effect of unitarity violation is significant?, and (2) how large can the $W^2$  corrections be? 
We address the questions (1) and (2) in sections~\ref{sec:with-without-UV} and \ref{sec:correction-terms}, respectively. 

\subsection{Comparison between the oscillation probabilities with and without unitarity violation } 
\label{sec:with-without-UV}

To know where the effect of unitarity violation is large, and how large it is, we calculate 
\begin{eqnarray} 
\Delta P (\nu_{\beta} \rightarrow \nu_{\alpha}) \equiv P(\nu_{\beta} \rightarrow \nu_{\alpha})_{ \text{standard} }- P(\nu_{\beta} \rightarrow \nu_{\alpha})_{ \text{non-unitary} }^{(0)}
\label{Delta-P-def}
\end{eqnarray}
as a function of $E$ and $L$, where 
$P(\nu_{\beta} \rightarrow \nu_{\alpha})_{ \text{standard} }$ and 
$P(\nu_{\beta} \rightarrow \nu_{\alpha})_{ \text{non-unitary} }^{(0)}$ imply, respectively, the oscillation probabilities calculated with the standard unitary mixing matrix and the leading order (i.e., $W^0$) one with non-unitarity. The probability leaking term $\mathcal{C}_{\alpha \beta}$ in eq.~\eqref{P-beta-alpha-final} as well as the $W^2$ correction terms in eq.~\eqref{W2_correction_terms} are not included in the analysis here. Therefore, the results given in section~\ref{sec:with-without-UV} apply to both high-scale unitarity violation as well as low-scale one in its leading order in $W$.\footnote{
%%%%%%%%%%%%%% footnote %%%%%%%%%%%%%%%
In fact, it is in agreement with the formulations in ref.~\cite{Blennow:2016jkn} with which we share the same evolution equation (\ref{Schroedinger-eq-0th}) in the vacuum mass eigenstate basis. See also \cite{Antusch:2006vwa}. However, it appears that the flavour basis formulation of neutrino evolution in matter in high-scale unitarity violation poses some nontrivial features such as non-Hermitian Hamiltonian \cite{Antusch:2006vwa}, or the evolution equation $i \frac{d}{dx} \nu_{\alpha} = \sum_{j} \left[ U \left( {\bf \Delta_{a} } + U^{\dagger} A U \right) U^{\dagger} \right]_{\alpha \beta} \nu_{\beta}$ \cite{Escrihuela:2016ube}. The latter is not equivalent to (\ref{Schroedinger-eq-0th}) in the vacuum mass eigenstate basis due to non-unitarity of the $U$ matrix.
}
On intuitive ground, at zeroth order in $W$ our system describes high-scale unitarity violation. There is no ``$W$ corrections'' in high-scale unitarity violation because the energy scale is so high that the high-mass sector is truncated. 
We examine the three channels $\nu_{\mu} \rightarrow \nu_{e}$, $\nu_{\mu} \rightarrow \nu_{\tau}$, and $\nu_{\mu} \rightarrow \nu_{\mu}$. However, we do not enter into any quantitative analyses, nor attempt to cover the whole parameter space. 

Here is a brief note on how the standard mixing and the unitarity violating parameters are chosen: 
We take the $(3+1)$ model in which the constraints on the parameters are best understood \cite{Kopp:2013vaa,deGouvea:2015euy,Fernandez-Martinez:2016lgt,TheIceCube:2016oqi}. In consistent with the current constraints we have chosen: 
$\sin^2\theta_{14} = 0.02$, $\sin^2\theta_{24} = 0.01$, and $\sin^2\theta_{34} = 0.1$ for $\Delta m^2_{41} = 0.1$ eV$^2$, and set all the CP phases to zero. 
Then, we cut out the $3\times3$ active neutrino mixing matrix, which is non-unitary.\footnote{
%%%%%%%%%%%%% 
It can be re-parameterized in terms of the ``$\alpha$ matrix parameterization'' defined in ref.~\cite{Escrihuela:2015wra}. 
The resultant values of $\alpha$ parameters are given as follows:
$\alpha_{11} = 0.990$,  
$\alpha_{21} = - 0.0141$,  
$\alpha_{22} = 0.995$,  
$\alpha_{31} = -0.0445$,  
$\alpha_{32} = -0.0316$,  
$\alpha_{33} = 0.949$.
}
For the standard leptonic mixing parameters in $U_{\text{\tiny PDG}}$, we take $\sin^2\theta_{12} = 0.3$, $\sin^2\theta_{23} = 0.5$, $\sin^2 (2\theta_{13}) = 0.09$, and the mass squared differences $\Delta m_{21}^2 = 7.4 \times 10^{-5}$ eV$^2$ and $\Delta m_{31}^2 = 2.4 \times 10^{-3}$ eV$^2$, and set the CP phase $\delta_{\text{CP}}$ to zero. 
The uniform matter density is taken as $\rho = 3.2~{\rm g\,cm}^{-3} $ over the entire baseline, which may not be realistic.\footnote{
%%%%%%%%%%%%%% footnote %%%%%%%%%%%%%%%
One can apply our formulas of $S$ matrix obtained under the constant matter density approximation to semi-realistic calculation for earth crossing neutrinos by using them in each shell (core, mantle, and crust regions, etc.) with proper connecting conditions at the boundaries. 
}

%%%%%%%%%%%%%%% FIG 1 %%%%%%%%%%%%%%%
\begin{figure}[h!]
\begin{center}
%\vspace{-4mm}
%\hspace{-18mm}
%\includegraphics[bb=0 0 792 600,width=0.68\textwidth]{Pmue_energy_dist_non_unitary_small_size.jpeg}
%\includegraphics[bb=0 0 792 600,width=0.68\textwidth]{Pmue_energy_dist_difference_small_size.jpeg}
\includegraphics[width=0.85\textwidth]{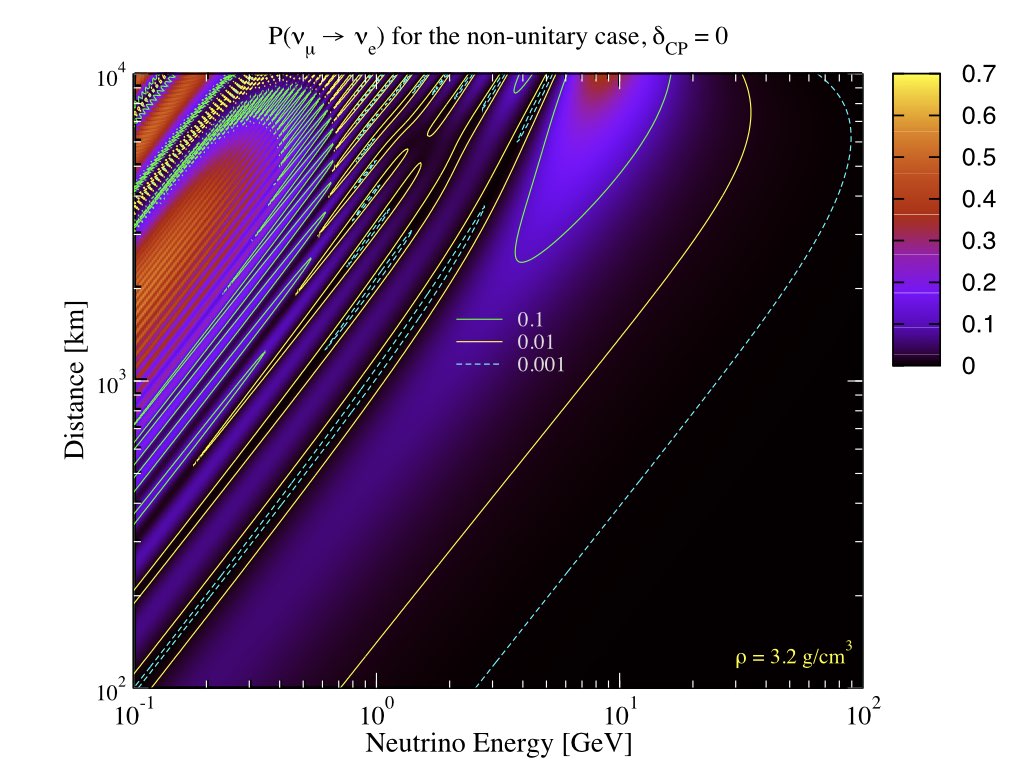}
\includegraphics[width=0.85\textwidth]{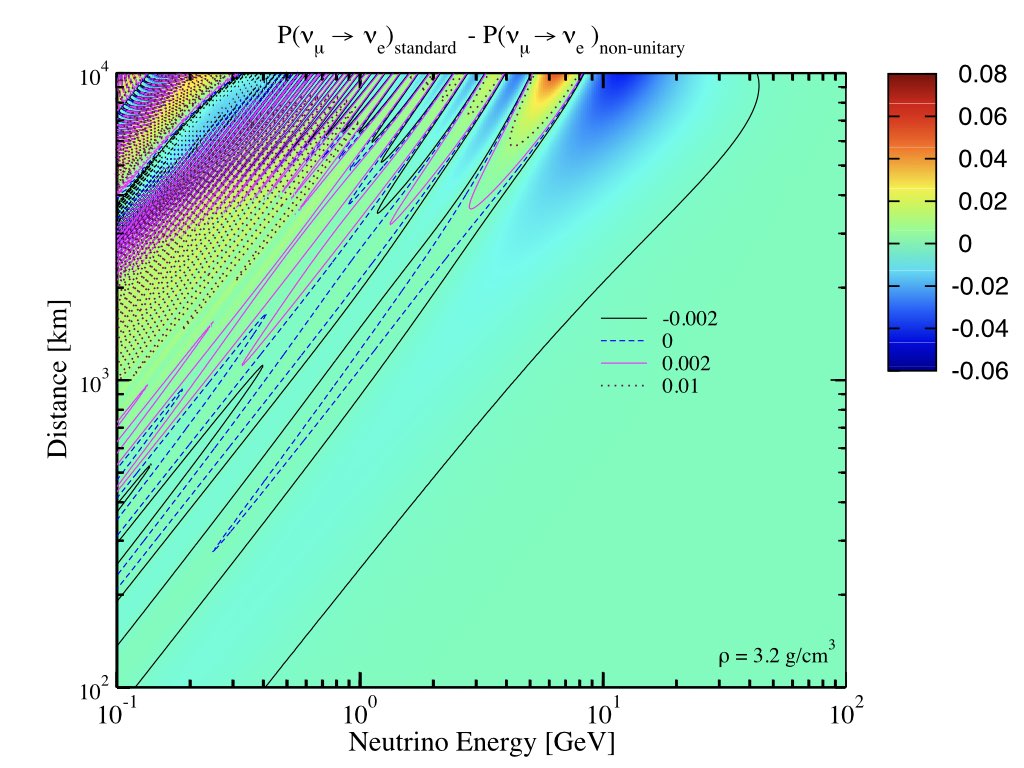}
\end{center}
\vspace{-3mm}
\caption{ 
In the upper panel (a), presented is the iso-contour of $P(\nu_{\mu} \rightarrow \nu_{e})_{ \text{non-unitary} }^{(0)}$ in space spanned by neutrino energy $E$ and baseline $L$. In the lower panel (b), the iso-contour of the difference $\Delta P (\nu_{\mu} \rightarrow \nu_{e}) \equiv P(\nu_{\mu} \rightarrow \nu_{e})_{ \text{standard} }- P(\nu_{\mu} \rightarrow \nu_{e})_{ \text{non-unitary} }^{(0)}$ is presented. For the values of unitarity-violating as well as the standard mixing parameters taken, see the text.
}
\label{fig:Pmue_energy_dist}
\end{figure}
%%%%%%%%%%%%%%% FIG 1 %%%%%%%%%%%%%%%
%
%%%%%%%%%%%%%%% FIG 2 %%%%%%%%%%%%%%%
\begin{figure}[h!]
\begin{center}
%\vspace{-4mm}
%\hspace{-18mm}
\includegraphics[width=0.85\textwidth]{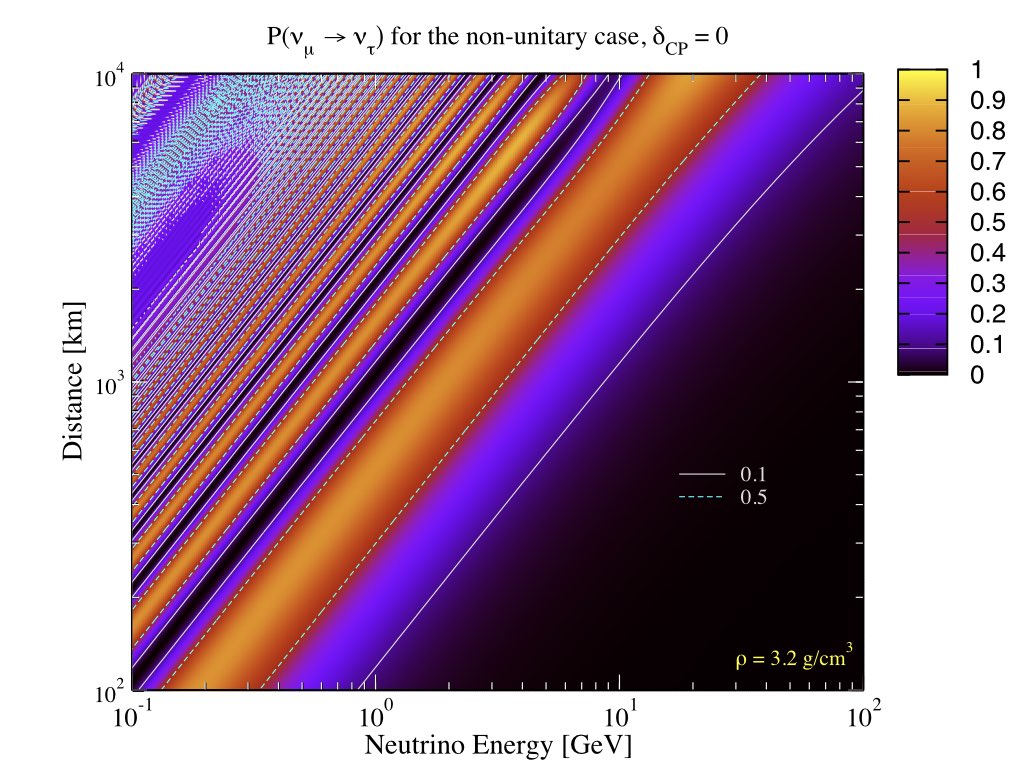}
\includegraphics[width=0.85\textwidth]{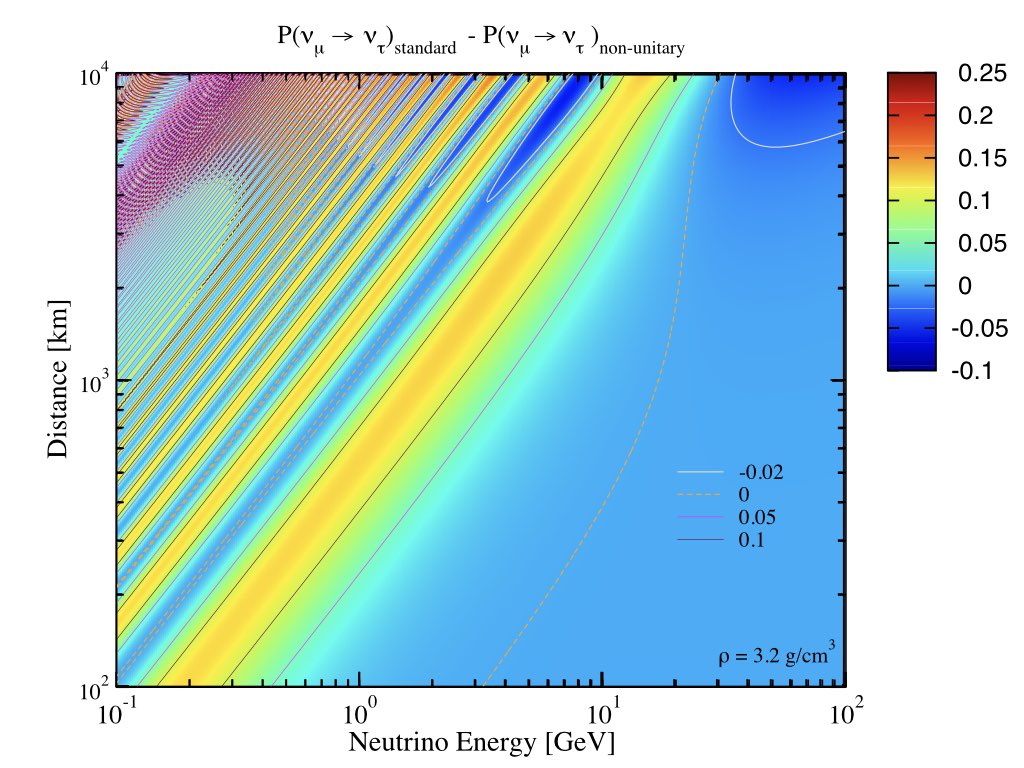}
\end{center}
\vspace{-3mm}
\caption{ 
In the upper panel (a), presented is the iso-contour of $P(\nu_{\mu} \rightarrow \nu_{\tau})_{ \text{non-unitary} }^{(0)}$ in $E-L$ space. In the lower panel (b), the iso-contour of the difference $\Delta P (\nu_{\mu} \rightarrow \nu_{\tau}) \equiv P(\nu_{\mu} \rightarrow \nu_{\tau})_{ \text{standard} }- P(\nu_{\mu} \rightarrow \nu_{\tau})_{ \text{non-unitary} }^{(0)}$ is presented. The parameters used are the same as in figure~\ref{fig:Pmue_energy_dist}. 
}
\label{fig:Pmutau_energy_dist}
\end{figure}
%%%%%%%%%%%%%%% FIG 2 %%%%%%%%%%%%%%%
%
%%%%%%%%%%%%%%% FIG 3 %%%%%%%%%%%%%%%
\begin{figure}[h!]
\begin{center}
%\vspace{-4mm}
%\hspace{-18mm}
\includegraphics[width=0.85\textwidth]{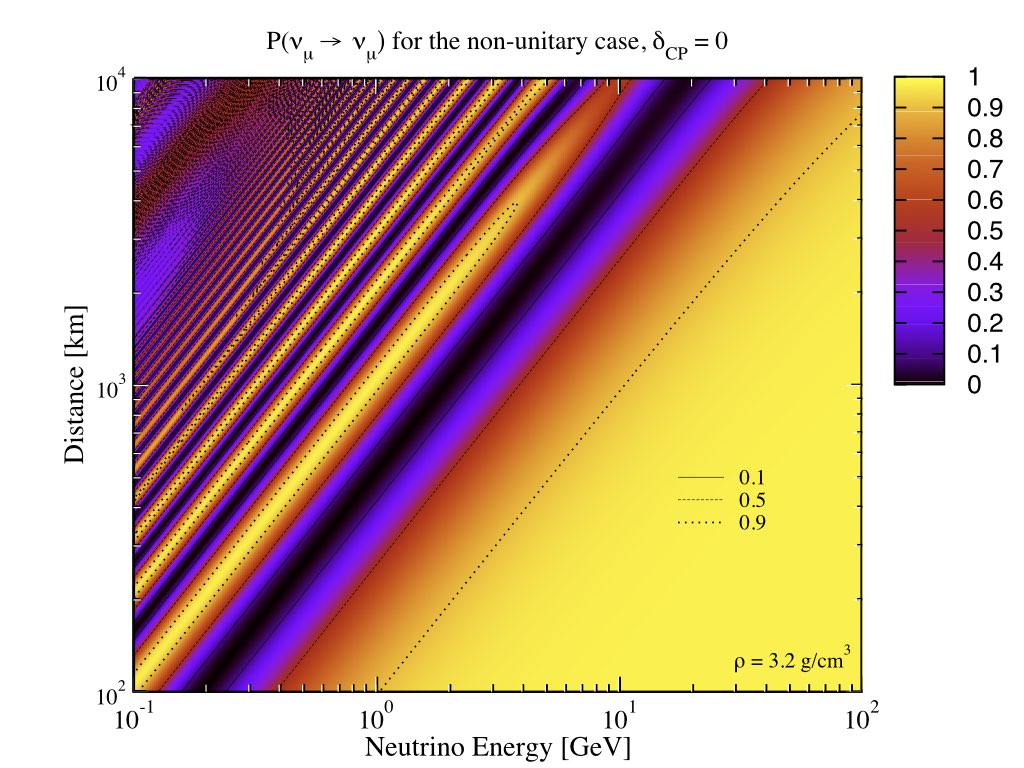}
\includegraphics[width=0.85\textwidth]{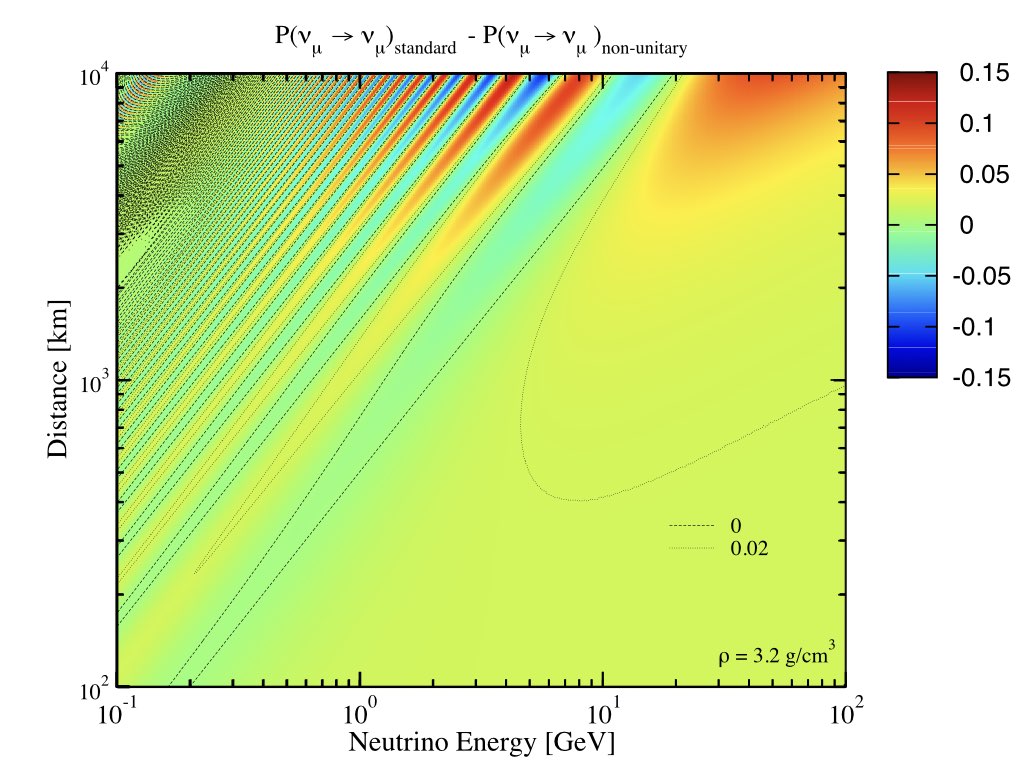}
\end{center}
\vspace{-3mm}
\caption{ 
In the upper panel (a), presented is the iso-contour of $P(\nu_{\mu} \rightarrow \nu_{\mu})_{ \text{non-unitary} }^{(0)}$ in $E-L$ space. In the lower panel (b), the iso-contour of the difference $\Delta P (\nu_{\mu} \rightarrow \nu_{\mu}) \equiv P(\nu_{\mu} \rightarrow \nu_{\mu})_{ \text{standard} }- P(\nu_{\mu} \rightarrow \nu_{\mu})_{ \text{non-unitary} }^{(0)}$ is presented. The parameters used are the same as in figure~\ref{fig:Pmue_energy_dist}. 
}
\label{fig:Pmumu_energy_dist}
\end{figure}
%%%%%%%%%%%%%%% FIG 3 %%%%%%%%%%%%%%%

\subsubsection{$P(\nu_{\mu} \rightarrow \nu_{e})$ } 

In figure~\ref{fig:Pmue_energy_dist}-(a) (upper panel) and (b) (lower panel), presented are the iso-contours of 
$P(\nu_{\mu} \rightarrow \nu_{e})_{ \text{non-unitary} }^{(0)}$ and 
$\Delta P (\nu_{\mu} \rightarrow \nu_{e}) \equiv P(\nu_{\mu} \rightarrow
\nu_{e})_{ \text{standard} }- P(\nu_{\mu} \rightarrow \nu_{e})_{
\text{non-unitary} }^{(0)}$ in $E - L$ space. Here, the superscript
$(0)$ implies that it is calculated in zeroth-order in $W$ by solving (\ref{Schroedinger-eq-0th}) with appropriate initial condition and final projection to flavour eigenstate. 
In most of the $E - L$ space 
$P(\nu_{\mu} \rightarrow \nu_{e})_{ \text{non-unitary} }^{(0)}$
is small. However, we identify the two regions where 
$P(\nu_{\mu} \rightarrow \nu_{e})_{ \text{non-unitary} }^{(0)}$
	is relatively large, $\gsim 0.3$. One of them is at low energy, $E \lsim \text{a few hundred}$ MeV, and baseline $L \gsim$1000 km. The other one is a region $E \sim 10$ GeV and $L \sim$ 10000 km. The former may be understood as due to the solar MSW enhancement, and the latter as the atmospheric MSW enhancement \cite{Mikheev:1986gs,Wolfenstein:1977ue}. Roughly speaking, the regions with relatively large $| \Delta P (\nu_{\mu} \rightarrow \nu_{e}) |$ overlap with these regions. 

\subsubsection{$P(\nu_{\mu} \rightarrow \nu_{\tau})$ and $P(\nu_{\mu} \rightarrow \nu_{\mu})$ } 

In figures~\ref{fig:Pmutau_energy_dist} and \ref{fig:Pmumu_energy_dist}, the same quantities (in each upper (a) and lower (b) panels) are presented but in $\nu_{\mu} \rightarrow \nu_{\tau}$ and $\nu_{\mu} \rightarrow \nu_{\mu}$ channels, respectively. 
In contrast to $\nu_{\mu} \rightarrow \nu_{e}$ channel, 
$P(\nu_{\mu} \rightarrow \nu_{\tau})_{ \text{non-unitary} }^{(0)}$
and 
$P(\nu_{\mu} \rightarrow \nu_{\mu})_{ \text{non-unitary} }^{(0)}$
contours are globally ``vacuum effect dominated'', apart from the solar MSW region, both in the standard (not shown) and the non-unitary cases. The first oscillation peak of 
$P(\nu_{\mu} \rightarrow \nu_{\tau})_{ \text{non-unitary} }^{(0)}$
scales roughly as the vacuum oscillation peak does, $L / 10^3 \,\text{km} = 0.33 E / 1 \, \text{GeV}$. This feature is more or less seen in 
$P(\nu_{\mu} \rightarrow \nu_{e})_{ \text{non-unitary} }^{(0)}$
, but 
$P(\nu_{\mu} \rightarrow \nu_{\tau})_{ \text{non-unitary} }^{(0)}$
 has a higher peak height $\simeq 0.7-0.8$, and the effect of atmospheric MSW enhancement is less prominent. 

For $P(\nu_{\mu} \rightarrow \nu_{\mu})_{ \text{non-unitary} }^{(0)}$, roughly speaking, the relation 
$P(\nu_{\mu} \rightarrow \nu_{\mu})_{ \text{non-unitary} }^{(0)} \approx 1 - P(\nu_{\mu} \rightarrow \nu_{\tau})_{ \text{non-unitary} }^{(0)} $
 holds in region where 
$P(\nu_{\mu} \rightarrow \nu_{e})_{ \text{non-unitary} }^{(0)}$
  is small. It must be the case in the unitary case, but even in non-unitary case the relation holds approximately because unitarity violation is small in our choice of the parameters. Therefore, 
$P(\nu_{\mu} \rightarrow \nu_{\mu})_{ \text{non-unitary} }^{(0)}$
  is large in region where 
$P(\nu_{\mu} \rightarrow \nu_{\tau})_{ \text{non-unitary} }^{(0)}$
  is small, and vice versa, as seen in figure~\ref{fig:Pmumu_energy_dist}. It appears that the anticorrelation is inherited to the relationship between $\Delta P (\nu_{\mu} \rightarrow \nu_{\mu})$ and $\Delta P (\nu_{\mu} \rightarrow \nu_{\tau})$. 
Relatively large $\Delta P (\nu_{\mu} \rightarrow \nu_{\tau})$ in first a few oscillation maxima, or similar large depletion of $\Delta P (\nu_{\mu} \rightarrow \nu_{\mu})$, would allow detection of non-unitarity if the detector has a good $\tau$ (in the former channel), or $\mu$ (in the latter channel) detection capabilities. If the detector can detect the both, anticorrelation between $\mu$ and $\tau$ yields must help. 

Some comments on observational aspects: 
In the two regions where $| \Delta P (\nu_{\mu} \rightarrow \nu_{e}) |$ is large, and $\Delta P (\nu_{\mu} \rightarrow \nu_{\mu})$ in energy region $E \lsim 10$~GeV may be explored by high-statistics atmospheric neutrino observation by Super-K, Hyper-K/HKK, or DUNE \cite{Abe:2017aap,Abe:2015zbg,Abe:2016ero,Acciarri:2015uup}. 
The atmospheric MSW enhanced region of $P(\nu_{\mu} \rightarrow \nu_{e})$ would be a good target for PINGU extensions of IceCube and KM3NeT-ORCA \cite{TheIceCube-Gen2:2016cap,Adrian-Martinez:2016zzs}. 
$P(\nu_{\mu} \rightarrow \nu_{\tau})$ and $P(\nu_{\mu} \rightarrow \nu_{\mu})$ would be explored by them, with possibility of seeing anticorrelation between $\mu$ and $\tau$ yields.  
Although it is very interesting to investigate these experimental prospects, a detailed examination of these questions is beyond the scope of this paper.

\subsection{The probability leaking and $W^2$ correction terms} 
\label{sec:correction-terms} 

\subsubsection{Low-scale versus high-scale unitarity violation } 
\label{sec:low-vs-high}

In leptonic unitarity test, a clear understanding of the relationship between low-scale and high-scale unitarity violation may be one of the key issues. We have stressed in our previous paper \cite{Fong:2016yyh} that observing the probability leaking term $\mathcal{C}_{\alpha \beta}$ in eq.~(\ref{Cab}) would testify for low-scale unitarity violation. As mentioned in section~\ref{sec:probability-4th} the leaking term is not dressed by the matter effect, which is perfectly natural for the effect of probability leakage. 
In this paper, we propose yet another way of distinguishing low-scale unitarity violation from high-scale one. That is, detection of the $W^2$ correction terms in eq.~\eqref{W2_correction_terms}. In this section~\ref{sec:correction-terms}, we give a brief sketch of how and where we might see visible effects of the probability leaking and the $W^2$ correction terms.

\subsubsection{How large are the $W^2$ corrections and $\mathcal{C}_{\alpha \beta}$? } 
\label{sec:how-large}

Let us go back to the expression of the oscillation probability to second order in $W$, eq.~(\ref{W2_correction_terms}), in section~\ref{sec:probability-2nd} to know where we might see visible effects. If we enter into the region $\rho E \gg 10 \, \text{ (g/cm}^3) \text{GeV}$ at around the first oscillation maximum, 
the first two terms in
%$2 \mbox{Re} \{ \cdot \cdot \cdot \}$ 
eq.~\eqref{W2_correction_terms} can become large apart from $W^2$ suppression, 
\begin{eqnarray} 
%&& 
\biggl | \frac{ AA L }{ ( \Delta_{J} - h_{i} ) } \biggr | 
&\sim& 
\biggl | \frac{ AA }{ ( h_{k} - h_{j} ) ( \Delta_{J} - h_{i} ) } \biggr | 
= 0.27 
%2.7 \times 10^{-2}
\left(\frac{ \Delta m^2_{J i} }{ 0.1 \,\mbox{eV}^2}\right)^{-1}
\left( \frac{ \rho E }{ 100 (\text{g/cm}^3) \,\mbox{GeV} }\right)^2. 
\label{enhanced-case}
\end{eqnarray}
Comparing to the conditions (\ref{denominator-size}), they can be larger than $W^4$ terms and hence cannot be neglected. After taking account of $W^2$ suppression of $\sim 0.01$ (assuming $W \simeq 0.1$), 
$| \frac{ AA L }{ ( \Delta_{J} - h_{i} ) } W^2 | \sim 3 \times 10^{-2}$ at $E \sim 100$ GeV, assuming $\Delta m^2_{J i} =0.1$ eV$^2$. 

\begin{figure}[h!]
\vspace{50mm}
%\begin{center}
\hspace{5mm}
\includegraphics[bb=0 0 792 600,width=1.08\textwidth]{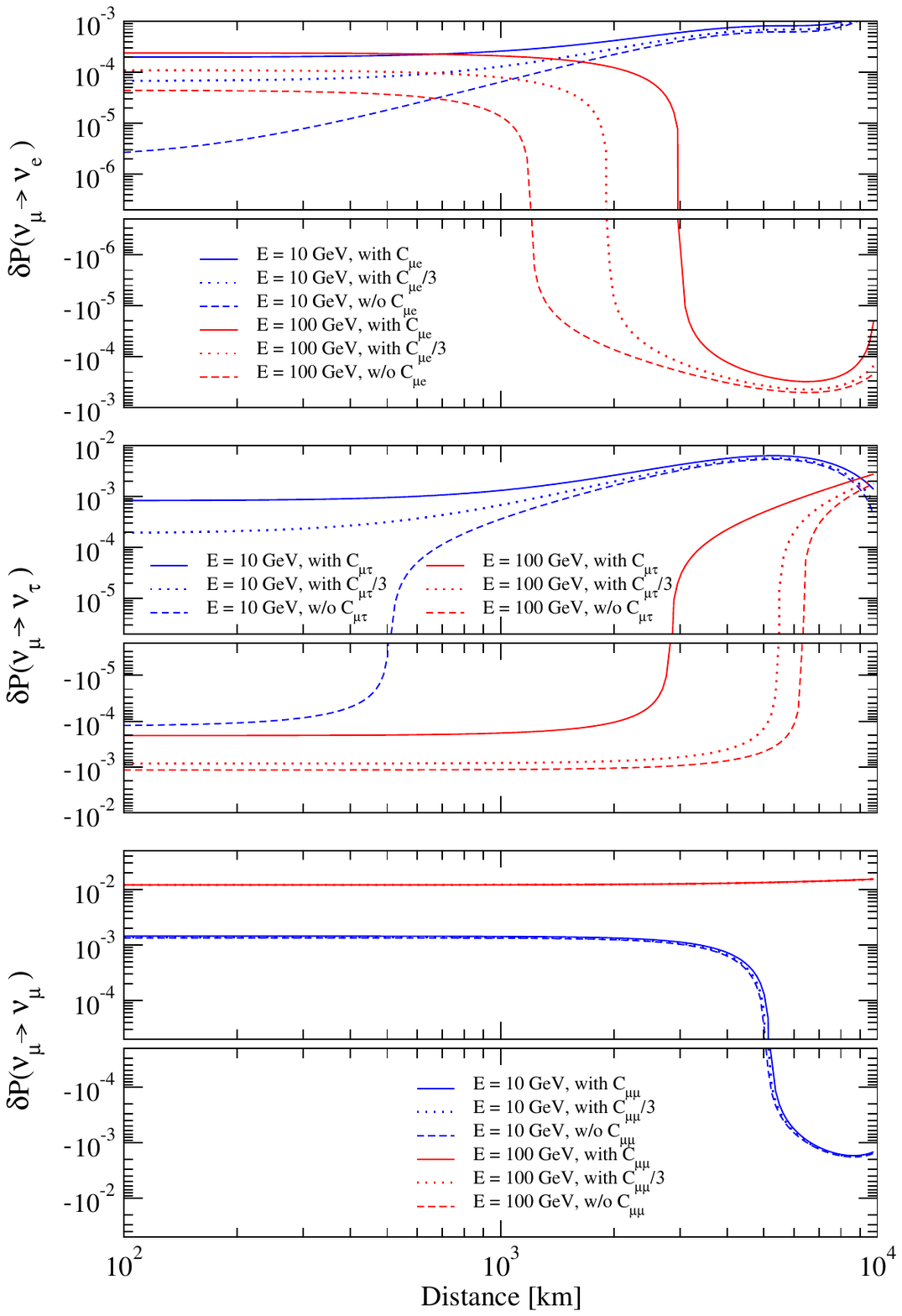}
%\end{center}
\vspace{-7mm}
\caption{ 
The sum of the order $W^2$ correction terms in eq.~\eqref{W2_correction_terms} plus the probability leaking
 term $\mathcal{C}_{\mu \alpha}$ (see eq.~(\ref{Cab}) for definition) in
 $P(\nu_{\mu} \rightarrow \nu_{\alpha})$, 
namely, $\delta P(\nu_{\mu} \rightarrow \nu_{\alpha})  \equiv \mathcal{C}_{\mu \alpha} + P(\nu_{\mu} \rightarrow \nu_{\alpha})^{(2)}$ is
plotted assuming a common $m_J^2 = 0.1$ eV$^2$. 
The top, middle and bottom panels are for $\alpha = e, \tau$, and $\mu$, respectively. 
In each panel the three cases are shown: 
$N=1$ case with maximal $\mathcal{C}_{\mu \alpha}$ (solid line), 
the universal scaling model with $N=3$ (dotted line), and 
the order $W^2$ correction terms only (dashed line). 
The last case corresponds to the universal scaling model with $N=\infty$. The blue lines are for $E=10$ GeV, and the red for $E=100$ GeV. 
The leaking terms in the $N=1$ model (shown without superscript $(N=1)$ in the legend) have values 
$\mathcal{C}_{e \mu} = 2 \times 10^{-4}$, $\mathcal{C}_{\tau \mu} = 9.5 \times 10^{-4}$, and $\mathcal{C}_{\mu \mu} = 9.6 \times 10^{-5}$. 
}
\label{fig:W-correction}
\vspace{-5mm}
\end{figure}

To know more quantitatively the sizes of $W^2$ corrections and their $E$ or $L$ dependences, we have to fix the $W$ matrix elements which have large arbitrariness. We defer this technical discussion to appendix~\ref{sec:parameter-choice}, which describes the recipe we took to fix them with a common $m_J^2 = 0.1$ eV$^2$.\footnote{
%%%%%%%%%%%%%% footnote %%%%%%%%%%%%%% 
We are aware that the assumption of equal sterile neutrino masses is contradictory to the assumption of no accidental degeneracy in the sterile mass spectrum we made in section~\ref{sec:probability-2nd}. It was done not to complicate term by term evaluation of the perturbative series, and to avoid using degenerate perturbation theory. Fortunately, we can remove this assumption to second order in $W$ in which no purely sterile state mass splitting denominator is involved.
}
We plot in figure~\ref{fig:W-correction}, $\delta P(\nu_{\mu} \rightarrow \nu_{\alpha}) \equiv P(\nu_{\mu} \rightarrow \nu_{\alpha})^{(2)} + \mathcal{C}_{\mu \alpha}$, that is, the order $W^2$ correction terms in $P(\nu_{\mu} \rightarrow \nu_{\alpha})$, eq.~\eqref{W2_correction_terms}, plus the probability leaking term $\mathcal{C}_{\mu \alpha}$, $\alpha = e$ (top panel), $\alpha = \tau$ (middle panel), and $\alpha = \mu$ (bottom panel). 
In other words, $\delta P(\nu_{\mu} \rightarrow \nu_{\alpha})$ 
is equal to the total probability minus $P(\nu_{\mu} \rightarrow \nu_{\alpha})^{(0)}$, if the fourth and the higher-order in $W$ correction terms with matter are neglected.
In each panel the three cases are examined. 
$N=1$ case with maximal $\mathcal{C}_{\mu \alpha}$ (solid line), 
the universal scaling model\footnote{
%%%%%%%%%%%%%% footnote %%%%%%%%%%%%%%
The universal scaling model is defined in appendix~\ref{sec:scaling-model}. It prescribes a way of distributing $W_{\alpha 4}$ matrix element in $3+1$ model to the W matrix elements in $3+N$ model in such a way that the size of order $W^2$ correction terms in \eqref{W2_correction_terms} remains unchanged when all the sterile masses are equal. 
}
with $N=3$ (dotted line), and 
the order $W^2$ correction only (dashed line). The last case corresponds to the universal scaling model with $N=\infty$. The blue lines are for $E=10$ GeV, and the red for $E=100$ GeV.

%%%%%%%%%%%%%%% FIG 4 %%%%%%%%%%%%%%%
%%%%%%%%%%%%%%% FIG 4 %%%%%%%%%%%%%%%

We will first focus on the appearance channels $\nu_\mu \to \nu_e$ and $\nu_\mu \to \nu_\tau$. 
At $E=10$ GeV (100 GeV) $\delta P$ depends very much on the above three cases, $N=1$, $N=3$, and $N=\infty$ for baseline $L$ of several 100 km ($L \gsim 1000$ km). The maximum value of $| \delta P |$ is always given by the case of maximal (minimal) $\mathcal{C}_{\mu \alpha}$ for positive (negative) $\delta P(\nu_{\mu} \rightarrow \nu_{\alpha})$ shown by the solid (dashed) lines. These maximal values of $| \delta P(\nu_{\mu} \rightarrow \nu_{\alpha}) |$ are, roughly speaking, $\simeq 10^{-3}$ for $\nu_{\mu} \rightarrow \nu_{\tau}$, and $\simeq \text{a few} \times 10^{-4}$ for $\nu_{\mu} \rightarrow \nu_{e}$. The effect might be visible for the former, though it might be challenging for the latter channel.\footnote{
%%%%%%%%%%%%%%% footnote %%%%%%%%%%%%
Of course, there is an issues of how to separate effects of $W^2$ correction terms from unitarity violation through $U$ matrix in leading order. 
}
For the disappearance channel $\nu_\mu \to \nu_\mu$, $|\delta P(\nu_{\mu} \rightarrow \nu_{\mu})| \sim 10^{-3} (10^{-2})$ for $E = 10$ GeV (100 GeV). In this case, the contribution from $\mathcal{C}_{\mu \mu}$ is subdominant compared to $W^2$ correction terms. 

At longer distance and in appearance channels, 
we see enhancement. At $E=10$ GeV, we observe a factor of several enhancement in $| \delta P (\nu_{\mu} \rightarrow \nu_{\alpha})|$ for both $\alpha=e$ and $\alpha = \tau$ in region $L \gsim 3000$ km.
They may provide a clear signature. The similar tendency exists at $E=100$ GeV, but in a less pronounced way. In this case, $\delta P(\nu_{\mu} \rightarrow \nu_{\alpha})$ flips sign at around $1000-3000$ km for $\alpha = e$ and $3000-6000$ km for $\alpha = \tau$ channels. It produces, assuming detector's sensitivity, a peculiar zenith angle dependence. 
The relevant energy region of $\rho E = 50 - 1000 \text{ (g/cm}^3) \text{GeV}$ may be explored, for example, by atmospheric neutrino observation by Deep Core, PINGU, or KM3NeT-ORCA \cite{Collaboration:2011ym,TheIceCube-Gen2:2016cap,Adrian-Martinez:2016zzs} as well as Super-K, Hyper-K/HKK, or DUNE \cite{Abe:2017aap,Abe:2015zbg,Abe:2016ero,Acciarri:2015uup} in relatively lower energy region. 

A final remark on $\mathcal{C}_{\alpha \beta}$ vs. $W^2$ corrections. 
Since $\mathcal{C}_{\alpha \beta}$ is a constant term in the oscillation probability, it can in principle be distinguished from the other normalization term which shares $U$ matrix element dependences with the oscillation terms. In particular, they can dominate for large $m_J^2$ since the $W^2$ correction terms are suppressed by at least $\sim 1/m_J^2$. In this case, they will be the sole indicator of low-scale unitarity violation. 
In general (though not in the $N=1$ model), the order $W^2$ terms depend upon details of the sterile sector, e.g., matrix structure of $W$. Therefore, once the effect is seen it would give us useful information on the structure of low-scale leptonic unitarity violation. 

\section{Some remaining theoretical issues and extending } 
\label{sec:theoretical}

In this section, we will give some remarks on the theoretical basis in our framework, basic one as well as on its perturbative aspects. They include our treatment of decoherence, generic structure of higher-order corrections and its relation to the ``Uniqueness theorem'' (see section~\ref{sec:nonunitarity-matter}), absence of enhancement due to small solar mass splitting denominator, and its relation to the other non-standard physics.

\subsection{Decoherence imposed onto coherent evolution system } 
\label{sec:decoherence}

We have started with the Schr\"odinger equation (\ref{evolution}) with Hamiltonian (\ref{hamiltonian}) assuming that all the neutrino states remain coherent. We have shown in this and the previous papers that the coherence between active and sterile, and sterile and sterile states are not maintained for sterile mass differences larger than $0.1$ eV$^2$. The effect of decoherence is taken into account by making average over the fast oscillations. We feel it desirable for the current treatment be replaced by the real quantum mechanical one using wave packets, in which the effect of decoherence would automatically come in. \blue{Yet,} we do believe that our present framework is able to describe effectively the right physics derived from such improved treatment.

\subsection{Smallness of expansion parameters and higher order corrections  }
\label{sec:higher-order}

Here, we discuss general structure of the perturbation series without recourse to averaging out the fast oscillations. The effective expansion parameters in our perturbative framework are the following four, 
\begin{eqnarray} 
\frac{A W }{ \Delta_{J} - h_{i} }, 
\hspace{10mm} 
\frac{A W }{ h_{j} - h_{i} }, 
\hspace{10mm} 
A L W, 
\hspace{6mm} 
\text{and}
\hspace{6mm} 
W. 
\label{expansion-parameters}
\end{eqnarray}
We already saw them, except for the last one, in the discussion in section~\ref{sec:energy-denominator}, and it can be seen by inspecting the expressions of the oscillation probabilities up to the fourth orders given in section~\ref{sec:probability-2nd} and appendix~\ref{sec:expression-probability-4th}. 
Formally, the expansion parameter is the first one in (\ref{expansion-parameters}) in view of (\ref{Omega-expand}) with $\Omega [1]$, the kernel, in (\ref{Omega-1st-order}). But, the spatial integration in (\ref{Omega-expand}) produces different effective expansion parameters, the second and the third ones in (\ref{expansion-parameters}). The extra factor of $W$'s without the kinematical factors is provided when transforming from the $\hat{S}$ to $S$ matrices, as seen in section~\ref{sec:S-matrix}.

For simplicity of the discussion in this section, we limit ourselves to the case of $|W| \simeq 0.1$. Under the same conditions we have imposed in section~\ref{sec:energy-denominator}, the first one in (\ref{expansion-parameters}) is 
$\simeq 7.6 \times 10^{-4}$ for $\Delta m^2_{J i} = 0.1$ eV$^2$ and $\rho E = 10 \text{ (g/cm}^3) \text{GeV}$ while the second and the third, which are comparable to each other at around the first oscillation maximum, are estimated to be $2.3 \times 10^{-2}$. Therefore, the smallness of the expansion parameter is ensured unless $\rho E \gg 10 \text{ (g/cm}^3) \text{GeV}$. In fact, a close examination of the order $W^4$ terms in the oscillation probability (see appendix~\ref{sec:expression-probability-4th}) shows that all the formally $W^4$ terms are actually further suppressed. The largest term in the fourth-order oscillation probabilities is of the one suppressed by a factor $\left\vert \left( \frac{A W }{ \Delta_{J} - h_{i} } \right) \left( A L W \right) W^2 \right\vert \lsim 1.7 \times 10^{-7}$, which is as small as $\sim10^{-4}$ even in the case $|W|=0.5$. 
Therefore, we expect that the formula for the oscillation probability in (\ref{P-beta-alpha-final}) works under much relaxed conditions than the one in (\ref{suppression-cond}).

\subsection{On Uniqueness theorem and matter-dependent dynamical phase } 
\label{sec:U-theorem}

We have shown in sections~\ref{sec:probability-2nd} and \ref{sec:probability-4th} that there is no surviving matter dependent correction term in the oscillation probability up to order $W^4$ after averaging out fast oscillations and using the suppression by large sterile state mass denominators. Should we expect that this feature is stable against higher order corrections beyond order $W^4$? We argue that the answer is {\em Yes}. 
Based on the feature of perturbative series we have learned, we postulate the following theorem: 

\vspace{1mm}
\noindent
{\bf Uniqueness theorem}

\begin{itemize}

\item
All the matter dependent perturbative corrections in $W$ in the oscillation probability either vanish or can be ignored after averaging over the fast oscillations and using the suppression due to the large sterile state mass denominators, leaving only the probability leakage term $\mathcal{C}_{\alpha \beta}$, the first term in eq.~\eqref{P-beta-alpha-final} with  (\ref{Cab}).

\end{itemize}

\noindent 
It must be remarked here that unitarity violation effects which are hidden in non-unitary active space mixing matrix $U$ produces zeroth- to higher order effects of $W$. The above theorem is only about the terms generated by explicit perturbative corrections in $W$.

We first note that higher-order corrections in terms of $W$ are computed by using $\Omega [1]$ as the kernel, as indicated in eq.~(\ref{Omega-expand}). Notice also that all the elements of $\Omega [1]$, except for $\Omega [1]_{JJ}$, carry the sterile state mass denominator, as shown in (\ref{Omega-1st-order}). Then, higher order correction terms are always accompanied by the sterile state mass denominators which are composed of some of the first three in (\ref{expansion-parameters}), and therefore they are suppressed. 
The unique exception for it is the terms generated only by $\Omega [1]_{JJ}$ which lacks the sterile state mass denominator.  
Therefore, apart from this special case, we have shown that higher-order corrections in $W$ does not produce the surviving terms after averaging over fast oscillation and using the sterile state mass denominator suppression. It is consistent with what we saw in our explicit computation to order $W^4$. This concludes our justification of the Uniqueness theorem. 

We need to clear up the issue of special type of perturbative correction terms which involve only $\Omega [1]_{JJ}$ as the kernel in (\ref{Omega-expand}). It produces the unique form of $\hat{S}_{JJ}$ as 
\begin{eqnarray} 
\hat{S}_{JJ} = 
e^{ - i \Delta_{J} x } 
\sum_{n} 
\frac{ (-ix)^n }{n !} \left\{ (W^{\dagger} A W)_{JJ} \right\}^n, 
\label{hatS-JJ-term}
\end{eqnarray}
a collection of terms of matter-dependent higher order renormalization to $\sum_{J} W_{\alpha J} W_{\beta J}^*$, the probability leaking term at the amplitude level. However, it exponentiates and has contribution to the $S$ matrix element as\footnote{
%%%%%%%%%%%%% footnote %%%%%%%%%%%%%
It might be easier to obtain the phase factor if we use a different decomposition of $\tilde{H}$ from (\ref{tilde-H0+H1}) by absorbing $W^{\dagger} A W$ into $\tilde{H}_{0}$. 
}
\begin{eqnarray} 
S_{\alpha \beta} = 
\sum_{J} 
W_{\alpha J} W_{\beta J}^* 
\exp { \left[ -i \left\{ \Delta_{J} + (W^{\dagger} A W )_{JJ} \right\} x \right] }. 
\label{S-alpha-beta-WW-term}
\end{eqnarray}
The unique form of $S$ matrix, in principle, raises an interesting issue
of dynamically generated phase produced jointly by unitarity violation
and the matter effect.\footnote{
%%%%%%%%%%%%% footnote %%%%%%%%%%%%%%
The phase itself needs not be small. Taking the matter potential of CC reaction and the earth diameter, $AL = 6.2 \left(\frac{\rho}{5 \text{g/cm}^3}\right) \left(\frac{L}{6,400 \mbox{km}}\right)$. Therefore, $ALW^2$ can be order unity for $|W| \simeq 0.4$.
}
In our setting, however, it either disappears from the amplitude squared, or has vanishing effect when the high frequency oscillation is averaged out. 

Finally, we should remark that our discussion to justify Uniqueness theorem in this section assumes the same kinematical region as in the treatment of order $W^2$ and $W^4$ correction terms in sections~\ref{sec:probability-2nd} and \ref{sec:probability-4th}, in particular, $\rho E \lesssim 10 \text{ (g/cm}^3) \text{GeV}$. However, as mentioned at the end of section~\ref{sec:higher-order}, it is likely that the region of validity of the probability formula (\ref{P-beta-alpha-final}) with vanishingly small higher order corrections is wider. At present, the precise boundary of kinematical region for its validity is not known to us.

\subsection{Absence of enhancement due to small solar mass splitting denominator } 
\label{sec:no-enhancement}

In perturbation theory one has to sum up intermediate states including off mass shell states. Therefore, even though we sit in the kinematic region where atmospheric-scale oscillations are large, the denominator can become small, to the order of solar $\Delta m^2$ mass splitting. Then, one might question whether the correction terms blow up at the small denominator, which would invalidate our perturbative treatment. 

Fortunately, one can show that the ``singularity'' which could be produced in the limit of small solar mass splitting always cancels against the small numerator of the similar size. This problem exists already in the second-order expression of the oscillation probability (\ref{P-beta-alpha-0th+2nd}). See the second term in second order (in $W$) term. If we denote $h_{l} - h_{k} \equiv \epsilon$ the term would have $1/\epsilon$ singularity in the limit of $\epsilon \rightarrow 0$. However, one can see by inspection by eye that the expression inside the square parenthesis is antisymmetric under $l \leftrightarrow k$, and hence it is of order $\epsilon$ or higher. Therefore, the singularity cancels. Notice that the antisymmetry under $l \leftrightarrow k$ is not required for the whole expression including the matrix element factor.

The situation is a little bit more complicated in the fourth-order expression of the oscillation probability given in appendix~\ref{sec:expression-probability-4th}. In addition to $1/\epsilon$ singularity similar to the one we already saw, there exist apparent singularity of $1/\epsilon^2$ type. See, for example, the second term in (\ref{P-beta-alpha-W4-H4-diag}) and the last term in (\ref{P-beta-alpha-W4-H4-single}). But, an explicit calculation shows that the $1/\epsilon^2$ singularity always cancels against order $\epsilon^2$ numerator in the limit of small solar splitting.

This phenomenon is reminiscent of the finiteness of the oscillation probability at the small solar mass splitting limit in helio-perturbation theory with the unique expansion parameter $\frac{\Delta m^2_{21}}{\Delta m^2_{31}}$ (or a renormalized one), see e.g., \cite{Minakata:2015gra} and the references therein. Possible interpretation of applicability of the perturbative framework to the region of solar level crossing has been discussed \cite{Xu:2015kma,Ge:2016dlx}. 
Another example for the similar phenomena is the one at the small atmospheric mass splitting limit with additional expansion parameter $\sin \theta_{13}$. In this case it is observed that near the atmospheric resonance region not only the oscillation probability is finite but also its accuracy improves when the higher order terms to fourth order in $\sin \theta_{13}$ is added \cite{Asano:2011nj}. 

Then, one might ask if our small unitarity violation perturbation theory gives quantitatively accurate result at around the denominator with small solar mass splitting. However, we note that this problem is not relevant in our case because all these terms with apparent singularities vanish after averaging over the high-frequency oscillations and using the suppression by the large sterile mass denominators. 
Yet, we must remark that if we investigate possible enhancement of the correction terms outside the condition (\ref{suppression-cond}), as done in section~\ref{sec:correction-terms}, the quantitative accuracy of the expression may become an issue. 

\subsection{Leptonic non-unitarity and the other non-standard physics  }
\label{sec:sterile-NSI}

This final subsection is to mention the related but different approaches, and to make some clarifying remarks. 
Our $3+N$ model has obvious relation with the various versions of active plus sterile neutrino models proposed in the context of LSND-MiniBooNE anomaly, 
as reviewed in \cite{Conrad:2013mka}, see also the references therein. The clear difference exists in the attitude of the treatment of the model, in our case seeking the conditions to make the predictions as model-independent as possible, while in the others pursuing the particular model which provide the best fit to the data.
Unless we use our model-independent simplified formula (\ref{P-beta-alpha-final}), we would have to marginalise over the huge parameter space of the $(3+N)$ model to obtain the bound on non-unitarity $3\times 3$ mixing matrix $U$.

It is pointed out that one can establish a mapping between parameters in the mass eigenstate basis which describe non-unitary leptonic mixing and the ones for non-standard neutrino interactions (NSI) under certain conditions between neutron and electron number densities~\cite{Blennow:2016jkn}. However, when we rotate back to the flavour basis, a non-unitary  mixing matrix is involved in unitarity violating case, but not in the NSI case, as far as propagation in matter is concerned. 

\section{Concluding remarks }
\label{sec:conclusion}

In this paper, we have presented a comprehensive treatment of the three active plus $N$ sterile neutrino model in the context of leptonic unitarity test. We have formulated an appropriate perturbative framework with expansion in small unitarity violating $W$ matrix elements, while keeping (non-$W$ suppressed) matter effect to all orders.

What we have done in this paper is mainly threefold:

\begin{itemize}

\item
We have shown that the oscillation probability in matter between active states can be made sterile-sector model independent, apart from $N$ dependence in the lower bound on probability leaking term $\mathcal{C}_{\alpha \beta}$ [see eq.~\eqref{Cab-bound}]. The property holds under the environment of active and sterile neutrino evolution with decoherence in active-sterile and sterile-sterile channels, which requires $0.1\, \text{eV}^2 \lsim m^2_{J} \lsim (1-10) \, \text{GeV}^2$ for the typical kinematical setting of LBL experiments. It leads to a very simple expression of the oscillation probability in matter, eq.~\eqref{P-beta-alpha-final}. 

The model-independent nature of the observable is demonstrated by showing that perturbative corrections to eq.~\eqref{P-beta-alpha-final} either vanish or are negligible after averaging over fast oscillations and using large sterile state mass denominator suppression. It is done by an explicit computation to fourth order in $W$ which lefts the unique non-vanishing vacuum term, the probability leaking term $\mathcal{C}_{\alpha \beta}$. We have argued by postulating the ``Uniqueness theorem'' that this feature prevails to all orders in $W$ perturbation theory.

\item 
We have used the oscillation probability formula, eq.~\eqref{P-beta-alpha-final}, to analyze $\nu_{\mu} \rightarrow \nu_{\alpha}$ channels ($\alpha=e,\mu,\tau$), to know in which region of energy and baseline the effect of unitarity violation is large. As a general tendency the effect is sizeable in regions where standard oscillation probability is large, with notable amplification in the two regions corresponding to the solar and the atmospheric MSW enhancement. 
We have observed relatively large effect in $\nu_{\mu} \rightarrow \nu_{\mu}$ and $\nu_{\mu} \rightarrow \nu_{\tau}$ channels, and pointed out, though qualitatively, that anticorrelation of signals between them would enhance the sensitivity to unitarity violating effects. 

\item
We have discussed the question of how to distinguish low-scale unitarity violation from high-scale one. We have pointed out that 
outside the region of validity of our Uniqueness theorem, $\rho E \gg 10 \text{ (g/cm}^3)~\text{GeV}$,
the second order $W$ correction [eq.~\eqref{W2_correction_terms}] to the leading order
could become large and, if detected, it would signal low-scale unitarity violation,
offering new way of discriminating between low- and high-scale unitarity violation.  
Then, it would allow us to probe structure of $W$ matrix elements which bridges between active and sterile sectors.
This is to add to the method of detecting the probability leaking term $\mathcal{C}_{\alpha\beta}$ discussed in \cite{Fong:2016yyh}, 
which may have a broader applicability by relying on the existence of ``sterile'', or undetectable but communicable, sector at low energy scales,
a generic feature beyond the $(3+N)$ model.

\end{itemize}

\noindent
Notice that in ``constraining mode'' of unitarity violation, the model-independence of the framework translates into a 
universal nature of the bounds, thereby making them more powerful. Whereas in ``discovery mode'' of unitarity violation, the model dependence, in particular through the $W$ dependent correction terms, is welcome because it serves for identifying the structure of the sterile sector. 

During the course of this work, we have obtained the new results and had some interesting observations including: 

\begin{itemize}

\item
We have obtained an exact solution, eq.~\eqref{S-alpha-beta-0th-final}, of $S$ matrix for the Hamiltonian (\ref{H0-def}) with uniform matter density. 
It describes neutrino evolution in low-scale unitarity violation in zeroth-order in $W$, which applies also to the case of high-scale unitarity violation. 
It has been utilized in section~\ref{sec:where-UV} to calculate the oscillation probability in the leading-order as well as its higher order corrections in $W$. 
When applied to each shell inside the earth, it could provide a semi-quantitative way of simulating non-unitary neutrino evolution for the terrestrial experiments. 

\item
The value of $\mathcal{C}_{\alpha \beta}$, if detected, could reveal structure of the hidden sterile sector. In this paper, this point is illustrated only in a toy model of equally distributed $W$ matrix elements within each flavour, as defined in appendix~\ref{sec:scaling-model}. In this model, the probability leaking term scales as $1/N$ depending upon number of sterile states. 

\end{itemize}

We emphasize that neutrino experiment is the most powerful way to execute leptonic unitarity test in scenarios of low-scale unitarity violation, 
though it is unlikely in the case of high-scale unitarity violation.
Nonetheless, we have to admit that our observations on what we could do for experimental detection of possible non-unitarity effects are rather qualitative to make any definitive claim for possible detection in the future. Clearly, more detailed analyses are called for.  

While we worked exclusively on the $(3+N)$ state unitary model as a model of low-scale unitarity violation, we do not know if it is the unique choice, or it merely reflects our ignorance. Even in the case there exist more generic class of models for low-scale unitarity violation, the phenomenon of probability leaking is likely to survive. It is because the probability leaking must take place whenever the extra light sector exists and communicates with the three active neutrinos. 

\acknowledgments

One of the authors (H.M.) thanks Enrique Fernandez-Martinez for interesting discussions about the relationship between high-scale and low-scale unitarity violation. 
He expresses a deep gratitude to Instituto F\'{\i}sica Te\'{o}rica, UAM/CSIC in Madrid, for its support via ``Theoretical challenges of new high energy, astro and cosmo experimental data''  project, Ref: 201650E082. 
This work has received funding/support from the European Union's Horizon 2020 research and innovation programme under the Marie Sklodowska-Curie grant agreement No 674896 and No 690575.
He had been a member of Yachay Tech for 14 months, at that time the first research-oriented university in Ecuador~\cite{Yachay-story}, during which he was warmly supported by Ecuadorian people. 
He thanks kind supports by ICTP-SAIFR (FAPESP grant 2016/01343-7), UNICAMP (FAPESP grant 2014/19164-6), and PUC-Rio (CNPq) which enabled him to visit these institutions where part of this work was done. 
C.S.F. is supported by FAPESP under grants 2013/01792-8 and 2012/10995-7.
H.N. is supported by CNPq.
This work was supported in part by the Fermilab Neutrino Physics Center. 

\appendix

\section{Neutrino evolution equation in flavour basis}
\label{sec:flavor-basis-evolution}

The Schr\"odinger equation takes the form with flavour basis Hamiltonian $H$ in (\ref{flavor-hamiltonian})
\begin{eqnarray}
i \frac{d}{dx} 
\left[
\begin{array}{c}
\nu_{a} \\
\nu_{s} \\
\end{array}
\right] = 
\left[
\begin{array}{cc}
U {\bf \Delta_{a} } U^{\dagger} + W {\bf \Delta_{s} } W^{\dagger} + A & 
U {\bf \Delta_{a} } Z^{\dagger} + W {\bf \Delta_{s} } V^{\dagger}  \\
Z {\bf \Delta_{a} } U^{\dagger} + V {\bf \Delta_{s} } W^{\dagger} & 
Z {\bf \Delta_{a} } Z^{\dagger} + V {\bf \Delta_{s} } V^{\dagger} \\
\end{array}
\right]
\left[
\begin{array}{c}
\nu_{a} \\
\nu_{s} \\
\end{array}
\right] ,
\label{Schroedinger-eq-flavor-basis}
\end{eqnarray}
where $\nu_{a}$ ($\nu_{s}$) denotes $3$ ($N$) component vector in active (sterile) space. Apparently, the system depends not only on $U$ and $W$, but also on $Z$ and $V$ matrix elements, which is not the case in our treatment using the mass eigenstate basis. 

Here, we show that the dependence on $Z$ and $V$ is superficial. Since there is no physical meaning of the particular basis for the sterile sector fermions we can redefine it by doing the transformation 
\begin{eqnarray}
\left[
\begin{array}{c}
\nu_{a} \\
\nu_{s} \\
\end{array}
\right] 
\rightarrow 
\left[
\begin{array}{cc}
1 & 0 \\
0 & Y \\
\end{array}
\right] 
\left[
\begin{array}{c}
\nu_{a} \\
\nu_{s} \\
\end{array}
\right] 
\equiv 
\left[
\begin{array}{c}
\nu_{a}' \\
\nu_{s}' \\
\end{array}
\right] ,
\end{eqnarray}
where $Y$ is a $N \times N$ unitary matrix. In the primed basis the Hamiltonian becomes 
\begin{eqnarray} 
H' \equiv 
\left[
\begin{array}{cc}
1 & 0 \\
0 & Y \\
\end{array}
\right] 
H
\left[
\begin{array}{cc}
1 & 0 \\
0 & Y^{\dagger} \\
\end{array}
\right] =
\left[
\begin{array}{cc}
( U {\bf \Delta_{a} } U^{\dagger} + W {\bf \Delta_{s} } W^{\dagger} + A ) & 
( U {\bf \Delta_{a} } Z^{\dagger} + W {\bf \Delta_{s} } V^{\dagger} ) Y^{\dagger} \\
Y ( Z {\bf \Delta_{a} } U^{\dagger} + V {\bf \Delta_{s} } W^{\dagger} ) & 
Y ( Z {\bf \Delta_{a} } Z^{\dagger} + V {\bf \Delta_{s} } V^{\dagger} ) Y^{\dagger} \\
\end{array}
\right]. 
\end{eqnarray}
We can arbitrarily choose $Y=V^{\dagger}$. Then, one can show by using unitarity (\ref{unitarity}) that 
\begin{eqnarray} 
H' = 
\left[
\begin{array}{cc}
( U {\bf \Delta_{a} } U^{\dagger} + W {\bf \Delta_{s} } W^{\dagger} + A ) & 
- U {\bf \Delta_{a} } U^{\dagger} W %Z^{\dagger} V 
+ W {\bf \Delta_{s} } ( {\bf 1} - W^{\dagger} W )% V^{\dagger} V \\ 
\\
- W^{\dagger} U % V^{\dagger} Z 
{\bf \Delta_{a} } U^{\dagger} + ( {\bf 1} - W^{\dagger} W ) %V^{\dagger} V 
{\bf \Delta_{s} } W^{\dagger} & 
W^{\dagger} U %V^{\dagger} Z 
{\bf \Delta_{a} } 
U^{\dagger} W  %Z^{\dagger} V 
+ ( {\bf 1} - W^{\dagger} W ) %V^{\dagger} V 
{\bf \Delta_{s} } 
( {\bf 1} - W^{\dagger} W ) %V^{\dagger} V \\
\end{array}
\right]. 
\nonumber \\ 
\end{eqnarray}
Therefore, our system depends only on $U$ and $W$, and the dependence on $Z$ and $V$ is superficial.

\section{$\hat{S}$ matrix elements}
\label{sec:hatS-elements}

The method of computing $\hat{S}$ matrix elements is outlined in section~\ref{sec:hatS-matrix} and here, we will collect the results.
%In this appendix,
%~\ref{sec:hatS-elements} 
%we carry out this task. With the expression of $H_{1}$ in (\ref{H1-matrix}) we compute perturbatively the matrix elements of $\Omega (x)$. Then, $\hat{S} (x) = e^{- i \hat{H}_{0} x} \Omega(x)$.
We denote computed results of the matrix elements of $\hat{S}$ as $\hat{S} [n]$ to indicate that it is the one that comes from $n$-th order contribution in $H_{1}$. Since the elements of $H_{1}$ are of order either $W$ or $W^2$, $\hat{S} [n]$ generally has order $W^n$ or higher. To show that a particular contribution is of order $W^m$ we use the superscript ``$(m)$''. That is, $\hat{S}^{(m)} [n]$ denotes contribution to $\hat{S}$ that arizes from $n$-th order perturbative contribution in $H_{1}$ and is of order $W^m$. 

In the following, we will denote
\begin{eqnarray}
%c_{iJ} &\equiv& \left[\left(U X\right)^\dagger A W\right]_{iJ}, \label{eq:c_iJ} \\
%d_{IJ} &\equiv& \left[W^\dagger A W\right]_{IJ} = d_{JI}^*, \label{eq:d_IJ} \\
c_{iJ} &\equiv& \left[\left(U X\right)^\dagger A W\right]_{iJ},
%\label{eq:c_iJ}, 
\hspace{5mm}
d_{IJ} \equiv \left[W^\dagger A W\right]_{IJ} = d_{JI}^*,
%\label{eq:d_IJ} \\
\label{eq:c_iJ-d_IJ} \\
e_i &\equiv& e^{-i h_i x}, \;\;\;\;
e_I \equiv e^{-i \Delta_I x}, \label{eq:e_a} \\
\Delta_{i j } &\equiv & \Delta_i - \Delta_j, \;\;\;\;
\Delta_{I j } \equiv \Delta_I - \Delta_j, \;\;\;\;
\Delta_{I J} \equiv \Delta_I - \Delta_J \label{eq:Delta_ab}.
\end{eqnarray}
\subsection{Contribution to $\hat{S}$ matrix elements from zeroth and first order in $H_{1}$}
\label{sec:hatS-0th-1st}

The zeroth and first order $\hat{S}$ matrix elements can be calculated as follows: 
\begin{eqnarray} 
\hat{S}_{ij}^{(0)}[0+1] &=& 
\left( e^{-i \hat{H}_{0} x} \right)_{i k} (\Omega_{k i}) +
\left( e^{-i \hat{H}_{0} x} \right)_{i K} (\Omega_{K i}) = 
e^{-i h_{i} x} (\Omega_{i i}) 
= \delta_{ij} e_i
\nonumber \\
\hat{S}_{i J}^{(1)}[0+1] &=& 
\left( e^{-i \hat{H}_{0} x} \right)_{i j} (\Omega_{j J}) +
\left( e^{-i \hat{H}_{0} x} \right)_{i K} (\Omega_{K J}) = 
e^{- i h_{i} x} (\Omega_{i J}) 
= c_{iJ} \frac{e_J - e_i}{\Delta_{Ji}}
\nonumber \\
\hat{S}_{J i}^{(1)}[0+1] &=& 
\left( e^{-i \hat{H}_{0} x} \right)_{J k} (\Omega_{k i}) +
\left( e^{-i \hat{H}_{0} x} \right)_{J K} (\Omega_{K i}) 
= c_{iJ}^* \frac{e_J - e_i}{\Delta_{Ji}}
\nonumber \\ 
\hat{S}_{J K}^{(2)} \vert_{J \neq K}[0+1] &=& 
\left( e^{-i \hat{H}_{0} x} \right)_{J i} (\Omega_{i K}) +
\sum_{I} \left( e^{-i \hat{H}_{0} x} \right)_{J I} (\Omega_{I K}) 
= d_{JK}\frac{e_K - e_J}{\Delta_{KJ}} 
\nonumber \\ 
\hat{S}_{J J}^{(0+2)}[0+1] &=& 
\sum_{i} \left( e^{-i \hat{H}_{0} x} \right)_{J i} (\Omega_{i J}) +
\sum_{I} \left( e^{-i \hat{H}_{0} x} \right)_{J I} (\Omega_{I J}) 
= e_J \left( 1 - i x d_{JJ} \right) .
\label{hat-S-elements-1st}
\end{eqnarray}
The terms above are invariant under generalized T transformation $\hat S_{p q} (U, W, X, A) \to \hat S_{q p} (U^*, W^*, X^*, A^*)$ [eqs.~\eqref{T-invariance}].

\subsection{Contribution to $\hat{S}$ matrix elements from second order in $H_{1}$}
\label{sec:hatS-2nd}

Likewise, $\hat{S}$ matrix elements can be calculated in second order in $\hat{H}_{1}$ by using the formula for $\Omega$ in (\ref{Omega-expand}) and $\hat{S}$-$\Omega$ relation in (\ref{hatS-Omega}) as 
\begin{eqnarray}
\hat{S}_{ij}^{\left(2\right)}[2] & = & \sum_{K}c_{iK}c_{jK}^{*}f_{ij,K}^{\left(2\right)},\\
\hat{S}_{iJ}^{\left(3\right)}[2] & = &
 \sum_{K}c_{iK}d_{KJ}f_{iJ,K}^{\left(2\right)},\hspace{5mm} 
\hat{S}_{Ij}^{\left(3\right)}[2]  =  \sum_{K}c_{jK}^{*}d_{IK}f_{Ij,K}^{\left(2\right)},\\
\hat{S}_{IJ}^{\left(2+4\right)}[2] & = & \sum_{k}c_{kJ}c_{kI}^{*}f_{IJ,k}^{\left(2\right)}+\sum_{K}d_{IK}d_{KJ}f_{IJ,K}^{\left(2\right)},
\end{eqnarray}
where
\begin{eqnarray}
f_{ij,K}^{\left(2\right)} & = & \begin{cases}
\frac{1}{\Delta{}_{Ki}}\left(ixe_{i}+\frac{e_{K}-e_{i}}{\Delta{}_{Ki}}\right) & {\rm for}\;j=i\\
\frac{1}{\Delta{}_{Kj}\Delta{}_{Ki}}\left(e_{K}+\frac{\Delta{}_{Kj}}{\Delta_{ji}}e_{i}-\frac{\Delta{}_{Ki}}{\Delta_{ji}}e_{j}\right) & {\rm for}\;j\neq i
\end{cases}, \nonumber \\
f_{iJ,K}^{\left(2\right)} & = & f_{Ji,K}^{\left(2\right)}=\begin{cases}
-\frac{1}{\Delta{}_{Ji}}\left(ixe_{J}+\frac{e_{J}-e_{i}}{\Delta{}_{Ji}}\right) & {\rm for}\;K=J\\
\frac{1}{\Delta{}_{Ji}\Delta{}_{Ki}\Delta_{KJ}}\left(e_{K}\Delta{}_{Ji}-e_{J}\Delta{}_{Ki}+e_{i}\Delta_{KJ}\right) & {\rm for}\;K\neq J
\end{cases}, \nonumber \\
f_{IJ,k}^{\left(2\right)} & = & \begin{cases}
-\frac{1}{\Delta{}_{Ik}}\left(ixe_{I}+\frac{e_{I}-e_{k}}{\Delta{}_{Ik}}\right) & {\rm for}\;J=I\\
-\frac{1}{\Delta{}_{Jk}\Delta_{IJ}\Delta{}_{Ik}}\left(e_{J}\Delta{}_{Ik}-e_{I}\Delta{}_{Jk}-e_{k}\Delta_{IJ}\right) & {\rm for}\;J\neq I
\end{cases}, \nonumber \\
f_{IJ,K}^{\left(2\right)} & = & \begin{cases}
-\frac{x^{2}}{2}e_{I} & {\rm for}\;J=I,\,K=I\\
-\frac{1}{\Delta_{IK}}\left(ixe_{I}+\frac{e_{I}-e_{K}}{\Delta_{IK}}\right) & {\rm for}\;J=I,\,K\neq I\\
\frac{1}{\Delta_{JI}}\left(ixe_{I}+\frac{e_{J}-e_{I}}{\Delta_{JI}}\right) & {\rm for}\;J\neq I,\,K=I\\
-\frac{1}{\Delta_{JI}}\left(ixe_{J}+\frac{e_{J}-e_{I}}{\Delta_{JI}}\right) & {\rm for}\;J\neq I,\,K=J\\
-\frac{1}{\Delta_{JK}\Delta_{IJ}\Delta_{IK}}\left(e_{J}\Delta_{IK}-e_{I}\Delta_{JK}-e_{K}\Delta_{IJ}\right) & {\rm for}\;J\neq I,\,K\neq I,J \nonumber
\end{cases} .
\end{eqnarray}
In the expressions above, the combinations of the couplings remain invariant taking the complex conjugate together with $p \leftrightarrow q$ while one can verify directly that $f_{pq,r}^{\left(2\right)}=f_{qp,r}^{\left(2\right)}$. Hence the expressions are T invariance [eqs.~\eqref{T-invariance}]. 

What we should do in the rest of appendix is to compute $\hat{S}$ matrix elements perturbatively to fourth order in $H_{1}$. In the rest of the appendix, we present only the terms which are required to compute $S$ matrix elements to order $W^4$. In view of the relations between $\hat{S}$ and $S$ matrix elements given in eq.~(\ref{Sab-hatSab}), $\hat{S}_{I J}^{(3)}$, $\hat{S}_{I J}^{(4)}$, and $\hat{S}_{i J}^{(4)}$ (and $\hat{S}_{J i}^{(4)}$) are all unnecessary. We only give the results of manifestly generalized T invariant form of $\hat{S}$ matrix elements with which it must be straightforward to prove generalized T invariance. 
\subsection{Contribution to $\hat{S}$ matrix elements from third order in $H_{1}$}
\label{sec:hatS-3rd}

For the third order terms in $H_{1}$, we have
\begin{eqnarray}
\hat{S}_{ij}^{\left(4\right)}[3] & = & \sum_{K,L}c_{iK}d_{KL}c_{jL}^{*}f_{ij,KL}^{\left(3\right)},\\
\hat{S}_{iJ}^{\left(3\right)}[3] & = &
 \sum_{k,L}c_{iL}c_{kL}^{*}c_{kJ}f_{iJ,kL}^{\left(3\right)}+{\cal
 O}\left(W^{5}\right), \hspace{3mm}
\hat{S}_{Ij}^{\left(3\right)}[3]  =
\sum_{k,L}c_{kL}c_{kI}^{*}c_{jL}^{*}f_{Ij,kL}^{\left(3\right)}+{\cal
O}\left(W^{5}\right),\hspace{7mm} \\
\hat{S}_{IJ}^{\left(4\right)}[3] & = & \sum_{k,L}c_{kL}^{*}c_{kJ}d_{IL}g_{IJ,kL}^{\left(3\right)}+\sum_{k,L}c_{kL}c_{kI}^{*}d_{LJ}h_{IJ,kL}^{\left(3\right)}+{\cal O}\left(W^{5}\right),
\end{eqnarray}
where
\begin{eqnarray}
f_{ij,KL}^{\left(3\right)} & = & \begin{cases}
-\frac{1}{\Delta{}_{Ki}\Delta{}_{Li}}\left(ixe_{i}-\frac{e_{L}-e_{i}}{\Delta{}_{Li}\Delta_{LK}}\Delta{}_{Ki}+\frac{e_{K}-e_{i}}{\Delta{}_{Ki}\Delta_{LK}}\Delta{}_{Li}\right) & {\rm for}\;j=i,\,K\neq L \nonumber \\
-\frac{1}{\Delta{}_{Li}^{2}}\left[ix\left(e_{L}+e_{i}\right)+2\frac{e_{L}-e_{i}}{\Delta{}_{Li}}\right] & {\rm for}\;j=i,\,K=L\\
\frac{1}{\Delta{}_{Ki}\Delta{}_{Kj}\Delta{}_{Li}\Delta{}_{Lj}}\left(\frac{e_{L}\Delta{}_{Ki}\Delta{}_{Kj}-e_{K}\Delta{}_{Li}\Delta{}_{Lj}}{\Delta_{LK}}-\frac{e_{j}\Delta{}_{Ki}\Delta{}_{Li}-e_{i}\Delta{}_{Kj}\Delta{}_{Lj}}{\Delta_{ij}}\right) & {\rm for}\;j\neq i,\,K\neq L \nonumber \\
-\frac{1}{\Delta{}_{Lj}\Delta{}_{Li}}\left(ixe_{L}+\frac{e_{i}-e_{j}}{\Delta_{ji}}+\frac{e_{L}-e_{i}}{\Delta{}_{Li}}+\frac{e_{L}-e_{j}}{\Delta{}_{Lj}}\right) & {\rm for}\;j\neq i,\,K=L
\end{cases},\\
f_{iJ,kL}^{\left(3\right)} & = & f_{Ji,kL}^{\left(3\right)}=\begin{cases}
-\frac{1}{\Delta{}_{Ji}\Delta{}_{Jk}\Delta{}_{Li}\Delta{}_{Lk}}\left(\frac{e_{L}\Delta{}_{Ji}\Delta{}_{Jk}-e_{J}\Delta{}_{Li}\Delta{}_{Lk}}{\Delta_{JL}}+\frac{e_{k}\Delta{}_{Ji}\Delta{}_{Li}-e_{i}\Delta{}_{Jk}\Delta{}_{Lk}}{\Delta_{ik}}\right) & {\rm for}\;k\neq i,\,L\neq J\\
-\frac{1}{\Delta{}_{Ji}\Delta{}_{Li}}\left(ixe_{i}+\frac{e_{J}-e_{i}}{\Delta{}_{Ji}}+\frac{e_{L}-e_{i}}{\Delta{}_{Li}}-\frac{e_{J}-e_{L}}{\Delta_{JL}}\right) & {\rm for}\;k=i,\,L\neq J \nonumber\\
-\frac{1}{\Delta{}_{Jk}\Delta{}_{Ji}}\left(ixe_{J}-\frac{e_{i}-e_{k}}{\Delta_{ik}}+\frac{e_{J}-e_{i}}{\Delta{}_{Ji}}+\frac{e_{J}-e_{k}}{\Delta{}_{Jk}}\right) & {\rm for}\;k\neq i,\,L=J \nonumber\\
-\frac{1}{\Delta{}_{Ji}^{2}}\left[ix\left(e_{i}+e_{J}\right)+2\frac{e_{J}-e_{i}}{\Delta{}_{Ji}}\right] & {\rm for}\;k=i,\,L=J
\end{cases}, \nonumber 
\end{eqnarray}
%%%
\begin{eqnarray}
g_{IJ,kL}^{\left(3\right)} & = & \begin{cases}
\frac{ixe_{I}}{\Delta{}_{Ik}\Delta_{LI}}+\frac{1}{\Delta{}_{Ik}\Delta_{LI}\Delta{}_{Lk}}\left(\frac{e_{L}-e_{I}}{\Delta_{LI}}\Delta{}_{Ik}+\frac{e_{I}-e_{k}}{\Delta_{Ik}}\Delta_{LI}\right) & {\rm for}\;J=I,\,L\neq I\\
\frac{1}{\Delta_{Ik}}\left(-\frac{x^{2}e_{I}}{2}+\frac{ix}{\Delta{}_{Ik}}e_{I}-\frac{e_{k}-e_{I}}{\Delta_{Ik}^{2}}\right) & {\rm for}\;J=I,\,L=I\\
\frac{ixe_{I}}{\Delta{}_{Ik}\Delta_{JI}}+\frac{1}{\Delta{}_{Jk}}\left(\frac{e_{J}-e_{I}}{\Delta_{JI}^{2}}+\frac{e_{I}-e_{k}}{\Delta_{Ik}^{2}}\right) & {\rm for}\;J\neq I,\,L=I\\
\frac{ixe_{J}}{\Delta{}_{Jk}\Delta_{IJ}}+\frac{1}{\Delta{}_{Ik}}\left(\frac{e_{I}-e_{J}}{\Delta_{IJ}^{2}}+\frac{e_{J}-e_{k}}{\Delta^{2}{}_{Jk}}\right) & {\rm for}\;J\neq I,\,L=J\\
\frac{e_{I}}{\Delta_{Ii}\Delta_{KI}\Delta_{JI}}+\frac{e_{J}}{\Delta_{Ji}\Delta_{KJ}\Delta_{IJ}}-\frac{e_{i}}{\Delta_{Ii}\Delta_{Ji}\Delta_{Ki}}+\frac{e_{K}}{\Delta_{KI}\Delta_{KJ}\Delta_{Ki}} & {\rm for}\;J\neq I,\,L\neq I,J
\end{cases}, \nonumber\\
h_{IJ,kL}^{\left(3\right)} & = & \begin{cases}
\frac{ixe_{I}}{\Delta{}_{Ik}\Delta_{LI}}+\frac{1}{\Delta{}_{Ik}\Delta_{LI}\Delta{}_{Lk}}\left(\frac{e_{L}-e_{I}}{\Delta_{LI}}\Delta{}_{Ik}+\frac{e_{I}-e_{k}}{\Delta_{Ik}}\Delta_{LI}\right) & {\rm for}\;J=I,\,L\neq I\\
\frac{1}{\Delta{}_{Ik}}\left(-\frac{x^{2}e_{I}}{2}+\frac{ix}{\Delta{}_{Ik}}e_{I}-\frac{e_{k}-e_{I}}{\Delta_{Ik}^{2}}\right) & {\rm for}\;J=I,\,L=I\\
\frac{ixe_{J}}{\Delta{}_{Jk}\Delta_{IJ}}+\frac{1}{\Delta{}_{Ik}}\left(\frac{e_{I}-e_{J}}{\Delta_{IJ}^{2}}+\frac{e_{J}-e_{k}}{\Delta{}_{Jk}^{2}}\right) & {\rm for}\;J\neq I,\,L=J\\
\frac{ixe_{I}}{\Delta{}_{Ik}\Delta_{JI}}+\frac{1}{\Delta{}_{Jk}}\left(\frac{e_{J}-e_{I}}{\Delta_{JI}^{2}}+\frac{e_{I}-e_{k}}{\Delta_{Ik}^{2}}\right) & {\rm for}\;J\neq I,\,L=I\\
\frac{e_{I}}{\Delta_{Ii}\Delta_{KI}\Delta_{JI}}+\frac{e_{J}}{\Delta_{Ji}\Delta_{KJ}\Delta_{IJ}}-\frac{e_{i}}{\Delta_{Ii}\Delta_{Ji}\Delta_{Ki}}+\frac{e_{K}}{\Delta_{KI}\Delta_{KJ}\Delta_{Ki}} & {\rm for}\;J\neq I,\,L\neq I,J  \nonumber
\end{cases}.
\end{eqnarray}
Notice that for $\hat S_{ij}^{\left(4\right)}[3]$, $\hat S_{iJ}^{\left(3\right)}[3]$
and $\hat S_{Ij}^{\left(3\right)}[3]$, the combinations of the couplings
remain invariant under complex conjugation together with $\left(p\leftrightarrow q\right)$ and
hence we need that $f_{ab,cd}^{\left(3\right)}=f_{ba,cd}^{\left(3\right)}$ as can be verified in the expressions above.
On the other hand, for $S_{IJ}^{\left(3\right)}$, we have $c_{kL}^{*}c_{kJ}d_{IL}\leftrightarrow c_{kL}c_{kI}^{*}d_{LJ}$
under complex conjugation with $\left(I\leftrightarrow J\right)$ and hence we need that $g_{IJ,kL}^{\left(3\right)}\leftrightarrow h_{IJ,kL}^{\left(3\right)}$
under $\left(I\leftrightarrow J\right)$ which can again be verified from the expressions above.

\subsection{Contribution to $\hat{S}$ matrix elements from fourth order in $H_{1}$}
\label{sec:hatS-4th}

For the fourth order terms in $H_{1}$, we have
\begin{eqnarray}
\hat{S}_{ij}^{\left(4\right)}[4] & = & \sum_{k,L,M}c_{iL}c_{kL}^{*}c_{kM}c_{jM}^{*}f_{ij,kLM}^{\left(4\right)}+{\cal O}\left(W^{5}\right),\\
\hat{S}_{iJ}^{\left(4\right)}[4] & = & {\cal O}\left(W^{5}\right),
\;\;\;
\hat{S}_{Ij}^{\left(4\right)}[4] = {\cal O}\left(W^{5}\right),\\
\hat{S}_{IJ}^{\left(4\right)}[4] & = & \sum_{k,l,M}c_{kI}^{*}c_{kM}c_{lM}^{*}c_{lJ}f_{IJ,klM}^{\left(4\right)}+{\cal O}\left(W^{5}\right),
\end{eqnarray}
where
\begin{eqnarray}
f_{ij,kLM}^{\left(4\right)} & = & \begin{cases}
-\frac{1}{\Delta{}_{Li}\Delta{}_{Lk}\Delta{}_{Mi}\Delta{}_{Mk}}\left[\frac{\left(e_{M}-e_{i}\right)\Delta{}_{Li}\Delta{}_{Lk}}{\Delta{}_{Mi}\Delta_{LM}}-\frac{\left(e_{L}-e_{i}\right)\Delta{}_{Mi}\Delta{}_{Mk}}{\Delta{}_{Li}\Delta_{LM}}\right. & {\rm for}\;j=i,\,k\neq i,\,M\neq L\\
\left.-\frac{\left(e_{k}-e_{i}\right)\Delta{}_{Li}\Delta{}_{Mi}}{\Delta_{ik}\Delta_{ik}}+ixe_{i}\frac{\Delta{}_{Lk}\Delta{}_{Mk}}{\Delta_{ik}}\right]\\
-\frac{1}{\Delta{}_{Li}\Delta{}_{Lk}}\left(ix\frac{e_{L}}{\Delta{}_{Li}}+ix\frac{e_{i}}{\Delta_{ik}}+ix\frac{e_{i}}{\Delta{}_{Li}}\right. & {\rm for}\;j=i,\,k\neq i,\,M=L\\
\left.+\frac{e_{L}-e_{i}}{\Delta{}_{Li}^{2}}-\frac{e_{k}-e_{i}}{\Delta_{ik}^{2}}+\frac{e_{L}-e_{i}}{\Delta_{Li}\Delta_{Lk}}+\frac{e_{k}-e_{i}}{\Delta_{ik}\Delta{}_{Lk}}+\frac{e_{L}-e_{i}}{\Delta{}_{Li}\Delta{}_{Lk}}\right)\\
-\frac{1}{\Delta{}_{Li}\Delta{}_{Mi}}\left(\frac{x^{2}}{2}e_{i}+ix\frac{e_{i}}{\Delta{}_{Li}}+ix\frac{e_{i}}{\Delta{}_{Mi}}\right. & {\rm for}\;j=i,\,k=i,\,M\neq L\\
\left.+\frac{e_{L}-e_{i}}{\Delta_{Li}^{2}}+\frac{e_{M}-e_{i}}{\Delta_{Mi}^{2}}+\frac{e_{M}-e_{i}}{\Delta{}_{Mi}\Delta_{LM}}-\frac{e_{L}-e_{i}}{\Delta{}_{Li}\Delta_{LM}}\right)\\
-\frac{1}{\Delta{}_{Li}^{2}}\left(\frac{x^{2}}{2}e_{i}+ix\frac{e_{L}}{\Delta{}_{Li}}+2ix\frac{e_{i}}{\Delta{}_{Li}}-\frac{e_{i}-e_{L}}{\Delta{}_{Li}^{2}}+2\frac{e_{L}-e_{i}}{\Delta{}_{Li}^{2}}\right) & {\rm for}\;j=i,\,k=i,\,M=L\\
\frac{1}{\Delta_{Lk}\Delta_{Mk}}\left(\frac{e_{L}\Delta_{Mk}}{\Delta_{Lj}\Delta_{Li}\Delta_{LM}}-\frac{e_{M}\Delta_{Lk}}{\Delta_{Mj}\Delta_{Mi}\Delta_{LM}}\right. & {\rm for}\;j\neq i,\,k\neq i,j,\,M\neq L\\
\left.+\frac{\Delta_{Lk}\Delta_{Mk}}{\Delta_{ji}\Delta_{ki}\Delta_{Li}\Delta_{Mi}}e_{i}-\frac{\Delta_{Lk}\Delta_{Mk}}{\Delta_{ji}\Delta_{kj}\Delta_{Lj}\Delta_{Mj}}e_{j}+\frac{e_{k}}{\Delta_{ki}\Delta_{kj}}\right)\\
\frac{1}{\Delta_{Lk}}\left[-ix\frac{e_{L}}{\Delta_{Li}\Delta_{Lj}}-\left(\frac{1}{\Delta_{Li}\Delta_{Lj}\Delta_{Lk}}+\frac{\Delta_{Li}+\Delta_{Lj}}{\Delta_{Li}^{2}\Delta_{Lj}^{2}}\right)e_{L}\right. & {\rm for}\;j\neq i,\,k\neq i,j,\,M=L\\
\left.+\frac{\Delta_{Lk}}{\Delta_{ji}\Delta_{ki}\Delta_{Li}^{2}}e_{i}-\frac{\Delta_{Lk}}{\Delta_{ji}\Delta_{kj}\Delta_{Lj}^{2}}e_{j}+\frac{1}{\Delta_{kj}\Delta_{ki}\Delta_{Lk}}e_{k}\right]\\
+\frac{1}{\Delta_{Li}\Delta_{Mi}}\left[ix\frac{e_{i}}{\Delta_{ji}}+\frac{\Delta_{Mi}}{\Delta_{Lj}\Delta_{Li}\Delta_{LM}}e_{L}-\frac{\Delta_{Li}}{\Delta_{Mj}\Delta_{Mi}\Delta_{LM}}e_{M}\right. & {\rm for}\;j\neq i,\,k=i,\,M\neq L\\
\left.-\left(\frac{1}{\Delta_{ji}^{2}}+\frac{1}{\Delta_{ji}\Delta_{Li}}+\frac{1}{\Delta_{ji}\Delta_{Mi}}\right)e_{i}+\frac{\Delta_{Li}\Delta_{Mi}}{\Delta_{Lj}\Delta_{Mj}\Delta_{ji}^{2}}e_{j}\right]\\
-\frac{1}{\Delta_{Lj}\Delta_{Mj}}\left[ix\frac{e_{j}}{\Delta_{ji}}+\frac{\Delta_{Lj}}{\Delta_{Mi}\Delta_{Mj}\Delta_{LM}}e_{M}-\frac{\Delta_{Mj}}{\Delta_{Li}\Delta_{Lj}\Delta_{LM}}e_{L}\right. & {\rm for}\;j\neq i,\,k=j,\,M\neq L\\
\left.+\left(\frac{1}{\Delta_{ji}^{2}}-\frac{1}{\Delta_{ji}\Delta_{Lj}}-\frac{1}{\Delta_{ji}\Delta_{Mj}}\right)e_{j}-\frac{\Delta_{Lj}\Delta_{Mj}}{\Delta_{Li}\Delta_{Mi}\Delta_{ji}^{2}}e_{i}\right]\\
-\frac{1}{\Delta_{Li}^{2}}\left(ix\frac{e_{L}}{\Delta_{Lj}}+ix\frac{e_{i}}{\Delta_{ij}}\right) & {\rm for}\;j\neq i,\,k=i,\,M=L\\
-\frac{1}{\Delta_{Li}^{2}}\left[\left(\frac{1}{\Delta_{ji}^{2}}+\frac{2}{\Delta_{ji}\Delta_{Li}}\right)e_{i}-\frac{\Delta_{Li}^{2}}{\Delta_{ji}^{2}\Delta_{Lj}^{2}}e_{j}+\frac{e_{L}}{\Delta_{Lj}}\left(\frac{1}{\Delta_{Lj}}+\frac{2}{\Delta_{Li}}\right)\right]\\
-\frac{1}{\Delta_{Lj}^{2}}\left(ix\frac{e_{L}}{\Delta_{Li}}+ix\frac{e_{j}}{\Delta_{ji}}\right) & {\rm for}\;j\neq i,\,k=j,\,M=L\\
-\frac{1}{\Delta_{Lj}^{2}}\left[\left(\frac{1}{\Delta_{ji}^{2}}-\frac{2}{\Delta_{ji}\Delta_{Lj}}\right)e_{j}-\frac{\Delta_{Lj}^{2}}{\Delta_{ji}^{2}\Delta_{Li}^{2}}e_{i}+\frac{e_{L}}{\Delta_{Li}}\left(\frac{1}{\Delta_{Li}}+\frac{2}{\Delta_{Lj}}\right)\right]
\end{cases}, \nonumber
\end{eqnarray}
%
%\newpage
%\vglue -1.5cm
\begin{eqnarray}
f_{IJ,klM}^{\left(4\right)} & = & \begin{cases}
\frac{e_{k}}{\Delta_{kl}\Delta_{kM}\Delta_{Ik}^{2}}+\frac{e_{l}}{\Delta_{lk}\Delta_{lM}\Delta_{Il}^{2}}+\frac{e_{M}}{\Delta_{Mk}\Delta_{Ml}\Delta_{IM}^{2}} & {\rm for}\;J=I,\,l\neq k,\,M\neq I\\
-\frac{ix}{\Delta_{Ik}\Delta_{Il}\Delta_{IM}}e_{I}+\left(\frac{1}{\Delta_{Mk}\Delta_{kl}\Delta_{Ik}^{2}}-\frac{1}{\Delta_{Ml}\Delta_{kl}\Delta_{Il}^{2}}-\frac{1}{\Delta_{Mk}\Delta_{Ml}\Delta_{IM}^{2}}\right)e_{I}\\
ix\left(\frac{e_{I}}{\Delta_{Ik}^{2}\Delta_{MI}}+\frac{e_{k}}{\Delta_{Ik}^{2}\Delta_{Mk}}\right)-\left(\frac{1}{\Delta_{Mk}^{2}\Delta_{Ik}^{2}}+\frac{2}{\Delta_{Mk}\Delta_{Ik}^{3}}\right)e_{k}+\frac{e_{M}}{\Delta_{IM}^{2}\Delta_{Mk}^{2}} & {\rm for}\;J=I,\,l=k,\,M\neq I\\
+\left(\frac{1}{\Delta_{Ik}^{2}\Delta_{Mk}^{2}}-\frac{1}{\Delta_{IM}^{2}\Delta_{Mk}^{2}}+\frac{2}{\Delta_{Ik}^{3}\Delta_{Mk}}\right)e_{I}\\
ix\left(\frac{1}{\Delta_{Ik}\Delta_{Il}^{2}}+\frac{1}{\Delta_{Ik}^{2}\Delta_{Il}}\right)e_{I}-\frac{x^{2}}{2\Delta_{Ik}\Delta_{Il}}e_{I} & {\rm for}\;J=I,\,l\neq k,\,M=I\\
+ \frac{e_{k}}{\Delta_{lk}\Delta_{Ik}^{3}}+\frac{e_{l}}{\Delta_{kl}\Delta_{Il}^{3}}+\left(\frac{1}{\Delta_{Ik}^{2}\Delta_{Il}^{2}}+\frac{1}{\Delta_{Ik}\Delta_{Il}^{3}}+\frac{1}{\Delta_{Il}\Delta_{Ik}^{3}}\right)e_{I}\\
\frac{ix}{\Delta_{Ik}^{3}}e_{k}+\frac{2ix}{\Delta_{Ik}^{3}}e_{I}-\frac{x^{2}}{2\Delta_{Ik}^{2}}e_{I}+\frac{3}{\Delta_{Ik}^{4}}\left(e_{I}-e_{k}\right) & {\rm for}\;J=I,\,l=k,\,M=I\\
\frac{e_{k}}{\Delta_{Ik}\Delta_{kk}\Delta_{Jk}\Delta_{Mk}}+\frac{e_{l}}{\Delta_{kl}\Delta_{Il}\Delta_{Jl}\Delta_{Ml}}+\frac{e_{M}}{\Delta_{IM}\Delta_{JM}\Delta_{Mk}\Delta_{Ml}} & {\rm for}\;J\neq I,\,l\neq k,\,M\neq J,I\\
+\frac{e_{I}}{\Delta_{Ik}\Delta_{Il}\Delta_{IM}\Delta_{IJ}}+\frac{e_{J}}{\Delta_{Jk}\Delta_{Jl}\Delta_{JM}\Delta_{JI}}\\
\frac{ix}{\Delta_{Ik}\Delta_{Jk}\Delta_{Mk}}e_{k}-\left(\frac{1}{\Delta_{Ik}^{2}\Delta_{Jk}\Delta_{Mk}}+\frac{1}{\Delta_{Ik}\Delta_{Jk}^{2}\Delta_{Mk}}+\frac{1}{\Delta_{Ik}\Delta_{Jk}\Delta_{Mk}^{2}}\right)e_{k} & {\rm for}\;J\neq I,\,l=k,\,M\neq J,I\\
+\frac{1}{\Delta_{IM}\Delta_{JM}\Delta_{Mk}^{2}}e_{M}+\frac{1}{\Delta_{Ik}^{2}\Delta_{IJ}\Delta_{IM}}e_{I}+\frac{1}{\Delta_{Jk}^{2}\Delta_{JI}\Delta_{JM}}e_{J}\\
\frac{ix}{\Delta_{kI}\Delta_{IJ}\Delta_{Il}}e_{I}+\frac{1}{\Delta_{kJ}\Delta_{Ik}^{2}\Delta_{kl}}e_{k}+\frac{1}{\Delta_{lk}\Delta_{Jl}\Delta_{Il}^{2}}e_{l}+\frac{1}{\Delta_{Jk}\Delta_{JI}^{2}\Delta_{Jl}}e_{J} & {\rm for}\;J\neq I,\,l\neq k,\,M=I\\
-\left(\frac{1}{\Delta_{Ik}\Delta_{IJ}\Delta_{Il}^{2}}+\frac{1}{\Delta_{Ik}\Delta_{JI}^{2}\Delta_{Il}}+\frac{1}{\Delta_{Ik}^{2}\Delta_{IJ}\Delta_{Il}}\right)e_{I}\\
\frac{ix}{\Delta_{kJ}\Delta_{JI}\Delta_{Jl}}e_{J}+\frac{1}{\Delta_{kI}\Delta_{Jk}^{2}\Delta_{kl}}e_{k}+\frac{1}{\Delta_{lk}\Delta_{Il}\Delta_{Jl}^{2}}e_{l}+\frac{1}{\Delta_{Ik}\Delta_{JI}^{2}\Delta_{Il}}e_{I} & {\rm for}\;J\neq I,\,l\neq k,\,M=J\\
-\left(\frac{1}{\Delta_{Jk}\Delta_{JI}\Delta_{Jl}^{2}}+\frac{1}{\Delta_{Jk}\Delta_{JI}^{2}\Delta_{Jl}}+\frac{1}{\Delta_{Jk}^{2}\Delta_{JI}\Delta_{Jl}}\right)e_{J}\\
ix\left(\frac{e_{k}}{\Delta_{Ik}^{2}\Delta_{Jk}}+\frac{e_{I}}{\Delta_{Ik}^{2}\Delta_{JI}}\right)-\left(\frac{1}{\Delta_{Ik}^{2}\Delta_{Jk}^{2}}+\frac{2}{\Delta_{Ik}^{3}\Delta_{Jk}}\right)e_{k} & {\rm for}\;J\neq I,\,l=k,\,M=I\\
-\left(\frac{1}{\Delta_{Ik}^{2}\Delta_{JI}^{2}}+\frac{2}{\Delta_{Ik}^{3}\Delta_{IJ}}\right)e_{I}+\frac{1}{\Delta_{Jk}^{2}\Delta_{JI}^{2}}e_{J}\\
ix\left(\frac{e_{k}}{\Delta_{Jk}^{2}\Delta_{Ik}}+\frac{e_{J}}{\Delta_{Jk}^{2}\Delta_{IJ}}\right)-\left(\frac{1}{\Delta_{Ik}^{2}\Delta_{Jk}^{2}}+\frac{2}{\Delta_{Jk}^{3}\Delta_{Ik}}\right)e_{k} & {\rm for}\;J\neq I,\,l=k,\,M=J\\
-\left(\frac{1}{\Delta_{Jk}^{2}\Delta_{JI}^{2}}+\frac{2}{\Delta_{Jk}^{3}\Delta_{JI}}\right)e_{J}+\frac{1}{\Delta_{Ik}^{2}\Delta_{JI}^{2}}e_{I}
\end{cases}. \nonumber
\end{eqnarray}
Notice that for $\hat S_{ij}^{\left(4\right)}[4]$, the combinations of couplings remain invariant under complex conjugation with $\left(i\leftrightarrow j\right)$
and $\left(L\leftrightarrow M\right)$ and hence we need that
$f_{ij,kLM}^{\left(4\right)}=f_{ji,kML}^{\left(4\right)}$ as can be verified from the expressions above. As for
$\hat S_{IJ}^{\left(4\right)}[4]$, the combinations of the couplings remain invariant under complex conjugation with $\left(I\leftrightarrow J\right)$ and $\left(k\leftrightarrow l\right)$ and hence we need that $f_{IJ,klM}^{\left(4\right)}=f_{JI,lkM}^{\left(4\right)}$ which can be verified from the expression above.
%\newpage
%\section{Structure of $S$ matrix elements}
%\label{sec:structure-S-matrix}
%

\section{Expression of the oscillation probability in fourth order in $W$}
\label{sec:expression-probability-4th}

%Here we present details in computation of the $S$ matrix elements. 

%\subsection{The $S$ matrix elements $S_{\alpha \beta}^{(4)}$}

For the $S$ matrix elements $S_{\alpha \beta}^{(4)}$, we decompose $S_{\alpha \beta}^{(4)}$ into the following three pieces (include both $\alpha \neq \beta$ and $\alpha = \beta$)
\begin{eqnarray} 
S_{\alpha \beta}^{(4)} &=& S_{\alpha \beta}^{(4)} [3+4] + S_{\alpha \beta}^{(4)} [3] + S_{\alpha \beta}^{(4)} [2] .
\label{Sab-4th}
\end{eqnarray}
%
%where $[n]$ implies that the term comes from $n$-th order perturbation of $H_{1}$. 
To prevent too long expression, we decompose the first term in (\ref{Sab-4th}) as 
\begin{eqnarray} 
S_{\alpha \beta}^{(4)} [3+4] 
&=& 
S_{\alpha \beta}^{(4)} [3]_{ \text{ diag } } + S_{\alpha \beta}^{(4)} [4]_{ \text{ diag } } + 
S_{\alpha \beta}^{(4)} [3]_{ \text{ offdiag } } + S_{\alpha \beta}^{(4)} [4]_{ \text{ offdiag } } ,
\label{Sab-4th-3+4}
\end{eqnarray}
%
%where ($n = 3, 4$) ``diag'' and ``offdiag'', respectively, implies 
where the $S$ matrix with subscript ``diag'' (``offdiag'') implies ($n = 3, 4$) 
\begin{eqnarray} 
S_{\alpha \beta}^{(4)} [n]_{ \text{ diag (offdiag)} } 
&=& 
\sum_{k (k\ne l)} (UX)_{ik} 
\left( \hat{S}_{kk(kl)}^{(4)} [n] \right)
\left\{ (UX)^{\dagger} \right\}_{kj(lj)}.
\end{eqnarray} 
%
%\begin{eqnarray} 
%S_{\alpha \beta}^{(4)} [n]_{ \text{ diag } } 
%&=& 
%\sum_{k} (UX)_{ik} 
%\left( \hat{S}_{kk}^{(4)} [n] \right)
%\left\{ (UX)^{\dagger} \right\}_{kj}, 
%%
%\nonumber \\
%S_{\alpha \beta}^{(4)} [n]_{ \text{ offdiag } }
%&=& 
%=
%\sum_{k \neq l} (UX)_{ik} 
%\left( \hat{S}_{kl}^{(4)} [n] \right)
%\left\{ (UX)^{\dagger} \right\}_{lj}. 
%\end{eqnarray}
%
The latter two terms in (\ref{Sab-4th}) are given, respectively, by 
\begin{eqnarray} 
S_{\alpha \beta}^{(4)} [3] 
&=& 
\sum_{k L} (UX)_{ik} \hat{S}_{kL}^{(3)} \left\{ (W^{\dagger}) \right\}_{L j} 
+ \sum_{K l} W_{iK} \hat{S}_{Kl}^{(3)} \left\{ (UX)^{\dagger} \right\}_{lj}, 
\nonumber \\
S_{\alpha \beta}^{(4)} [2] &=& 
\sum_{K} W_{iK} \hat{S}_{KK}^{(2)} \left\{ (W^{\dagger}) \right\}_{K j} + 
\sum_{K \neq L} W_{iK} \hat{S}_{KL}^{(2)} \left\{ (W^{\dagger}) \right\}_{L j}. 
\label{S-alpha-beta-4th-[3]}
\end{eqnarray}

We do not display explicitly the expression of each term in (\ref{Sab-4th}). But, the notation of $S_{\alpha \beta}^{(4)} [n]_{ \text{ diag } }$ and $S_{\alpha \beta}^{(4)} [n]_{ \text{ offdiag } }$ will be transported to the notation for the oscillation probability such that 
$2 \mbox{Re} \left[ \left( S^{(0)}_{\alpha \beta} \right)^{*} S_{\alpha \beta}^{(4)} [n]_{ \text{ diag } } \right]$. 
Similarly, to make the equation fit to a single page we present the first and the second terms of $S_{\alpha \beta}^{(4)} [3]$ in (\ref{S-alpha-beta-4th-[3]}) separately, 
$S_{\alpha \beta}^{(4)} [3]_\text{First} = \sum_{k L} (UX)_{\alpha k} W^*_{\beta L} \hat{S}_{kL}^{(3)}$ and 
$S_{\alpha \beta}^{(4)} [3]_\text{Second} = \sum_{L k} W_{\alpha L} (UX)^*_{\beta k} \hat{S}_{L k}^{(3)}$, whose notations are also transported to the oscillation probability.

The oscillation probability to second order in $W$ is given in eq.~(\ref{P-beta-alpha-0th+2nd}) in section~\ref{sec:probability-2nd}. What is left is, therefore, the expressions of the oscillation probability in fourth order in $W$, the explicit form of the two terms in (\ref{P-beta-alpha-4th-def}), 
$P(\nu_\beta \rightarrow \nu_\alpha) = 
\left| S^{(2)}_{\alpha \beta} \right|^2 
+ 2 \mbox{Re} \left[ \left( S^{(0)}_{\alpha \beta} \right)^{*} S^{(4)}_{\alpha \beta} \right] $. 

Besides using the notations defined in eqs. \eqref{eq:c_iJ-d_IJ} -- \eqref{eq:Delta_ab}, we will further define the following quantities
%%%
\begin{eqnarray}
% \tilde U & \equiv & U X, \\
e_{ij} & \equiv & e^{-i (h_i - h_j)x},\;\;\;\; 
e_{Ij} \equiv e^{-i (\Delta_I - h_j)x}, \;\;\;\;
e_{IJ} \equiv e^{-i (\Delta_I - \Delta_J)x}. \label{eq:e_ab}
\end{eqnarray}
%%%

\subsection{Second order $S$ matrix squared term: $\left| S^{(2)}_{\alpha \beta} \right|^2$}
\label{sec:second-order-square}

The $S$ matrix element $S^{(2)}_{\alpha \beta}$ in eq.~(\ref{S-alpha-beta-2nd}) contains four terms. To prevent too long expressions, we divide $\left| S^{(2)}_{\alpha \beta} \right|^2$  into the two terms, one sum of each term squared and the other one composed of cross terms. The first one is given by 
\begin{eqnarray} 
&& \left| S^{(2)}_{\alpha \beta} \right|^2_{\text{1st}} = 
%
%\nonumber \\ &=& 
\sum_{k, K} \sum_{l, L} 
\frac{ (UX)_{\alpha k} (UX)^*_{\beta k} 
	c_{kK} c_{kK}^*
%	\times 
	(UX)_{\alpha l}^* (UX)_{\beta l} 
	c_{lL} c_{lL}^* }
{ \Delta_{Kk} \Delta_{Ll}  } 
\nonumber \\ 
&\times&
\biggl[
x^2 e_{kl} 
- (ix) \frac{e_{Kl} - e_{kl} }{ \Delta_{Kk}  } 
+ (ix) \frac{e_{kL} - e_{kl} }{ \Delta_{Ll}  } 
+ 
\frac{ 1 }{ \Delta_{Kk} \Delta_{Ll} } 
\biggl(
e_{KL} + e_{kl} 
- e_{Kl} - e_{kL} 
\biggr)
\biggr] 
%
%\nonumber \\ 
%&\times&
%(UX)_{\alpha k} (UX)^*_{\beta k} 
%c_{kK} c_{kK}^*
%\times 
%(UX)_{\alpha l}^* (UX)_{\beta l} 
%c_{lL} c_{lL}^*
%%
\nonumber \\ 
&+& 
\sum_{k \neq m} \sum_{K}  \sum_{l \neq n} \sum_{L}  
\frac{ (UX)_{\alpha k} (UX)^*_{\beta m} c_{kK} c_{mK}^*
	%
%	\times
	(UX)_{\alpha l}^* (UX)_{\beta n} c_{nL} c_{lL}^* }
{ \Delta_{mk} \Delta_{Kk} \Delta_{Km} 
	\Delta_{nl} \Delta_{Ll} \Delta_{Ln} } 
\nonumber \\
&\times& 
\biggl[
\Delta_{Kk}  e_m 
- \Delta_{Km} e_k  
- \Delta_{mk} e_K 
\biggr]
\biggl[
\Delta_{Ll} e_n^* 
- \Delta_{Ln} e_l^*
- \Delta_{nl} e_L^*
\biggr]
%
%\nonumber \\
%&\times&
%(UX)_{\alpha k} (UX)^*_{\beta m} c_{kK} c_{mK}^*
%%
%\times
%(UX)_{\alpha l}^* (UX)_{\beta n} c_{nL} c_{lL}^*
%%
%\nonumber \\
%&+& 
+\sum_{k, K} \sum_{l, L} 
\frac{\left( e_K - e_k \right) 
	\left( e_L^* - e_l^* \right)}
{ \Delta_{Kk} \Delta_{Ll} } 
\nonumber \\
&\times&
\biggl[ 
(UX)_{\alpha k} W^*_{\beta K} c_{kK}  
+
W_{\alpha K} (UX)^*_{\beta k} c_{kK}^*
\biggr]
\biggl[ 
(UX)_{\alpha l}^* W_{\beta L} c_{lL}^*
+
W_{\alpha L}^* (UX)_{\beta l} c_{lL}
\biggr]
\nonumber \\
&+& 
\sum_{K} 
\vert W_{\alpha K} \vert^2 \vert W_{\beta K} \vert^2
+ \sum_{K \neq L} 
e_{KL} 
W_{\alpha K} W^*_{\beta K} W_{\alpha L}^* W_{\beta L}.  
\label{S(2)-squared-1}
\end{eqnarray}
%
%Except for the first term in (\ref{S(2)-squared-1}) we did not try to unify the two exponential factors because the expressions become cumbersome. 
%
Apart from the last line in (\ref{S(2)-squared-1}) all the terms are suppressed by the two sterile state mass denominators with $\Delta m^2_{J k}$ which doubly suppress the active-sterile state transition. The first term in the last line is the probability leaking term mentioned in section~\ref{sec:nonunitarity-vacuum-small-matter}.

The second term of $\left| S^{(2)}_{\alpha \beta} \right|^2$ (interference terms) is given by 
\begin{eqnarray} 
&& \left| S^{(2)}_{\alpha \beta} \right|^2_{\text{2nd}} = 
%
%\nonumber \\ &=& 
- 2 \mbox{Re} 
\biggl\{ 
\sum_{k, K} 
\sum_{l \neq m} \sum_{L} 
\frac{ (UX)_{\alpha l} (UX)^*_{\beta m} 
	(UX)_{\alpha k}^* (UX)_{\beta k} 
	\, c_{lL} c_{mL}^* c_{kK} c_{kK}^* }
{ \Delta_{Kk} \Delta_{Ll} \Delta_{Lm} \Delta_{ml} } 
\nonumber \\
&\times& 
\biggl[
\Delta_{Ll} e_m 
- \Delta_{Lm} e_l 
- \Delta_{ml} e_L
\biggr]
\left[
- (ix) e_k^* + \frac{e_K^* - e_k^* }{ \Delta_{Kk}  } 
\right]
%
%\nonumber \\
%&\times&
%(UX)_{\alpha l} (UX)^*_{\beta m} 
%(UX)_{\alpha k}^* (UX)_{\beta k} 
%%
%c_{lL} c_{mL}^* c_{kK} c_{kK}^*
\biggl\}
%
%\nonumber \\
%&+& 
+
2 \mbox{Re} 
\biggl\{ 
\sum_{k, K} \sum_{l, L} 
\left[
- (ix) e_k^* + \frac{e_K^* - e_k^*}{ \Delta_{Kk} } 
\right]
\nonumber \\ 
&\times&
\frac{e_L - e_l }{ \Delta_{Kk} \Delta_{Ll} } 
(UX)_{\alpha k}^* (UX)_{\beta k} c_{kK} c_{kK}^*
%
%\times
%
\biggl[ 
(UX)_{\alpha l} W^*_{\beta L} c_{lL}
+
W_{\alpha L} (UX)^*_{\beta l} c_{lL}^*
\biggr]
\biggr\}
\nonumber \\
&+& 
2 \mbox{Re} 
\biggl\{
\sum_{k, K} \sum_{L} 
\left[
- (ix) e_{Lk} + \frac{ e_{LK} - e_{Lk} }{ \Delta_{Kk} } 
\right] 
\frac{ (UX)_{\alpha k}^* (UX)_{\beta k} 
	W_{\alpha L} W^*_{\beta L} 
	c_{k K} c_{k K}^* }
{ \Delta_{Kk} } 
%
%\nonumber \\ 
%&\times&
%(UX)_{\alpha k}^* (UX)_{\beta k} 
%W_{\alpha L} W^*_{\beta L} 
%c_{k K} c_{k K}^*
\biggr\}
\nonumber \\
&-& 
2 \mbox{Re} 
\biggl\{
\sum_{k \neq m} \sum_{K} \sum_{l, L} 
\frac{ 1 }{ \Delta_{mk} \Delta_{Kk} \Delta_{Km} } 
\biggl[
\Delta_{Kk} e_i^* 
- \Delta_{Km} e_k^* 
- \Delta_{mk} e_K^* 
\biggr]
\frac{e_L - e_l }{ \Delta_{Ll} } 
\nonumber \\
&\times& 
(UX)^*_{\alpha k} (UX)_{\beta m} c_{mK} c_{kK}^*
\times
\biggl[ 
(UX)_{\alpha l} W^*_{\beta L} c_{lL}
+
W_{\alpha L} (UX)^*_{\beta l} c_{lL}^*
\biggr]
\biggr\}
\nonumber \\
&-& 
2 \mbox{Re} 
\biggl\{
\sum_{k \neq m} \sum_{K} \sum_{L} 
\frac{ (UX)^*_{\alpha k} (UX)_{\beta m} 
	W_{\alpha L} W^*_{\beta L} 
	c_{mK} c_{kK}^* }
{ \Delta_{mk} \Delta_{Kk} \Delta_{Km} } 
\biggl[
\Delta_{Kk}  e_{Lm} 
- \Delta_{Km} e_{Lk}
- \Delta_{mk} e_{LK}
\biggr]
%
%\nonumber \\
%&\times& 
%(UX)^*_{\alpha k} (UX)_{\beta m} 
%W_{\alpha L} W^*_{\beta L} 
%c_{mK} c_{kK}^*
%%
\biggr\}
\nonumber \\
&+& 
2 \mbox{Re} 
\biggl\{
\sum_{K} \sum_{l, L} 
\frac{e_{LK} - e_{lK} }{ \Delta_{Ll} } 
%
%\nonumber \\
%&\times&
\biggl[ 
W^*_{\alpha K} W_{\beta K} 
(UX)_{\alpha l} W^*_{\beta L} c_{lL} 
+
W^*_{\alpha K} W_{\beta K} 
W_{\alpha L} (UX)^*_{\beta l} c_{lL}^*
\biggr]
\biggr\}.
\label{S(2)-squared-2}
\end{eqnarray}
\subsection{Interference terms of the type $2 \mbox{Re} \left[ \left( S^{(0)}_{\alpha \beta} \right)^{*} S^{(4)}_{\alpha \beta} \right]$}
\label{sec:interference}

We classify the fourth order in $W$ contribution of the interference terms into 8 terms: 
\begin{eqnarray} 
P(\nu_\beta \rightarrow \nu_\alpha)^{(4)}_{ \text{interference} }
&=& P(\nu_\beta \rightarrow \nu_\alpha)^{(4)}_{\rm 1st} 
+ P(\nu_\beta \rightarrow \nu_\alpha)^{(4)}_{\rm 2nd} 
+ P(\nu_\beta \rightarrow \nu_\alpha)^{(4)}_{\rm 3rd} 
\nonumber \\
&+&
P(\nu_\beta \rightarrow \nu_\alpha)^{(4)}_{\rm 4th-s} 
+ P(\nu_\beta \rightarrow \nu_\alpha)^{(4)}_{\rm 4th-d} 
%
%\nonumber 
\\
&+& 
P(\nu_\beta \rightarrow \nu_\alpha)^{(4)}_{\rm 5th-1st}
+ P(\nu_\beta \rightarrow \nu_\alpha)^{(4)}_{\rm 5th-2nd} 
+ P(\nu_\beta \rightarrow \nu_\alpha)^{(4)}_{\rm 6th}. 
\nonumber 
\end{eqnarray}
The nature of each term is explicitly indicated as follows:
\begin{eqnarray} 
&& P(\nu_\beta \rightarrow \nu_\alpha)^{(4)}_{\rm 1st} 
\equiv 
2 \mbox{Re} \left[ \left( 
S^{(0)}_{\alpha \beta} \right)^{*} 
S_{\alpha \beta}^{(4)} [3]_{ \text{ diag } } 
\right]
\nonumber \\
&=& 
2 \mbox{Re} 
\biggl\{
%%%%%%%%%%%%%%%%%%%%%%%
- \sum_{K} \sum_{k} 
\sum_{m} 
\biggl[ 
\frac{ 1 }{ \Delta_{Kk}^2 } 
\left\{ (ix) + \frac{ 2 }{ \Delta_{Kk} } \right\}
\left( e_{Km} + e_{km} \right) 
\biggr]
\nonumber \\ 
&\times& 
(UX)^*_{\alpha m} (UX)_{\beta m} 
(UX)_{\alpha k} (UX)^*_{\beta k}
c_{kK} 
%\left\{ (UX)^{\dagger} A W \right\}_{k K} 
d_{KK}
%\left\{ W ^{\dagger} A W \right\}_{K K} 
c_{kK}^*
%\left\{ W ^{\dagger} A (UX) \right\}_{K k} 
%%\left\{ (UX)^{\dagger} \right\}_{k \beta}
%
\nonumber \\ 
&+& 
\sum_{K \neq L} \sum_{k} 
\sum_{m} 
\biggl[ 
\frac{ 1 }{ \Delta_{Kk}^2 \Delta_{KL} } 
e_{Km} 
-
\frac{ 1 }{ \Delta_{Lk}^2 \Delta_{KL} } 
e_{Lm} 
+ 
\frac{ 1 }{ \Delta_{Kk}^2 \Delta_{Lk}^2  }  
\left( \Delta_{K} + \Delta_{L} - 2 h_{k} \right) 
e_{km} 
\biggr]
\nonumber \\ 
&\times& 
(UX)^*_{\alpha m} (UX)_{\beta m} 
(UX)_{\alpha k} (UX)^*_{\beta k}
c_{kK}
%\left\{ (UX)^{\dagger} A W \right\}_{k K} 
d_{KL}
%\left\{ W ^{\dagger} A W \right\}_{K L} 
c_{kL}^*
%\left\{ W ^{\dagger} A (UX) \right\}_{L k} 
\biggr\}. 
%%%%%%%%%%%%%%%%%%%%%%%
%
\label{P-beta-alpha-W4-H3-diag}
\end{eqnarray}
\begin{eqnarray} 
&& P(\nu_\beta \rightarrow \nu_\alpha)^{(4)}_{\rm 2nd} 
\equiv 
2 \mbox{Re} \left[ \left( 
S^{(0)}_{\alpha \beta} \right)^{*} 
S_{\alpha \beta}^{(4)} [3]_{ \text{ offdiag } } 
\right]
\nonumber \\
&=& 
2 \mbox{Re} 
\biggl\{ 
%%%%%%%%%%%%%%%%%%%%%%%%%%%%%%%
\sum_{m} 
%e^{ +i h_{m} x}
\sum_{k \neq l } \sum_{K} 
\biggl[ 
- \frac{ (ix) }{ \Delta_{Kk} \Delta_{Kl} } 
e_{Km} 
+ \frac{ 1 }{ \Delta_{lk} \Delta_{Kk}^2 \Delta_{Kl}^2 } 
\nonumber \\ 
&\times& 
\biggl\{
\Delta_{lk} ( h_{l} + h_{k} - 2  \Delta_{K} ) 
e_{Km} 
%e^{- i \Delta_{K} x} + 
+ \Delta_{Kk}^2 e_{lm} 
- \Delta_{Kl}^2 e_{km} 
\biggr\}
\biggr]
\nonumber \\ 
&\times&
(UX)_{\alpha k} (UX)^*_{\beta l} 
(UX)^*_{\alpha m} (UX)_{\beta m} 
c_{kK}
%\left\{ (UX)^{\dagger} A W \right\}_{k K} 
d_{KL}
%\left\{ W ^{\dagger} A W \right\}_{K K} 
c_{lK}^*
%\left\{ W ^{\dagger} A (UX) \right\}_{K l} 
%
\nonumber \\ 
&+& 
\sum_{m} 
%e^{ +i h_{m} x}
\sum_{k \neq l } \sum_{K \neq L } 
\frac{ 1 }{ \Delta_{lk} \Delta_{LK} \Delta_{Kk} 
	\Delta_{Kl} \Delta_{Lk} \Delta_{Ll}   } 
\nonumber \\ 
&\times& 
\biggl[ 
\Delta_{lk} 
\biggl\{ \Delta_{Kk} \Delta_{Kl} 
e_{Lm} - 
%e^{- i \Delta_{L} x} - 
\Delta_{Lk} \Delta_{Ll}  
e_{Km}
%e^{- i \Delta_{K} x} 
\biggr\} 
+ 
\Delta_{LK}
\biggl\{ \Delta_{Kk} \Delta_{Lk} e_{lm} - 
\Delta_{Kl} \Delta_{Ll} e_{km} \biggr\} 
\biggl]
\nonumber \\ 
&\times&
(UX)_{\alpha k} (UX)^*_{\beta l} 
(UX)^*_{\alpha m} (UX)_{\beta m} 
c_{kK}
%\left\{ (UX)^{\dagger} A W \right\}_{k K} 
d_{KL}
%\left\{ W ^{\dagger} A W \right\}_{K L} 
c_{lL}^*
%\left\{ W ^{\dagger} A (UX) \right\}_{L l} 
%
\biggr\}.
\label{P-beta-alpha-W4-H3-offdiag}
\end{eqnarray}
\begin{eqnarray} 
&& P(\nu_\beta \rightarrow \nu_\alpha)^{(4)}_{\rm 3rd} 
\equiv 
2 \mbox{Re} \left[ \left( 
S^{(0)}_{\alpha \beta} \right)^{*} 
S_{\alpha \beta}^{(4)} [4]_{ \text{ diag } } 
\right] 
\nonumber \\
&=& 
2 \mbox{Re} 
\biggl\{ 
%%%%%%%%%%%%%%%%%%%%%%%%%%%%%%%
\sum_{n} \sum_{k} \sum_{K} 
\biggl[
- \frac{x^2}{2} \frac{ 1 }{ \Delta_{Kk}^2 } 
e_{kn} 
- 
\frac{ 2 (ix) }{ \Delta_{Kk}^3 } 
e_{kn} 
-
\frac{ (ix) }{ \Delta_{Kk}^3 } 
e_{Kn}
-
\frac{ 3 }{ \Delta_{Kk}^4 } 
\left( e_{Kn} - e_{kn} \right)
\biggr]
\nonumber \\ 
&\times&
(UX)_{\alpha k} (UX)^*_{\beta k} 
(UX)^*_{\alpha n} (UX)_{\beta n} 
c_{kK} c_{kK}^* c_{kK} c_{kK}^*
%\left\{ (UX)^{\dagger} A W \right\}_{k K} 
%\left\{ W ^{\dagger} A (UX) \right\}_{K k} 
%\left\{ (UX)^{\dagger} A W \right\}_{k K} 
%\left\{ W ^{\dagger} A (UX) \right\}_{K k} 
%
\nonumber \\
&+& 
\sum_{n} \sum_{k} \sum_{K} \sum_{m \neq k} 
\biggl[
\frac{ (ix) }{ \Delta_{Kk}^2 \Delta_{mk} } 
e_{kn} 
-
\frac{ (ix) }{ \Delta_{Kk}^2 \Delta_{Km} } 
e_{Kn} 
+
\frac{ ( h_{k} + 2 h_{m} - 3 \Delta_{K} ) }{ \Delta_{Kk}^3 \Delta_{Km}^2 } 
e_{Kn} 
\nonumber \\
&+& 
\frac{ 1 }{ \Delta_{Km}^2 \Delta_{mk}^2 } 
e_{mn}
- 
\frac{ \left( \Delta_{K} + 2 h_{m} - 3 h_{k} \right) }{ \Delta_{Kk}^3 \Delta_{mk}^2 } 
e_{kn}
\biggr]
\nonumber \\ 
&\times&
(UX)_{\alpha k} (UX)^*_{\beta k} 
(UX)^*_{\alpha n} (UX)_{\beta n} 
c_{kK} c_{mK}^* c_{mK} c_{kK}^*
%\left\{ (UX)^{\dagger} A W \right\}_{k K} 
%\left\{ W ^{\dagger} A (UX) \right\}_{K m} 
%\left\{ (UX)^{\dagger} A W \right\}_{m K} 
%\left\{ W ^{\dagger} A (UX) \right\}_{K k} 
%
\nonumber \\
&+& 
\sum_{n} \sum_{k} \sum_{K \neq L} 
\biggl[
- \frac{x^2}{2} 
\frac{ 1 }{ \Delta_{Kk} \Delta_{Lk} } 
e_{kn} 
- (ix) 
\frac{ \left( \Delta_{K} + \Delta_{L} - 2 h_{k} \right) }{ \Delta_{Kk}^2 \Delta_{Lk}^2 } 
e_{kn} 
-
\frac{ 1 }{ \Delta_{Kk}^3 \Delta_{LK} } 
e_{Kn} 
\nonumber \\
&+& 
\frac{ 1 }{ \Delta_{Lk}^3 \Delta_{LK} } 
e_{Ln}
+ 
\frac{ 1 }{ \Delta_{Kk}^3 \Delta_{Lk}^3 } 
\biggl\{
\Delta_{L}^2 + \Delta_{L} \Delta_{K} + \Delta_{K}^2 - 3 h_{k} ( \Delta_{L} + \Delta_{K} ) + 3 h_{k}^2  
\biggr\}
e_{kn}
\biggr]
\nonumber \\ 
&\times&
(UX)_{\alpha k} (UX)^*_{\beta k} 
(UX)^*_{\alpha n} (UX)_{\beta n} 
c_{kK} c_{kK}^* c_{kL} c_{kL}^*
%\left\{ (UX)^{\dagger} A W \right\}_{k K} 
%\left\{ W ^{\dagger} A (UX) \right\}_{K k} 
%\left\{ (UX)^{\dagger} A W \right\}_{k L} 
%\left\{ W ^{\dagger} A (UX) \right\}_{L k} 
%
\nonumber \\
&+& 
\sum_{n} \sum_{k} \sum_{K \neq L} \sum_{m \neq k} 
\biggl[
\frac{ (ix) }{ \Delta_{Kk} \Delta_{Lk} \Delta_{mk} }
e_{kn} 
\nonumber \\
&-& 
\frac{1}{ \Delta_{Kk}^2 \Delta_{Lk}^2 \Delta_{mk}^2 } 
\biggl\{ 
\Delta_{K} \Delta_{L} + ( h_{m} - 2 h_{k} ) ( \Delta_{K} + \Delta_{L} ) + 3 
h_{k}^2 - 2 h_{m} h_{k}
\biggr\}
e_{kn} 
\nonumber \\
&+&
\frac{1}{ \Delta_{Km} \Delta_{Lm} \Delta_{mk}^2 } 
e_{mn}
+
\frac{1}{ \Delta_{KL} \Delta_{Kk}^2 \Delta_{Km} } e_{Kn} 
- \frac{1}{ \Delta_{KL} \Delta_{Lk}^2 \Delta_{Lm} } 
e_{Ln}
\biggr]
\nonumber \\ 
&\times&
(UX)_{\alpha k} (UX)^*_{\beta k} 
(UX)^*_{\alpha n} (UX)_{\beta n} 
c_{kK} c_{mK}^* c_{mL} c_{kL}^*
%\left\{ (UX)^{\dagger} A W \right\}_{k K} 
%\left\{ W ^{\dagger} A (UX) \right\}_{K m} 
%\left\{ (UX)^{\dagger} A W \right\}_{m L} 
%\left\{ W ^{\dagger} A (UX) \right\}_{L k} 
\biggr\}. 
\label{P-beta-alpha-W4-H4-diag}
\end{eqnarray}
\begin{eqnarray} 
&& P(\nu_\beta \rightarrow \nu_\alpha)^{(4)}_{\rm 4th-s} 
\equiv 
2 \mbox{Re} \left[ \left( 
S^{(0)}_{\alpha \beta} \right)^{*} 
S_{\alpha \beta}^{(4)} [4]_{ \text{ offdiag } } (\text{single})
\right]
\nonumber \\
&=& 
2 \mbox{Re} 
\biggl\{
%%%%%%%%%%%%%%%%%%%%%%%%%%%%%%%
\sum_{n} 
\sum_{k \neq l } 
\sum_{K}
\biggl[
\frac{ (ix)e_{kn} }{ \Delta_{Kk}^2 \Delta_{lk} } 
- 
\frac{ (ix) e_{Kn}}{ \Delta_{Kk}^2 \Delta_{Kl} } 
+
\frac{ e_{ln} }{ \Delta_{Kl}^2 \Delta_{lk}^2 } 
- \frac{ \left( \Delta_{K} + 2 h_{l} - 3 h_{k} \right) 
e_{kn}
 }{ \Delta_{Kk}^3 \Delta_{lk}^2 } 
\nonumber \\
&+&
\frac{ \left( h_{k} + 2 h_{l}  - 3 \Delta_{K} \right) 
e_{Kn} 
 }{ \Delta_{Kk}^3 \Delta_{Kl}^2 } 
\biggr]
%
%\nonumber \\ 
%&\times&
(UX)_{\alpha k} (UX)^*_{\beta l}
(UX)^*_{\alpha n} (UX)_{\beta n} 
c_{kK} c_{cK}^* c_{kK} c_{lK}^*
%\left\{ (UX)^{\dagger} A W \right\}_{k K} 
%\left\{ W ^{\dagger} A (UX) \right\}_{K k} 
%\left\{ (UX)^{\dagger} A W \right\}_{k K} 
%\left\{ W ^{\dagger} A (UX) \right\}_{K l} 
%
\nonumber \\
&+& 
\sum_{n} 
\sum_{k \neq l }  
\sum_{K}
\biggl[
- 
\frac{ (ix)e_{ln} }{ \Delta_{Kl}^2 \Delta_{lk} } 
- 
\frac{ (ix)e_{Kn} }{ \Delta_{Kl}^2 \Delta_{Kk} } 
+
\frac{ \left( h_{l} + 2 h_{k} - 3 \Delta_{K} \right) e_{Kn} }{ \Delta_{Kk}^2 \Delta_{Kl}^3 } 
-\frac{ \left( \Delta_{K} + 2 h_{k} - 3 h_{l}  \right) e_{ln} }{ \Delta_{Kl}^3 \Delta_{lk}^2 } 
\nonumber \\
&+& 
\frac{ e_{kn} }{ \Delta_{Kk}^2 \Delta_{lk}^2 } 
\biggr]
%
%\nonumber \\ 
%&\times&
(UX)_{\alpha k} (UX)^*_{\beta l} 
(UX)^*_{\alpha n} (UX)_{\beta n} 
c_{kK} c_{lK}^* c_{lK} c_{lK}^*
%\left\{ (UX)^{\dagger} A W \right\}_{k K} 
%\left\{ W ^{\dagger} A (UX) \right\}_{K l} 
%\left\{ (UX)^{\dagger} A W \right\}_{l K} 
%\left\{ W ^{\dagger} A (UX) \right\}_{K l} 
%
\nonumber \\ 
&+& 
\sum_{n} 
\sum_{k \neq l } 
\sum_{K} \sum_{m \neq k, l} 
\biggl[
\frac{ (ix)e_{Kn} }{ \Delta_{Kk} \Delta_{Kl} \Delta_{Km} } 
-
\frac{ 
\left\{ 
3 \Delta_{K}^2 - 2 \Delta_{K} \left( h_{k} + h_{l} + h_{m} \right) + \left( h_{k} h_{l}+ h_{l} h_{m} + h_{m} h_{k} \right)
\right\}
e_{Kn}
%e^{- i \Delta_{K} x} 
}{ \Delta_{Kk}^2 \Delta_{Kl}^2 \Delta_{Km}^2 } 
%
%\nonumber \\
%&\times&
%
\nonumber \\
&+& 
\frac{ 1 }{ \Delta_{Km}^2 \Delta_{mk} \Delta_{ml} } 
e_{mn}
+ 
\frac{ 1 }{ \Delta_{Kl}^2 \Delta_{lm} \Delta_{lk} } 
e_{ln}
- 
\frac{ 1 }{ \Delta_{Kk}^2 \Delta_{km} \Delta_{lk} } 
e_{kn}
\biggr]
\nonumber \\ 
&\times&
(UX)_{\alpha k} (UX)^*_{\beta l} 
(UX)^*_{\alpha n} (UX)_{\beta n} 
c_{kK} c_{mK}^* c_{mK} c_{lK}^*
%\left\{ (UX)^{\dagger} A W \right\}_{k K} 
%\left\{ W ^{\dagger} A (UX) \right\}_{K m} 
%\left\{ (UX)^{\dagger} A W \right\}_{m K} 
%\left\{ W ^{\dagger} A (UX) \right\}_{K l} 
\biggr\}. 
\label{P-beta-alpha-W4-H4-single}
\end{eqnarray}
\begin{eqnarray} 
&& P(\nu_\beta \rightarrow \nu_\alpha)^{(4)}_{\rm 4th-d} 
\equiv 
2 \mbox{Re} \left[ \left( 
S^{(0)}_{\alpha \beta} \right)^{*} 
S_{\alpha \beta}^{(4)} [4]_{ \text{ offdiag } } (\text{double})  
\right] 
%\red{1st and 2nd terms }
%
\nonumber \\
&=& 
2 \mbox{Re} 
\biggl\{
%%%%%%%%%%%%%%%%%%%%%%%%
\sum_{n} 
%e^{ +i h_{n} x}
\sum_{k \neq l } 
\sum_{K \neq L} 
\biggl[ 
%e^{- i h_{k} x} 
\frac{ (ix) }{ \Delta_{Kk} \Delta_{Lk} \Delta_{lk} } 
e_{kn}
-
\frac{ 1 }{ \Delta_{KL} \Delta_{Lk}^2 \Delta_{Ll} } 
e_{Ln}
+ \frac{ 1 }{ \Delta_{KL} \Delta_{Kk}^2 \Delta_{Kl} } 
e_{Kn}
\nonumber \\
&+& 
\frac{ 1 }{ \Delta_{Kl} \Delta_{Ll} \Delta_{lk}^2 } 
e_{ln}
%e^{- i h_{l} x} 
%
-
\frac{ 1 }{ \Delta_{Kk}^2 \Delta_{Lk}^2 \Delta_{lk}^2 } 
\biggl\{
3 h_{k}^2 - 2 h_{k} h_{l} + \left( h_{l} - 2 h_{k} \right)  (\Delta_{K} + \Delta_{L} ) + \Delta_{K} \Delta_{L} 
\biggr\}
e_{kn}
%e^{- i h_{k} x} 
\biggr] 
\nonumber \\ 
&\times&
(UX)_{\alpha k} (UX)^*_{\beta l} 
(UX)^*_{\alpha n} (UX)_{\beta n} 
c_{cK} c_{kK}^* c_{kL} c_{lL}^*
%\left\{ (UX)^{\dagger} A W \right\}_{k K} 
%\left\{ W ^{\dagger} A (UX) \right\}_{K k} 
%\left\{ (UX)^{\dagger} A W \right\}_{k L} 
%\left\{ W ^{\dagger} A (UX) \right\}_{L l} 
%
\nonumber \\
&+&
\sum_{n} 
%e^{ +i h_{n} x}
\sum_{k \neq l } 
\sum_{K \neq L} 
\biggl[
- 
%e^{- i h_{l} x} 
\frac{ (ix) }{ \Delta_{Kl} \Delta_{Ll} \Delta_{lk} } 
e_{ln}
+
\frac{ 1 }{ \Delta_{KL} \Delta_{Kk} \Delta_{Kl}^2 } 
e_{Kn}
- \frac{ 1 }{ \Delta_{KL} \Delta_{Lk} \Delta_{Ll}^2 } 
e_{Ln}
+
\frac{ 1 }{ \Delta_{Kk} \Delta_{Lk} \Delta_{lk}^2 } 
e_{kn}
%e^{- i h_{k} x} 
%
\nonumber \\
&-& 
\frac{ 1 }{ \Delta_{Kl}^2 \Delta_{Ll}^2 \Delta_{lk}^2 } 
\biggl\{
3 h_{l}^2 - 2 h_{k} h_{l} - \left( 2 h_{l} - h_{k} \right)  (\Delta_{K} + \Delta_{L} ) + \Delta_{K} \Delta_{L} 
\biggr\}
e_{ln}
%e^{- i h_{l} x} 
\biggr] 
\nonumber \\ 
&\times&
(UX)_{\alpha k} (UX)^*_{\beta l} 
(UX)^*_{\alpha n} (UX)_{\beta n} 
c_{kK} c_{lK}^* c_{lL} c_{lL}^*
%\left\{ (UX)^{\dagger} A W \right\}_{k K} 
%\left\{ W ^{\dagger} A (UX) \right\}_{K l} 
%\left\{ (UX)^{\dagger} A W \right\}_{l L} 
%\left\{ W ^{\dagger} A (UX) \right\}_{L l} 
%
\nonumber \\
&+& 
\sum_{n} 
%e^{ +i h_{n} x}
\sum_{k \neq l } 
\sum_{K \neq L} \sum_{m \neq k, l} 
\biggl[ 
\frac{
\Delta_{ml} \Delta_{Kl} \Delta_{Ll} 
e_{kn}
%e^{ - i h_{k} x} 
-
\Delta_{mk} \Delta_{Kk} \Delta_{Lk} 
e_{ln}
%e^{ - i h_{l} x} 
}{ \Delta_{mk} \Delta_{ml} \Delta_{lk} \Delta_{Kk} 
	\Delta_{Kl} \Delta_{Lk} \Delta_{Ll} } 
+\frac{1}{ \Delta_{mk} \Delta_{ml} \Delta_{Km} \Delta_{Lm} } 
e_{mn}
\nonumber \\
&+& 
%e^{ - i h_{m} x}
%
\frac{
\Delta_{Kk} \Delta_{Kl} \Delta_{Km} 
e_{Ln}
- \Delta_{Lk} \Delta_{Ll} \Delta_{Lm} 
e_{Kn}
}{ \Delta_{Kk} \Delta_{Kl} \Delta_{Km} \Delta_{Lk}  \Delta_{Ll} \Delta_{Lm} \Delta_{LK} } 
%
%\nonumber \\
%&\times&
\biggr]
\nonumber \\ 
&\times&
(UX)_{\alpha k} (UX)^*_{\beta l} 
(UX)^*_{\alpha n} (UX)_{\beta n} 
c_{kK} c_{mK}^* c_{mL} c_{lL}^*
%\left\{ (UX)^{\dagger} A W \right\}_{k K} 
%\left\{ W ^{\dagger} A (UX) \right\}_{K m} 
%\left\{ (UX)^{\dagger} A W \right\}_{m L} 
%\left\{ W ^{\dagger} A (UX) \right\}_{L l} 
\biggr\}.
\label{P-beta-alpha-W4-H4-double}
\end{eqnarray}
\begin{eqnarray} 
&& P(\nu_\beta \rightarrow \nu_\alpha)^{(4)}_{\rm 5th-1st} 
\equiv 
2 \mbox{Re} \left[ \left( 
S^{(0)}_{\alpha \beta} \right)^{*} 
S_{\alpha \beta}^{(4)} [3]_\text{First} 
\right] 
\nonumber \\
&=& 
2 \mbox{Re} 
\biggl\{
%%%%%%%%%%%%%%%%%%%%%%%%
- \sum_{n} 
%e^{ +i h_{n} x} 
\sum_{k L} 
\frac{ 1 }{ \Delta_{Lk} } 
\left[ (ix) e_{Ln} + \frac{ e_{Ln} - e_{kn} }{ \Delta_{Kk} } 
\right]
%
%\nonumber \\
%&\times&
%
(UX)_{\alpha k} W^*_{\beta L} 
(UX)^*_{\alpha n} (UX)_{\beta n} 
c_{kL} d_{LL}
%\left\{ (UX)^{\dagger} A W \right\}_{k L} 
%\left\{ W^{\dagger} A W \right\}_{L L} 
%
\nonumber \\
&+& 
\sum_{n} 
%e^{ +i h_{n} x} 
\sum_{k L} 
\sum_{K \neq L} 
\frac{
[
\Delta_{Kk} 
e_{Ln}
%e^{- i \Delta_{L} x} 
- \Delta_{Lk} - h_{k} 
e_{Kn}
%e^{- i \Delta_{K} x} 
- \Delta_{KL} 
e_{kn}
%e^{- i h_{k} x} 
]
}{ \Delta_{LK} \Delta_{Lk} \Delta_{Kk} } 
%
%
%\nonumber \\
%&\times&
(UX)_{\alpha k} W^*_{\beta L} 
(UX)^*_{\alpha n} (UX)_{\beta n} 
c_{kK} d_{KL}
%\left\{ (UX)^{\dagger} A W \right\}_{k K} 
%\left\{ W^{\dagger} A W \right\}_{K L} 
%
\nonumber \\ 
&-& 
\sum_{n} 
%e^{ +i h_{n} x} 
\sum_{k L} 
\frac{ 1 }{ \Delta_{Lk}^2 } 
\biggl[
(ix) \left( e_{kn} + e_{Ln} \right)
+ 2 \frac{e_{Ln} - e_{kn} }{ \Delta_{Lk}} 
\biggr] 
%
%\nonumber \\ 
%&\times& 
(UX)_{\alpha k} W^*_{\beta L} 
(UX)^*_{\alpha n} (UX)_{\beta n} 
c_{kL} c_{kL}^* c_{kL}
%\left\{ (UX)^{\dagger} A W \right\}_{k L} 
%\left\{ W^{\dagger} A (UX) \right\}_{L k} 
%\left\{ (UX)^{\dagger} A W \right\}_{k L} 
%
\nonumber \\ 
&+& 
\sum_{n} 
%e^{ +i h_{n} x} 
\sum_{k L} 
\sum_{m \neq k} 
\biggl[
- \frac{ (ix) e_{Ln} }{ \Delta_{Lk} \Delta_{Lm} } 
+ \frac{ 
\Delta_{Lm}^2 
e_{kn} 
%e^{- i h_{k} x} 
- \Delta_{Lk}^2  
e_{mn}
%e^{- i h_{m} x} 
+ \Delta_{km} ( h_{k} + h_{m} - 2 \Delta_{L} ) e_{Ln} 
}{ \Delta_{km} \Delta_{Lk}^2 \Delta_{Lm}^2 } \biggr]
\nonumber \\ 
&\times&
(UX)_{\alpha k} W^*_{\beta L} 
(UX)^*_{\alpha n} (UX)_{\beta n} 
c_{kL} c_{mL}^* c_{mL}
%\left\{ (UX)^{\dagger} A W \right\}_{k L} 
%\left\{ W^{\dagger} A (UX) \right\}_{L m} 
%\left\{ (UX)^{\dagger} A W \right\}_{m L} 
%
%\nonumber \\ 
%&+& 
+
\sum_{n} 
%e^{ +i h_{n} x} 
\sum_{k L} 
\sum_{K \neq L} 
\biggl[
- \frac{ (ix) e_{kn} }{ \Delta_{Lk} \Delta_{Kk} }   
+ \frac{ 1 }{ \Delta_{LK} \Delta_{Lk}^2 \Delta_{Kk}^2 }
\nonumber \\ 
&\times&
\biggl\{
\Delta_{Kk}^2 
e_{Ln} 
%e^{- i \Delta_{L} x} 
- \Delta_{Lk}^2  
e_{Kn}
%e^{- i \Delta_{K} x} 
+ \Delta_{LK} (\Delta_{L} + \Delta_{K} - 2 h_{k} ) 
e_{kn} 
%e^{- i h_{k} x} 
\biggr\}
\biggr]
%\nonumber \\ 
%&\times&
(UX)_{\alpha k} W^*_{\beta L} 
(UX)^*_{\alpha n} (UX)_{\beta n} 
c_{kK} c_{kK}^* c_{kL}
%\left\{ (UX)^{\dagger} A W \right\}_{k K} 
%\left\{ W^{\dagger} A (UX) \right\}_{K k} 
%\left\{ (UX)^{\dagger} A W \right\}_{k L} 
%
\nonumber \\ 
&+& 
\sum_{n} 
%e^{ +i h_{n} x} 
\sum_{k L} 
\sum_{K \neq L} 
\sum_{m \neq k} 
\frac{ 
[
\Delta_{mk} 
\{\Delta_{Kk} \Delta_{Km} 
e_{Ln} 
%e^{- i \Delta_{L} x} 
- \Delta_{Lk} \Delta_{Lm}  
e_{Kn}
%e^{- i \Delta_{K} x}
\}
-
\Delta_{LK}  
\{ \Delta_{Lm} \Delta_{Km}  
e_{kn} 
%e^{- i h_{k} x} 
- \Delta_{Lk} \Delta_{Kk} 
e_{mn} 
%e^{- i h_{m} x}
\}
]
}{ \Delta_{LK} \Delta_{mk} \Delta_{Lk} 
	\Delta_{Lm} \Delta_{Kk} \Delta_{Km} } 
%
%\nonumber \\ 
%&\times&
%
\nonumber \\ 
&\times&
(UX)_{\alpha k} W^*_{\beta L} 
(UX)^*_{\alpha n} (UX)_{\beta n} 
c_{kK} c_{mK}^* c_{mL}
%\left\{ (UX)^{\dagger} A W \right\}_{k K} 
%\left\{ W^{\dagger} A (UX) \right\}_{K m} 
%\left\{ (UX)^{\dagger} A W \right\}_{m L} 
\biggr\}. 
\label{P-beta-alpha-W4-H3-First} 
\end{eqnarray}
\begin{eqnarray} 
&& P(\nu_\beta \rightarrow \nu_\alpha)^{(4)}_{\rm 5th-2nd} 
\equiv 
2 \mbox{Re} \left[ \left( 
S^{(0)}_{\alpha \beta} \right)^{*} 
S_{\alpha \beta}^{(4)} [3]_\text{Second} 
\right] 
\nonumber \\
&=& 
2 \mbox{Re} 
\biggl\{
%%%%%%%%%%%%%%%%%%%%%%%%
- \sum_{n} 
%e^{ +i h_{n} x} 
\sum_{L k} 
\frac{ 1 }{ \Delta_{Lk}} 
\left[
(ix) e_{Ln} 
%e^{- i \Delta_{L} x} 
+ 
\frac{ e_{Ln} - e_{kn} }{ \Delta_{Lk} } 
\right]
%
%\nonumber \\
%&\times& 
%
W_{\alpha L} (UX)^*_{\beta k} 
(UX)^*_{\alpha n} (UX)_{\beta n} 
d_{LL} c_{kL}^*
%\left\{ W^{\dagger} A W \right\}_{L L} 
%\left\{ W^{\dagger} A (UX) \right\}_{L k} 
%
\nonumber \\
&+& 
\sum_{n} 
%e^{ +i h_{n} x} 
\sum_{L k} 
\sum_{K \neq L} 
\frac{
\Delta_{Kk}  e_{Ln} 
- \Delta_{Lk} e_{Kn} 
- \Delta_{KL}  e_{kn} 
}{ \Delta_{LK} \Delta_{Lk} \Delta_{Kk} } 
%
%%\biggl[ \Delta_{Kk}  e_{Ln} 
%e^{- i \Delta_{L} x} 
%%- \Delta_{Lk} e_{Kn} 
%e^{- i \Delta_{K} x} 
%%- \Delta_{KL}  e_{kn} 
%e^{- i h_{k} x} 
%%\biggr]
%
%\nonumber \\ 
%&\times&
W_{\alpha L} (UX)^*_{\beta k} 
(UX)^*_{\alpha n} (UX)_{\beta n} 
d_{LK} c_{kK}^*
%\left\{ W^{\dagger} A W \right\}_{L K}  
%\left\{ W^{\dagger} A (UX) \right\}_{K k} 
%
\nonumber \\ 
&-& 
\sum_{n} 
%e^{ +i h_{n} x} 
\sum_{L k} 
\frac{ 1 }{ \Delta_{Lk}^2 } 
\biggl[
(ix) \left( e_{kn} + e_{Ln}  \right)
+ 2 \frac{ e_{Ln} - e_{kn} }{ \Delta_{Lk} } 
\biggr]
%
%\nonumber \\ 
%&\times& 
W_{\alpha L} (UX)^*_{\beta k} 
(UX)^*_{\alpha n} (UX)_{\beta n} 
c_{kL}^* c_{kL} c_{kL}^*
%\left\{ W^{\dagger} A (UX) \right\}_{L k} 
%\left\{ (UX)^{\dagger} A W \right\}_{k L} 
%\left\{ W^{\dagger} A (UX) \right\}_{L k} 
%
\nonumber \\ 
&+& 
\sum_{n} 
%e^{ +i h_{n} x} 
\sum_{L k} 
\sum_{m \neq k} 
\biggl[
- \frac{ (ix) e_{Ln} }{ \Delta_{Lk} \Delta_{Lm} } 
+ \frac{ 1 }{ \Delta_{km} \Delta_{Lk}^2 \Delta_{Lm}^2 } 
%
%\nonumber \\ 
%&\times& 
\biggl\{
\Delta_{Lm}^2  
e_{kn} 
%e^{- i h_{k} x} 
- \Delta_{Lk}^2  
e_{mn} 
%e^{- i h_{m} x} 
+ \Delta_{km} ( h_{k} + h_{m} - 2 \Delta_{L} ) 
e_{Ln} 
\biggr\}
\biggr]
\nonumber \\ 
&\times& 
W_{\alpha L} (UX)^*_{\beta k} 
(UX)^*_{\alpha n} (UX)_{\beta n} 
c_{mL}^* c_{mL} c_{kL}^*
%\left\{ W^{\dagger} A (UX) \right\}_{L m} 
%\left\{ (UX)^{\dagger} A W \right\}_{m L} 
%\left\{ W^{\dagger} A (UX) \right\}_{L k} 
%
%\nonumber \\ 
%&+& 
+
\sum_{n} 
%e^{ +i h_{n} x} 
\sum_{L k} 
\sum_{K \neq L} 
\biggl[
- \frac{ (ix) e_{kn} }{ \Delta_{Lk} \Delta_{Kk} }  
+ \frac{ 1 }{ \Delta_{LK} \Delta_{Lk}^2 \Delta_{Kk}^2 }
\nonumber \\ 
&\times&
\biggl\{
\Delta_{Kk}^2  
e_{Ln} 
%e^{- i \Delta_{L} x} 
- \Delta_{Lk}^2 
e_{Kn} 
%e^{- i \Delta_{K} x} 
+ 
%e^{- i h_{k} x} 
\Delta_{LK} (\Delta_{L} + \Delta_{K} - 2 h_{k} ) 
e_{kn}
\biggr\}
\biggr]
%
%\nonumber \\ 
%&\times&
W_{\alpha L} (UX)^*_{\beta k} 
(UX)^*_{\alpha n} (UX)_{\beta n} 
c_{kL}^* c_{kK} c_{kK}^*
%\left\{ W^{\dagger} A (UX) \right\}_{L k} 
%\left\{ (UX)^{\dagger} A W \right\}_{k K} 
%\left\{ W^{\dagger} A (UX) \right\}_{K k} 
%
\nonumber \\ 
&+& 
\sum_{n} 
%e^{ +i h_{n} x} 
\sum_{L k} 
\sum_{K \neq L} 
\sum_{m \neq k} 
\frac{ 
[
\Delta_{mk} 
\{ \Delta_{Kk} \Delta_{Km} 
e_{Ln} 
%e^{- i \Delta_{L} x} 
- \Delta_{Lk} \Delta_{Lm}  
e_{Kn} 
%e^{- i \Delta_{K} x}
\}
-
\Delta_{LK}  
\{ \Delta_{Lm} \Delta_{Km}  
e_{kn}
%e^{- i h_{k} x} 
- \Delta_{Lk} \Delta_{Kk}  
e_{mn}
%e^{- i h_{m} x}
\}
]
}{ \Delta_{LK} \Delta_{mk} \Delta_{Lk}  
	\Delta_{Lm} \Delta_{Kk} \Delta_{Km} } 
%
%%\nonumber \\ 
%%&\times&
%%\biggl[
%%\Delta_{mk} 
%%\biggl\{ \Delta_{Kk} \Delta_{Km} 
%%e_{Ln} 
%e^{- i \Delta_{L} x} 
%%- \Delta_{Lk} \Delta_{Lm}  
%%e_{Kn} 
%e^{- i \Delta_{K} x}
%%\biggr\}
%
%%-
%%\Delta_{LK}  
%%\biggl\{ \Delta_{Lm} \Delta_{Km}  
%%e_{kn}
%e^{- i h_{k} x} 
%%- \Delta_{Lk} \Delta_{Kk}  
%%e_{mn}
%e^{- i h_{m} x}
%%\biggr\}
%%\biggr]
%
\nonumber \\ 
&\times&
W_{\alpha L} (UX)^*_{\beta k} 
(UX)^*_{\alpha n} (UX)_{\beta n} 
c_{mL}^* c_{mK} c_{kK}^*
%\left\{ W^{\dagger} A (UX) \right\}_{L m} 
%\left\{ (UX)^{\dagger} A W \right\}_{m K} 
%\left\{ W^{\dagger} A (UX) \right\}_{K k} 
\biggr\}. 
\label{P-beta-alpha-W4-H3-Second} 
\end{eqnarray}
\begin{eqnarray} 
&& P(\nu_\beta \rightarrow \nu_\alpha)^{(4)}_{\rm 6th} 
\equiv 
2 \mbox{Re} \left[ \left( 
S^{(0)}_{\alpha \beta} \right)^{*} 
S_{\alpha \beta}^{(4)} [2] 
\right]
\nonumber \\
&=& 
2 \mbox{Re} 
\biggl\{
%%%%%%%%%%%%%%%%%%%%%%%%
- \sum_{n} 
%e^{ +i h_{n} x} 
\sum_{K} \sum_{k} 
\frac{ 1 }{ \Delta_{Kk} } 
\left[ (ix) e_{Kn} 
%e^{- i \Delta_{K} x} 
+ 
\frac{e_{kn} - e_{Kn} }{ \Delta_{kK} } 
\right]
%
%\nonumber \\
%&\times& 
%
(UX)^*_{\alpha n} (UX)_{\beta n} 
W_{\alpha K} W^*_{\beta K} 
c_{kK}^* c_{kK}
%\left\{ W^{\dagger} A (UX) \right\}_{K k}  
%\left\{ (UX)^{\dagger} A W \right\}_{k K} 
%
\nonumber \\
&+& 
\sum_{n} 
%e^{ +i h_{n} x} 
\sum_{K \neq L} 
\sum_{k} 
\frac{ 
\Delta_{L} e_{Kn} 
%e^{- i \Delta_{K} x} 
- \Delta_{K} e_{Ln} 
%e^{- i \Delta_{L} x} 
+ \left( e_{Ln} - e_{Kn} \right) h_{k} 
-  \Delta_{LK} 
e_{kn} 
%e^{- i h_{k} x} 
 }{ \Delta_{LK} \Delta_{Lk} \Delta_{Kk} } 
\nonumber \\
&\times& 
(UX)^*_{\alpha n} (UX)_{\beta n} 
W_{\alpha K} W^*_{\beta L} c_{kK}^* c_{kL}
%\left\{ W^{\dagger} A (UX) \right\}_{K k} 
%\left\{ (UX)^{\dagger} A W \right\}_{k L} 
\biggr\}. 
\label{P-beta-alpha-W4-H2} 
\end{eqnarray}
%
%%%%%%%%%%%%%%%%%%%%%%
%
\section{A note on the parameter choice } 
\label{sec:parameter-choice}

To discuss $W$ correction and the probability leaking term we have to determine the $W$ matrix. Given the non-unitary $U$ matrix there is a way to construct the $W$ matrix. 
In general, it is given by 
\begin{equation} 
W = S \sqrt{w} R, 
\label{W-construction}
\end{equation}
where $S$ is a $3\times 3$ matrix which diagonalizes ${\bf 1}_{3\times 3} - UU^\dagger$, $w$ is diagonal matrix which consists of eigenvalues of ${\bf 1}_{3\times 3} - UU^\dagger$, and $R$ is an arbitrary $3\times N$ complex matrix obeying $RR^\dagger = \bf{1}_{3 \times 3}$. The construction makes sense for $N \geq 3$. Therefore, for a given $N$ there is a large arbitrariness on the choice of the $W$ matrix, and hence on the sizes of the $W$ corrections and $\mathcal{C}_{\alpha \beta}$. 

Lacking a guiding principle of how to choose the $R$ matrix in (\ref{W-construction}), we examine the cases with largest and smallest possible values of $\mathcal{C}_{\alpha \beta}$ for given values of unitarity violation $1 - \sum_{j=1}^3 |U_{\alpha j}|^2$ ($\alpha=e,\mu,\tau$). It is shown that in the $(3+N)$ model $\mathcal{C}_{\alpha \beta}$ is bounded from above and below as \cite{Fong:2016yyh} 
\begin{equation}
\frac{1}{N}
\biggl( 1 - \sum_{j=1}^3 |U_{\alpha j}|^2 \biggr) 
\biggl( 1 - \sum_{j=1}^3 |U_{\beta j}|^2 \biggr) 
\leq
\mathcal{C}_{\alpha \beta}
\leq
\biggl( 1 - \sum_{j=1}^3 |U_{\alpha j}|^2 \biggr)
\biggl( 1 - \sum_{j=1}^3 |U_{\beta j}|^2 \biggr).
\label{Cab-bound}
\end{equation}
In the $(3+1)$ model, the $W$ matrix elements are unique, with the upper and lower bound being equal. 
For the numbers given in section~\ref{sec:with-without-UV}, we have $W_{e4} = 0.141$, $W_{\mu 4} = 0.099$, and $W_{\tau 4} = 0.141$ assuming that they are real. Then, the leaking terms have the unique values, 
$\mathcal{C}_{e \mu}^{(N=1)} = 2 \times 10^{-4}$, $\mathcal{C}_{\mu \mu}^{(N=1)} = 9.6 \times 10^{-5}$, and $\mathcal{C}_{\tau \mu}^{(N=1)} = 9.5 \times 10^{-4}$.
The lower bound is realized in the ``universal scaling'' model described in appendix~\ref{sec:scaling-model}, which predicts $W_{\alpha J} = \frac{ 1 }{ \sqrt{N} } W_{\alpha 4}^{(N=1)}$ ($J=4,5, \cdot \cdot \cdot, 3+N$).\footnote{
%%%%%%%%%%%%%% footnote %%%%%%%%%%%%%%%%
This feature must be obvious if one goes back to the derivation of bound on $\mathcal{C}_{\alpha \beta}$ in \cite{Fong:2016yyh}. 
}
It is shown in appendix~\ref{sec:scaling-model} that under the assumption of equal sterile state masses the universal scaling model predicts the same $W^2$ correction terms as those of the $(3+1)$ model.

\section{Universal scaling model of $N$ sterile sector }
\label{sec:scaling-model}

Suppose that we obtain a particular parametrization of $U$ matrix by taking $N=1$ sterile sector, as we did in section~\ref{sec:with-without-UV}. 
In this $(3+1)$ model, the $W$ matrix elements are completely determined, up to phase, by unitarity for a given $U$ matrix 
\begin{eqnarray}
\vert W_{\alpha 4} \vert^2 = 1 - \sum_{j=1}^3 |U_{\alpha j}|^2. 
\end{eqnarray}
Now, we attempt to create a toy model of $N$ sterile sector by ``universal scaling''. We postulate that all the $W$ matrix elements are real and equal:
\begin{eqnarray}
W_{\alpha 4} = W_{\alpha 5} = \cdot \cdot \cdot  W_{\alpha N+3} =
\frac{ 1 }{ \sqrt{N} }
\biggl( 1 - \sum_{j=1}^3 |U_{\alpha j}|^2 \biggr)^{1/2} ,
\end{eqnarray}
which is consistent with $(3+N)$ space unitarity. In this universal scaling model, the order $W^2$ correction terms in \eqref{W2_correction_terms} remains unchanged provided that we further assume that all the sterile masses are equal.\footnote{
%%%%%%%%%%%%%% footnote %%%%%%%%%%%%%%
This statement applies also to the original expression (\ref{P-beta-alpha-0th+2nd}). 
}
It is because the $W$ matrix elements enter into the $W^2$ terms in the form 
\begin{eqnarray}
\sum_{K} W_{\alpha K} W^{\dagger}_{K \beta} 
\frac{ 1 }{ ( \Delta_{K} - h_{k} )^n  }, 
\label{W2-scaling} 
\end{eqnarray}
where $n=1$ or 2.

However, the leaking term $\mathcal{C}_{\alpha \beta}$ becomes smaller by a factor of $N$ in the universal scaling model.
In the $(3+1)$ model, $\mathcal{C}_{\alpha \beta}$ takes the largest value, the upper limit in eq.~(\ref{Cab-bound}). Because $\mathcal{C}_{\alpha \beta}$ is fourth order in $W$ it is evident that in the universal scaling model, 
\begin{equation}
\mathcal{C}_{\alpha \beta} = 
\frac{1}{N}
\biggl( 1 - \sum_{j=1}^3 |U_{\alpha j}|^2 \biggr) 
\biggl( 1 - \sum_{j=1}^3 |U_{\beta j}|^2 \biggr), 
\label{Cab-USM}
\end{equation}
which is the lower limit of (\ref{Cab-bound}).


\begin{thebibliography}{99}

%\cite{Maki:1962mu}
\bibitem{Maki:1962mu}
  Z.~Maki, M.~Nakagawa and S.~Sakata,
``Remarks on the unified model of elementary particles,''
  Prog.\ Theor.\ Phys.\  {\bf 28} (1962) 870.
  doi:10.1143/PTP.28.870
  %%CITATION = doi:10.1143/PTP.28.870;%%

%\cite{Fukuda:1998mi}
\bibitem{Fukuda:1998mi}
  Y.~Fukuda {\it et al.} [Super-Kamiokande Collaboration],
 ``Evidence for oscillation of atmospheric neutrinos,''
  Phys.\ Rev.\ Lett.\  {\bf 81} (1998) 1562
  doi:10.1103/PhysRevLett.81.1562
  [hep-ex/9807003].
  %%CITATION = doi:10.1103/PhysRevLett.81.1562;%%

%\cite{Mikheev:1986gs}
\bibitem{Mikheev:1986gs}
  S.~P.~Mikheev and A.~Y.~Smirnov,
``Resonance Amplification of Oscillations in Matter and Spectroscopy of Solar Neutrinos,''
  Sov.\ J.\ Nucl.\ Phys.\  {\bf 42} (1985) 913
   [Yad.\ Fiz.\  {\bf 42} (1985) 1441].
  %%CITATION = SJNCA,42,913;%%

%\cite{Wolfenstein:1977ue}
\bibitem{Wolfenstein:1977ue}
  L.~Wolfenstein,
``Neutrino Oscillations in Matter,''
  Phys.\ Rev.\ D {\bf 17} (1978) 2369.
  doi:10.1103/PhysRevD.17.2369
  %%CITATION = doi:10.1103/PhysRevD.17.2369;%%

%\cite{Eguchi:2002dm}
\bibitem{Eguchi:2002dm}
  K.~Eguchi {\it et al.} [KamLAND Collaboration],
 ``First results from KamLAND: Evidence for reactor anti-neutrino disappearance,''
  Phys.\ Rev.\ Lett.\  {\bf 90} (2003) 021802
  doi:10.1103/PhysRevLett.90.021802
  [hep-ex/0212021].
  %%CITATION = doi:10.1103/PhysRevLett.90.021802;%%

%\cite{Ahmad:2002jz}
\bibitem{Ahmad:2002jz}
  Q.~R.~Ahmad {\it et al.} [SNO Collaboration],
 ``Direct evidence for neutrino flavor transformation from neutral current interactions in the Sudbury Neutrino Observatory,''
  Phys.\ Rev.\ Lett.\  {\bf 89} (2002) 011301
  doi:10.1103/PhysRevLett.89.011301
  [nucl-ex/0204008].
  %%CITATION = doi:10.1103/PhysRevLett.89.011301;%%

%\cite{Cleveland:1998nv}
\bibitem{Cleveland:1998nv}
  B.~T.~Cleveland, T.~Daily, R.~Davis, Jr., J.~R.~Distel, K.~Lande, C.~K.~Lee, P.~S.~Wildenhain and J.~Ullman,
 ``Measurement of the solar electron neutrino flux with the Homestake chlorine detector,''
  Astrophys.\ J.\  {\bf 496} (1998) 505.
  doi:10.1086/305343
  %%CITATION = doi:10.1086/305343;%%

%\cite{Hirata:1991ub}
\bibitem{Hirata:1991ub}
  K.~S.~Hirata {\it et al.} [Kamiokande-II Collaboration],
 ``Real time, directional measurement of B-8 solar neutrinos in the Kamiokande-II detector,''
  Phys.\ Rev.\ D {\bf 44} (1991) 2241
   Erratum: [Phys.\ Rev.\ D {\bf 45} (1992) 2170].
  doi:10.1103/PhysRevD.44.2241, 10.1103/PhysRevD.45.2170
  %%CITATION = doi:10.1103/PhysRevD.44.2241, 10.1103/PhysRevD.45.2170;%%

%\cite{Hampel:1998xg}
\bibitem{Hampel:1998xg}
  W.~Hampel {\it et al.} [GALLEX Collaboration],
 ``GALLEX solar neutrino observations: Results for GALLEX IV,''
  Phys.\ Lett.\ B {\bf 447} (1999) 127.
  doi:10.1016/S0370-2693(98)01579-2
  %%CITATION = doi:10.1016/S0370-2693(98)01579-2;%%

%\cite{Abdurashitov:2002nt}
\bibitem{Abdurashitov:2002nt}
  J.~N.~Abdurashitov {\it et al.} [SAGE Collaboration],
 ``Solar neutrino flux measurements by the Soviet-American Gallium Experiment (SAGE) for half the 22 year solar cycle,''
  J.\ Exp.\ Theor.\ Phys.\  {\bf 95} (2002) 181
   [Zh.\ Eksp.\ Teor.\ Fiz.\  {\bf 122} (2002) 211]
  doi:10.1134/1.1506424
  [astro-ph/0204245].
  %%CITATION = doi:10.1134/1.1506424;%%

%\cite{Fukuda:2001nj}
\bibitem{Fukuda:2001nj}
  S.~Fukuda {\it et al.} [Super-Kamiokande Collaboration],
 ``Solar B-8 and hep neutrino measurements from 1258 days of Super-Kamiokande data,''
  Phys.\ Rev.\ Lett.\  {\bf 86} (2001) 5651
  doi:10.1103/PhysRevLett.86.5651
  [hep-ex/0103032].
  %%CITATION = doi:10.1103/PhysRevLett.86.5651;%%

\bibitem{Aharmim:2011vm}
  B.~Aharmim {\it et al.} [SNO Collaboration],
``Combined Analysis of all Three Phases of Solar Neutrino Data from the Sudbury Neutrino Observatory,''
  Phys.\ Rev.\ C {\bf 88} (2013) 025501
  doi:10.1103/PhysRevC.88.025501
  [arXiv:1109.0763 [nucl-ex]].
  %%CITATION = doi:10.1103/PhysRevC.88.025501;%%

%\cite{An:2016ses}
\bibitem{An:2016ses}
  F.~P.~An {\it et al.} [Daya Bay Collaboration],
 ``Measurement of electron antineutrino oscillation based on 1230 days of operation of the Daya Bay experiment,''
  Phys.\ Rev.\ D {\bf 95} (2017) no.7,  072006
  doi:10.1103/PhysRevD.95.072006
  [arXiv:1610.04802 [hep-ex]].
  %%CITATION = doi:10.1103/PhysRevD.95.072006;%%

%\cite{RENO:2015ksa}
\bibitem{RENO:2015ksa}
  J.~H.~Choi {\it et al.} [RENO Collaboration],
``Observation of Energy and Baseline Dependent Reactor Antineutrino Disappearance in the RENO Experiment,''
  Phys.\ Rev.\ Lett.\  {\bf 116} (2016) no.21,  211801
  doi:10.1103/PhysRevLett.116.211801
  [arXiv:1511.05849 [hep-ex]].
  %%CITATION = doi:10.1103/PhysRevLett.116.211801;%%

%\cite{Schoppmann:2016iww}
\bibitem{Schoppmann:2016iww}
  S.~Schoppmann [Double Chooz Collaboration],
``Latest results of Double Chooz,''
  PoS HQL {\bf 2016} (2017) 010.
  %%CITATION = POSCI,HQL2016,010;%%


%\cite{Abe:2017vif}
\bibitem{Abe:2017vif}
 K.~Abe {\it et al.} [T2K Collaboration],
``Measurement of neutrino and antineutrino oscillations by the T2K experiment including a new additional sample of $\nu_e$ interactions at the far detector,''
  Phys.\ Rev.\ D {\bf 96} (2017) no.9,  092006
   Erratum: [Phys.\ Rev.\ D {\bf 98} (2018) no.1,  019902]
  doi:10.1103/PhysRevD.96.092006, 10.1103/PhysRevD.98.019902
  [arXiv:1707.01048 [hep-ex]].
  %%CITATION = doi:10.1103/PhysRevD.96.092006, 10.1103/PhysRevD.98.019902;%%


%\cite{Adamson:2016tbq}
\bibitem{Adamson:2016tbq}
  P.~Adamson {\it et al.} [NOvA Collaboration],
 ``First measurement of electron neutrino appearance in NOvA,''
  Phys.\ Rev.\ Lett.\  {\bf 116} (2016) no.15,  151806
  doi:10.1103/PhysRevLett.116.151806
  [arXiv:1601.05022 [hep-ex]].
  %%CITATION = doi:10.1103/PhysRevLett.116.151806;%%

\bibitem{Hartz-KEK-colloquium}
M.~Hartz (for the T2K Collaboration), 
``T2K Neutrino Oscillation Results with Data up to 2017 Summer'', 
KEK Colloquium, August 4, 2017. 

%\cite{Olive:2016xmw}
\bibitem{Olive:2016xmw}
  C.~Patrignani {\it et al.} [Particle Data Group],
 ``Review of Particle Physics,''
  Chin.\ Phys.\ C {\bf 40} (2016) no.10,  100001 and 2017 update.
  doi:10.1088/1674-1137/40/10/100001
  %%CITATION = doi:10.1088/1674-1137/40/10/100001;%%

%\cite{Fong:2016yyh}
\bibitem{Fong:2016yyh}
  C.~S.~Fong, H.~Minakata and H.~Nunokawa,
``A framework for testing leptonic unitarity by neutrino oscillation experiments,''
  JHEP {\bf 1702} (2017) 114
  doi:10.1007/JHEP02(2017)114
  [arXiv:1609.08623 [hep-ph]].
  %%CITATION = doi:10.1007/JHEP02(2017)114;%%

%%%% High-SCALE UV

%\cite{Antusch:2006vwa}
\bibitem{Antusch:2006vwa}
  S.~Antusch, C.~Biggio, E.~Fernandez-Martinez, M.~B.~Gavela and J.~Lopez-Pavon,
 ``Unitarity of the Leptonic Mixing Matrix,''
  JHEP {\bf 0610} (2006) 084
  doi:10.1088/1126-6708/2006/10/084
  [hep-ph/0607020].
  %%CITATION = doi:10.1088/1126-6708/2006/10/084;%%

%\cite{FernandezMartinez:2007ms}
\bibitem{FernandezMartinez:2007ms}
  E.~Fernandez-Martinez, M.~B.~Gavela, J.~Lopez-Pavon and O.~Yasuda,
``CP-violation from non-unitary leptonic mixing,''
  Phys.\ Lett.\ B {\bf 649} (2007) 427
  doi:10.1016/j.physletb.2007.03.069
  [hep-ph/0703098].
  %%CITATION = doi:10.1016/j.physletb.2007.03.069;%%

%\cite{Antusch:2009pm}
\bibitem{Antusch:2009pm}
  S.~Antusch, M.~Blennow, E.~Fernandez-Martinez and J.~Lopez-Pavon,
 ``Probing non-unitary mixing and CP-violation at a Neutrino Factory,''
  Phys.\ Rev.\ D {\bf 80} (2009) 033002
  doi:10.1103/PhysRevD.80.033002
  [arXiv:0903.3986 [hep-ph]].
  %%CITATION = doi:10.1103/PhysRevD.80.033002;%%

%\cite{Antusch:2009gn}
\bibitem{Antusch:2009gn}
  S.~Antusch, S.~Blanchet, M.~Blennow and E.~Fernandez-Martinez,
``Non-unitary Leptonic Mixing and Leptogenesis,''
  JHEP {\bf 1001} (2010) 017
  doi:10.1007/JHEP01(2010)017
  [arXiv:0910.5957 [hep-ph]].
  %%CITATION = doi:10.1007/JHEP01(2010)017;%%

%\cite{Antusch:2014woa}
\bibitem{Antusch:2014woa}
  S.~Antusch and O.~Fischer,
``Non-unitarity of the leptonic mixing matrix: Present bounds and future sensitivities,''
  JHEP {\bf 1410} (2014) 094
  doi:10.1007/JHEP10(2014)094
  [arXiv:1407.6607 [hep-ph]].
  %%CITATION = doi:10.1007/JHEP10(2014)094;%%

%\cite{Escrihuela:2015wra}
\bibitem{Escrihuela:2015wra}
  F.~J.~Escrihuela, D.~V.~Forero, O.~G.~Miranda, M.~Tórtola and J.~W.~F.~Valle,
``On the description of non-unitary neutrino mixing,''
  Phys.\ Rev.\ D {\bf 92} (2015) no.5,  053009
  doi:10.1103/PhysRevD.92.053009
  [arXiv:1503.08879 [hep-ph]].
  %%CITATION = doi:10.1103/PhysRevD.92.053009;%%

%\cite{Fernandez-Martinez:2016lgt}
\bibitem{Fernandez-Martinez:2016lgt}
  E.~Fernandez-Martinez, J.~Hernandez-Garcia and J.~Lopez-Pavon,
``Global constraints on heavy neutrino mixing,''
  JHEP {\bf 1608} (2016) 033
  doi:10.1007/JHEP08(2016)033
  [arXiv:1605.08774 [hep-ph]].
  %%CITATION = doi:10.1007/JHEP08(2016)033;%%

%\cite{Blennow:2016jkn}
\bibitem{Blennow:2016jkn}
  M.~Blennow, P.~Coloma, E.~Fernandez-Martinez, J.~Hernandez-Garcia and J.~Lopez-Pavon,
``Non-Unitarity, sterile neutrinos, and Non-Standard neutrino Interactions,''
  JHEP {\bf 1704} (2017) 153
  doi:10.1007/JHEP04(2017)153
  [arXiv:1609.08637 [hep-ph]].
  %%CITATION = doi:10.1007/JHEP04(2017)153;%%

%\cite{Escrihuela:2016ube}
\bibitem{Escrihuela:2016ube}
  F.~J.~Escrihuela, D.~V.~Forero, O.~G.~Miranda, M.~Tórtola and J.~W.~F.~Valle,
``Probing CP violation with non-unitary mixing in long-baseline neutrino oscillation experiments: DUNE as a case study,''
  New J.\ Phys.\  {\bf 19} (2017) no.9,  093005
  doi:10.1088/1367-2630/aa79ec
  [arXiv:1612.07377 [hep-ph]].
  %%CITATION = doi:10.1088/1367-2630/aa79ec;%%


%%% CP confusion 

%\cite{Klop:2014ima}
\bibitem{Klop:2014ima}
  N.~Klop and A.~Palazzo,
 ``Imprints of CP violation induced by sterile neutrinos in T2K data,''
  Phys.\ Rev.\ D {\bf 91} (2015) no.7,  073017
  doi:10.1103/PhysRevD.91.073017
  [arXiv:1412.7524 [hep-ph]].
  %%CITATION = doi:10.1103/PhysRevD.91.073017;%%

%\cite{Gandhi:2015xza}
\bibitem{Gandhi:2015xza}
  R.~Gandhi, B.~Kayser, M.~Masud and S.~Prakash,
``The impact of sterile neutrinos on CP measurements at long baselines,''
  JHEP {\bf 1511} (2015) 039
  doi:10.1007/JHEP11(2015)039
  [arXiv:1508.06275 [hep-ph]].
  %%CITATION = doi:10.1007/JHEP11(2015)039;%%

%\cite{Agarwalla:2016mrc}
\bibitem{Agarwalla:2016mrc}
  S.~K.~Agarwalla, S.~S.~Chatterjee, A.~Dasgupta and A.~Palazzo,
``Discovery Potential of T2K and NOvA in the Presence of a Light Sterile Neutrino,''
  JHEP {\bf 1602} (2016) 111
  doi:10.1007/JHEP02(2016)111
  [arXiv:1601.05995 [hep-ph]].
  %%CITATION = doi:10.1007/JHEP02(2016)111;%%

%\cite{Miranda:2016wdr}
\bibitem{Miranda:2016wdr}
  O.~G.~Miranda, M.~Tortola and J.~W.~F.~Valle,
 ``New ambiguity in probing CP violation in neutrino oscillations,''
  Phys.\ Rev.\ Lett.\  {\bf 117} (2016) no.6,  061804
  doi:10.1103/PhysRevLett.117.061804
  [arXiv:1604.05690 [hep-ph]].
  %%CITATION = doi:10.1103/PhysRevLett.117.061804;%%

%\cite{Ge:2016xya}
\bibitem{Ge:2016xya}
  S.~F.~Ge, P.~Pasquini, M.~Tortola and J.~W.~F.~Valle,
 ``Measuring the leptonic CP phase in neutrino oscillations with nonunitary mixing,''
  Phys.\ Rev.\ D {\bf 95} (2017) no.3,  033005
  doi:10.1103/PhysRevD.95.033005
  [arXiv:1605.01670 [hep-ph]].
  %%CITATION = doi:10.1103/PhysRevD.95.033005;%%

%\cite{Abe:2017jit}
\bibitem{Abe:2017jit}
  Y.~Abe, Y.~Asano, N.~Haba and T.~Yamada,
``Heavy neutrino mixing in the T2HK, the T2HKK and an extension of the T2HK with a detector at Oki Islands,''
  Eur.\ Phys.\ J.\ C {\bf 77} (2017) no.12,  851
  doi:10.1140/epjc/s10052-017-5294-7
  [arXiv:1705.03818 [hep-ph]].
  %%CITATION = doi:10.1140/epjc/s10052-017-5294-7;%%

%\cite{Dutta:2016vcc} 
\bibitem{Dutta:2016vcc}
  D.~Dutta and P.~Ghoshal,
 ``Probing CP violation with T2K, NO$\nu$A and DUNE in the presence of non-unitarity,''
  JHEP {\bf 1609} (2016) 110
  doi:10.1007/JHEP09(2016)110
  [arXiv:1607.02500 [hep-ph]].
  %%CITATION = doi:10.1007/JHEP09(2016)110;%%

%\cite{Dutta:2016czj}
\bibitem{Dutta:2016czj}
  D.~Dutta, P.~Ghoshal and S.~Roy,
 ``Effect of Non Unitarity on Neutrino Mass Hierarchy determination at DUNE, NO$\nu$A and T2K,''
  Nucl.\ Phys.\ B {\bf 920} (2017) 385
  doi:10.1016/j.nuclphysb.2017.04.018
  [arXiv:1609.07094 [hep-ph]].
  %%CITATION = doi:10.1016/j.nuclphysb.2017.04.018;%%

%\cite{Pas:2016qbg}
\bibitem{Pas:2016qbg}
  H.~Päs and P.~Sicking,
``Discriminating sterile neutrinos and unitarity violation with CP invariants,''
  Phys.\ Rev.\ D {\bf 95} (2017) no.7,  075004
  doi:10.1103/PhysRevD.95.075004
  [arXiv:1611.08450 [hep-ph]].
  %%CITATION = doi:10.1103/PhysRevD.95.075004;%%

%\cite{Rout:2017udo}
\bibitem{Rout:2017udo}
  J.~Rout, M.~Masud and P.~Mehta,
``Can we probe intrinsic CP and T violations and nonunitarity at long baseline accelerator experiments?,''
  Phys.\ Rev.\ D {\bf 95} (2017) no.7,  075035
  doi:10.1103/PhysRevD.95.075035
  [arXiv:1702.02163 [hep-ph]].
  %%CITATION = doi:10.1103/PhysRevD.95.075035;%%

%\cite{Nelson:2007yq}
\bibitem{Nelson:2007yq}
  A.~E.~Nelson and J.~Walsh,
``Short Baseline Neutrino Oscillations and a New Light Gauge Boson,''
  Phys.\ Rev.\ D {\bf 77} (2008) 033001
  doi:10.1103/PhysRevD.77.033001
  [arXiv:0711.1363 [hep-ph]].
  %%CITATION = doi:10.1103/PhysRevD.77.033001;%%

%\cite{Pospelov:2012gm}
\bibitem{Pospelov:2012gm}
  M.~Pospelov and J.~Pradler,
``Elastic scattering signals of solar neutrinos with enhanced baryonic currents,''
  Phys.\ Rev.\ D {\bf 85} (2012) 113016
   Erratum: [Phys.\ Rev.\ D {\bf 88} (2013) no.3,  039904]
  doi:10.1103/PhysRevD.85.113016, 10.1103/PhysRevD.88.039904
  [arXiv:1203.0545 [hep-ph]].
  %%CITATION = doi:10.1103/PhysRevD.85.113016, 10.1103/PhysRevD.88.039904;%%

%\cite{Harnik:2012ni}
\bibitem{Harnik:2012ni}
  R.~Harnik, J.~Kopp and P.~A.~N.~Machado,
``Exploring nu Signals in Dark Matter Detectors,''
  JCAP {\bf 1207} (2012) 026
  doi:10.1088/1475-7516/2012/07/026
  [arXiv:1202.6073 [hep-ph]].
  %%CITATION = doi:10.1088/1475-7516/2012/07/026;%%

\bibitem{JUNO} 
  F.~An {\it et al.} [JUNO Collaboration],
 ``Neutrino Physics with JUNO,''
  J.\ Phys.\ G {\bf 43}, no. 3, 030401 (2016)
  doi:10.1088/0954-3899/43/3/030401
  [arXiv:1507.05613 [physics.ins-det]].
  %%CITATION = doi:10.1088/0954-3899/43/3/030401;%%

%\cite{Parke:2015goa}
\bibitem{Parke:2015goa}
  S.~Parke and M.~Ross-Lonergan,
 ``Unitarity and the three flavor neutrino mixing matrix,''
  Phys.\ Rev.\ D {\bf 93} (2016) no.11,  113009
  doi:10.1103/PhysRevD.93.113009
  [arXiv:1508.05095 [hep-ph]].
  %%CITATION = doi:10.1103/PhysRevD.93.113009;%%

%\cite{Tang:2017khg}
\bibitem{Tang:2017khg}
  J.~Tang, Y.~Zhang and Y.~F.~Li,
 ``Probing Direct and Indirect Unitarity Violation in Future Accelerator Neutrino Facilities,''
  Phys.\ Lett.\ B {\bf 774} (2017) 217
  doi:10.1016/j.physletb.2017.09.055
  [arXiv:1708.04909 [hep-ph]].

%\cite{Abe:2017aap}
\bibitem{Abe:2017aap}
K.~Abe {\it et al.} [Super-Kamiokande Collaboration],
``Atmospheric neutrino oscillation analysis with external constraints in Super-Kamiokande I-IV,''
  Phys.\ Rev.\ D {\bf 97} (2018) no.7,  072001
  doi:10.1103/PhysRevD.97.072001
  [arXiv:1710.09126 [hep-ex]].
  %%CITATION = doi:10.1103/PhysRevD.97.072001;%%
  %18 citations counted in INSPIRE as of 06 Aug 2018

%\cite{Collaboration:2011ym}
\bibitem{Collaboration:2011ym}
  R.~Abbasi {\it et al.} [IceCube Collaboration],
 ``The Design and Performance of IceCube DeepCore,''
  Astropart.\ Phys.\  {\bf 35} (2012) 615
  doi:10.1016/j.astropartphys.2012.01.004
  [arXiv:1109.6096 [astro-ph.IM]].
  %%CITATION = doi:10.1016/j.astropartphys.2012.01.004;%%

%\cite{Abe:2015zbg}
\bibitem{Abe:2015zbg}
  K.~Abe {\it et al.} [Hyper-Kamiokande Proto- Collaboration],
 ``Physics potential of a long-baseline neutrino oscillation experiment using a J-PARC neutrino beam and Hyper-Kamiokande,''
  PTEP {\bf 2015} (2015) 053C02
  doi:10.1093/ptep/ptv061
  [arXiv:1502.05199 [hep-ex]].
  %%CITATION = doi:10.1093/ptep/ptv061;%%

%\cite{Abe:2016ero}
\bibitem{Abe:2016ero}
 K.~Abe {\it et al.} [Hyper-Kamiokande Collaboration],
``Physics potentials with the second Hyper-Kamiokande detector in Korea,''
  PTEP {\bf 2018} (2018) no.6,  063C01
  doi:10.1093/ptep/pty044
  [arXiv:1611.06118 [hep-ex]].
  %%CITATION = doi:10.1093/ptep/pty044;%%

%\cite{Acciarri:2015uup} 
\bibitem{Acciarri:2015uup}
  R.~Acciarri {\it et al.} [DUNE Collaboration],
 ``Long-Baseline Neutrino Facility (LBNF) and Deep Underground Neutrino Experiment (DUNE) : Volume 2: The Physics Program for DUNE at LBNF,''
  arXiv:1512.06148 [physics.ins-det].
  %%CITATION = ARXIV:1512.06148;%%

%\cite{TheIceCube-Gen2:2016cap}
\bibitem{TheIceCube-Gen2:2016cap}
  M.~G.~Aartsen {\it et al.} [IceCube Collaboration],
 ``PINGU: A Vision for Neutrino and Particle Physics at the South Pole,''
  J.\ Phys.\ G {\bf 44} (2017) no.5,  054006
  doi:10.1088/1361-6471/44/5/054006
  [arXiv:1607.02671 [hep-ex]].
  %%CITATION = doi:10.1088/1361-6471/44/5/054006;%%

%\cite{Adrian-Martinez:2016zzs}
\bibitem{Adrian-Martinez:2016zzs}
  S.~Adrián-Martínez {\it et al.},
``Intrinsic limits on resolutions in muon- and electron-neutrino charged-current events in the KM3NeT/ORCA detector,''
  JHEP {\bf 1705} (2017) 008
  doi:10.1007/JHEP05(2017)008
  [arXiv:1612.05621 [physics.ins-det]].
  %%CITATION = doi:10.1007/JHEP05(2017)008;%%

%\cite{Minakata:2015gra}
\bibitem{Minakata:2015gra}
  H.~Minakata and S.~J.~Parke,
 ``Simple and Compact Expressions for Neutrino Oscillation Probabilities in Matter,''
  JHEP {\bf 1601} (2016) 180
  doi:10.1007/JHEP01(2016)180
  [arXiv:1505.01826 [hep-ph]].
  %%CITATION = doi:10.1007/JHEP01(2016)180;%%

%\cite{Nunokawa:2003ep}
\bibitem{Nunokawa:2003ep}
  H.~Nunokawa, O.~L.~G.~Peres and R.~Zukanovich Funchal,
``Probing the LSND mass scale and four neutrino scenarios with a neutrino telescope,''
  Phys.\ Lett.\ B {\bf 562} (2003) 279
  doi:10.1016/S0370-2693(03)00603-8
  [hep-ph/0302039].
  %%CITATION = doi:10.1016/S0370-2693(03)00603-8;%%

%\cite{Kimura:2002wd}
\bibitem{Kimura:2002wd}
  K.~Kimura, A.~Takamura and H.~Yokomakura,
 ``Exact formulas and simple CP dependence of neutrino oscillation probabilities in matter with constant density,''
  Phys.\ Rev.\ D {\bf 66} (2002) 073005
  doi:10.1103/PhysRevD.66.073005
  [hep-ph/0205295].
  %%CITATION = doi:10.1103/PhysRevD.66.073005;%%

%\cite{Kopp:2013vaa}
\bibitem{Kopp:2013vaa}
  J.~Kopp, P.~A.~N.~Machado, M.~Maltoni and T.~Schwetz,
``Sterile Neutrino Oscillations: The Global Picture,''
  JHEP {\bf 1305} (2013) 050
  doi:10.1007/JHEP05(2013)050
  [arXiv:1303.3011 [hep-ph]].
  %%CITATION = doi:10.1007/JHEP05(2013)050;%%

%\cite{deGouvea:2015euy}
\bibitem{deGouvea:2015euy} 
  A.~de Gouvêa and A.~Kobach,
``Global Constraints on a Heavy Neutrino,''
  Phys.\ Rev.\ D {\bf 93}, no. 3, 033005 (2016)
  doi:10.1103/PhysRevD.93.033005
  [arXiv:1511.00683 [hep-ph]].
  %%CITATION = doi:10.1103/PhysRevD.93.033005;%%

%\cite{TheIceCube:2016oqi}
\bibitem{TheIceCube:2016oqi}
  M.~G.~Aartsen {\it et al.} [IceCube Collaboration],
``Searches for Sterile Neutrinos with the IceCube Detector,''
  Phys.\ Rev.\ Lett.\  {\bf 117} (2016) no.7,  071801
  doi:10.1103/PhysRevLett.117.071801
  [arXiv:1605.01990 [hep-ex]].
  %%CITATION = doi:10.1103/PhysRevLett.117.071801;%%


%\cite{Xu:2015kma}
\bibitem{Xu:2015kma}
  X.~J.~Xu,
``Why is the neutrino oscillation formula expanded in $\Delta m^2_{21} / \Delta m^2_{31}$ still accurate near the solar resonance in matter?,''
  JHEP {\bf 1510} (2015) 090
  doi:10.1007/JHEP10(2015)090
  [arXiv:1502.02503 [hep-ph]].
  %%CITATION = doi:10.1007/JHEP10(2015)090;%%

%\cite{Ge:2016dlx}
\bibitem{Ge:2016dlx}
  S.~F.~Ge and A.~Y.~Smirnov,
``Non-standard interactions and the CP phase measurements in neutrino oscillations at low energies,''
  JHEP {\bf 1610} (2016) 138
  doi:10.1007/JHEP10(2016)138
  [arXiv:1607.08513 [hep-ph]].
  %%CITATION = doi:10.1007/JHEP10(2016)138;%%

%\cite{Asano:2011nj}
\bibitem{Asano:2011nj}
  K.~Asano and H.~Minakata,
``Large-Theta(13) Perturbation Theory of Neutrino Oscillation for Long-Baseline Experiments,''
  JHEP {\bf 1106} (2011) 022
  doi:10.1007/JHEP06(2011)022
  [arXiv:1103.4387 [hep-ph]].
  %%CITATION = doi:10.1007/JHEP06(2011)022;%%

\bibitem{Conrad:2013mka} 
  J.~M.~Conrad, W.~C.~Louis and M.~H.~Shaevitz,
 ``The LSND and MiniBooNE Oscillation Searches at High $\Delta m^2$,''
  Ann.\ Rev.\ Nucl.\ Part.\ Sci.\  {\bf 63}, 45 (2013)
  doi:10.1146/annurev-nucl-102711-094957
  [arXiv:1306.6494 [hep-ex]].
  %%CITATION = doi:10.1146/annurev-nucl-102711-094957;%%

%\cite{Yachay-story}
\bibitem{Yachay-story}
Emiliano Rodríguez Mega, 
``Plans for a research powerhouse in the Andes begin to unravel'', 
  Science (2017)
doi:10.1126/science.aan7140. 


\end{thebibliography}
\end{document}